\documentclass[prb,eqsecnum,superscriptaddress,twocolumn]{revtex4}
\usepackage{ifthen}
\newcommand{\usepdfs}{false}
\ifthenelse{\equal{\usepdfs}{true}}{
  \usepackage[pdftex]{graphicx}
}
{
  \usepackage[dvips]{graphicx}
}
\ifthenelse{\equal{\usepdfs}{true}}{
  \newcommand{\ifig}[2][]{\includegraphics[#1]{#2.pdf}}
}
{
  \newcommand{\ifig}[2][]{\includegraphics[#1]{#2.eps}}
}
\usepackage{latexsym}
\usepackage{amsmath}
\preprint{cond-mat/0408329}
\begin{document}
\title{Putting competing orders in their place near the Mott transition}

\author{Leon Balents}
\affiliation{Department of Physics, University of California,
Santa Barbara, CA 93106-4030}

\author{Lorenz Bartosch}
\affiliation{Department of Physics, Yale University, P.O. Box
208120, New Haven, CT 06520-8120}

\affiliation{Institut f\"ur Theoretische Physik, Universit\"at
Frankfurt, Postfach 111932, 60054 Frankfurt, Germany}

\author{Anton Burkov}
\affiliation{Department of Physics, University of California,
Santa Barbara, CA 93106-4030}

\author{Subir Sachdev}
\affiliation{Department of Physics, Yale University, P.O. Box
208120, New Haven, CT 06520-8120}

\author{Krishnendu Sengupta}
\affiliation{Department of Physics, Yale University, P.O. Box
208120, New Haven, CT 06520-8120}

\date{August 13, 2004} 

\begin{abstract}
  We describe the localization transition of superfluids on
  two-dimensional lattices into commensurate Mott insulators with
  average particle density $p/q$ ($p$, $q$ relatively prime integers)
  per lattice site. For bosons on the square lattice, we argue that
  the superfluid has at least $q$ degenerate species of vortices which
  transform under a projective representation of the square lattice
  space group (a PSG). The formation of a single vortex condensate
  produces the Mott insulator, which is required by the PSG to have
  density wave order at wavelengths of $q/n$ lattice sites ($n$
  integer) along the principle axes; such a second-order transition is
  forbidden in the Landau-Ginzburg-Wilson framework. We also discuss
  the superfluid-insulator transition in the direct boson
  representation, and find that an interpretation of the quantum
  criticality in terms of deconfined fractionalized bosons is only
  permitted at special values of $q$ for which a {\em permutative\/}
  representation of the PSG exists. We argue (and demonstrate in
  detail in a companion paper: L. Balents {\em et al.}, cond-mat/0409470)
  that our results apply essentially unchanged to electronic
  systems with short-range pairing, with the PSG determined by the
  particle density of Cooper pairs. We also describe the effect of
  static impurities in the superfluid: the impurities locally break
  the degeneracy between the $q$ vortex species, and this induces
  density wave order near each vortex. We suggest that such a theory
  offers an appealing
  rationale for the local density of states modulations observed by
  Hoffman {\em et al\/}, Science {\bf 295}, 466 (2002), in scanning
  tunnelling microscopy (STM) studies of the vortex lattice of
  Bi$_2$Sr$_2$CaCu$_2$O$_{8+\delta}$, and allows a unified description
  of the nucleation of density wave order in zero and finite magnetic
  fields. We note signatures of our theory that may be tested by
  future STM experiments.

\end{abstract}

\maketitle

\section{Introduction}
\label{sec:intro}

One of the central debates in the study of cuprate
superconductivity is on the nature of the electronic correlations
in the `underdoped' materials. At these low hole densities, the
superconductivity is weak and appears only below a low critical
temperature ($T_c$), if at all. Nevertheless, there is no evidence
of metallic behavior and an underlying Fermi surface of long-lived
quasiparticles, as would be expected in a conventional BCS theory.
This absence of Fermi liquid physics, and the many unexplained
experimental observations, constitute key open puzzles of the
field.

In very general and most basic terms, the passage from optimal
doping to the underdoped region may be characterized by an
evolution from conducting (even superconducting) to insulating
behavior.  Indeed, most thinking on the cuprates is informed by
their proximity to a Mott insulating state, in which
interaction-induced carrier localization is the primary driving
influence.  Many recent attempts to describe the underdoped region
have, by contrast, focused on the presence or absence of
conventional orders (magnetism, stripe, superconductivity, etc.)
as a means of characterization.  This focus has practical merit,
such orders being directly measured by existing experimental
probes. Theoretically, however, the immense panoply of competing
potential orders reduces the predictive power of this approach,
which moreover obscures the basic conducting to insulating
transition in action in these materials.  In this paper, we
describe a theoretical approach in which Mott localization is
``back in the driver's seat''.  Less colloquially, we characterize
the behavior near to quantum critical points (QCPs) whose {\sl
dual} ``order parameter'' primarily describes the emergence of an
insulating state from a superconducting one. Remarkably, a proper
quantum mechanical treatment of this Mott transition leads
naturally to the appearance of competing {\em conventional\/}
orders. The manner in which this occurs is described in detail
below.

Among recent experiments, two classes are especially noteworthy
for the issues to be discussed in our paper. Measurements of
thermoelectric transport in a number of cuprates \cite{nernst}
have uncovered a large Nernst response for a significant range of
temperatures above $T_c$. This response is much larger than would
be expected in a Fermi liquid description. However, a model of
vortex fluctuations over a background of local superfluid order
does appear to lead to a satisfactory description of the data
\cite{nernst}. These experiments suggest a description of the
underdoped state in terms of a fluctuating superconducting order
parameter, $\Psi_{sc}$. Numerous such theories \cite{phase} have
been considered in the literature, under guises such as
``preformed pairs'' and ``phase fluctuations''. Intriguing new
evidence in support of such a ``bosonic Cooper pair'' approach has
been presented in recent work by Kapitulnik and collaborators
\cite{aharon}.

On the other hand, a somewhat different view of the underdoped
state has emerged from a second class of experiments. Scanning
tunnelling microscopy (STM) studies
\cite{krmp,fang,ali,mcelroy,hanaguri} of
Bi$_2$Sr$_2$CaCu$_2$O$_{8+\delta}$ and
Ca$_{2-x}$Na$_x$CuO$_2$Cl$_2$ show periodic modulations in the
electronic density of states, suggesting the appearance of a state
with density wave order. Such density wave order can gap out
portions of the Fermi surface, and so could be responsible for
onset of a spin-gap observed in the underdoped regime. However,
the strength of the density wave modulations is quite weak, and so
it is implausible that they produce the large spin gap. The
precise microscopic nature of the density wave modulations also
remains unclear, and they could represent spatial variations in
the local charge density, spin exchange energy, and pairing or
hopping amplitudes. Using symmetry considerations alone, all of
these modulations are equivalent, because they are associated with
observables which are invariant under time-reversal and spin
rotations. It is useful, therefore, to discuss a generic density
wave order, and to represent it by density wave order parameters
$\rho_{{\bf Q}}$ which determine the modulations in the `density'
by
\begin{equation}
\delta \rho ({\bf r}) = \sum_{{\bf Q}} \rho_{{\bf Q}} e^{i {\bf Q}
\cdot {\bf r}} \label{e1}
\end{equation}
where ${\bf Q} \neq 0$ extends over some set of wavevectors.

We will argue here that the apparent conflict between these two
classes of experiments is neatly resolved by our approach. The
strong vortex fluctuations suggest we examine a theory of
superfluidity in the vicinity of a Mott transition. We will show
that in a quantum theory of such a transition, which carefully
accounts for the complex Berry phase terms, fluctuations of
density wave order emerge naturally, and can become visible after
pinning by impurities.

\begin{figure}
\centering
\ifig[width=3in]{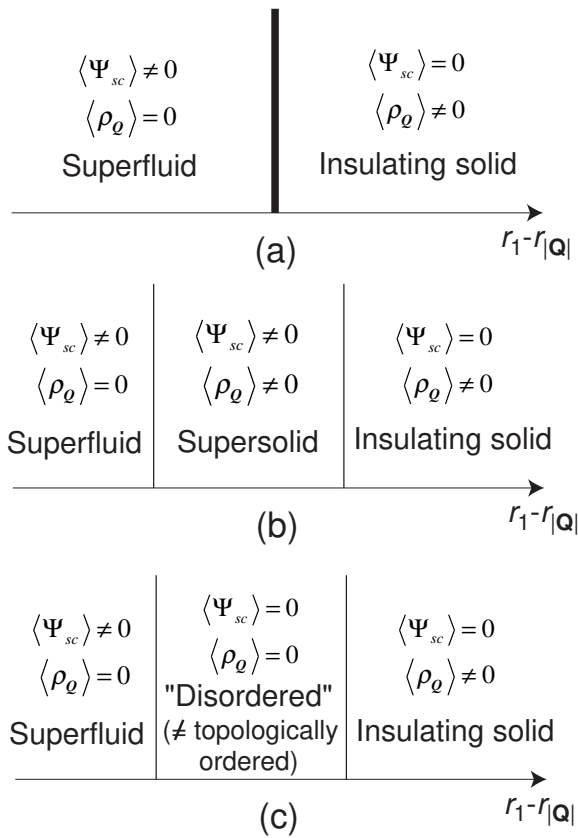}
\caption{LGW phase diagrams of the competition between the
superconducting and density wave orders. The choice between the
cases (a), (b), and (c) is made by the sign of $v_{|{\bf Q}|}$ and
by the values of the higher order couplings in (\protect\ref{e5}).
The thick line is a first order transition, and the thin lines are
second order transitions. The ``disordered'' phase of LGW theory
preserves all symmetries but is not the `topologically ordered'
phase which also preserves all symmetries: LGW theory does not
predict the degeneracies of the latter state on topologically
non-trivial spatial manifolds. It is believed there is no such
``disordered'' quantum state in the underlying quantum system at
$T=0$.} \label{liu}
\end{figure}

Within the conventional Landau-Ginzburg-Wilson (LGW) theory
\cite{fisherliu,sd}, with the order parameters $\Psi_{sc}$ and
$\rho_{{\bf Q}}$ at hand, a natural next step is to couple them to
each other. Here, one begins by writing down the most general
effective free energy consistent with the underlying symmetries. A
key point is that $\Psi_{sc}$ and $\rho_{{\bf Q}}$ have
non-trivial transformations under entirely distinct symmetries.
The density wave order transforms under space group operations
{\em e.g.\/} $T_{{\bf a}}$, translation by the lattice vector
${\bf a}$
\begin{equation}
T_{{\bf a}}: \rho_{{\bf Q}} \rightarrow \rho_{{\bf Q}} e^{i {\bf
Q} \cdot {\bf a}}~~;~~\Psi_{sc} \rightarrow \Psi_{sc}, \label{e2}
\end{equation}
while $\Psi_{sc}$ transforms under the electromagnetic gauge
transformation, $G$,
\begin{equation}
G: \rho_{{\bf Q}} \rightarrow \rho_{{\bf Q}} ~~;~~\Psi_{sc}
\rightarrow \Psi_{sc} e^{i \theta}. \label{e3}
\end{equation}
We could combine the two competing orders in a common `superspin'
order parameter (as in the `SO(5)' theory \cite{so5}), but neither
of the symmetry operations above rotate between the two orders,
and so there is no a priori motivation to use such a language. The
transformations in Eqs.~(\ref{e2}) and (\ref{e3}) place important
constraints on the LGW functional of $\Psi_{sc}$ and $\rho_{{\bf
Q}}$, which can be written as
\begin{equation}
\mathcal{F} = \mathcal{F}_1 \left[ \Psi_{sc} \right] +
\mathcal{F}_2 \left[ \rho_{{\bf Q}} \right] + \mathcal{F}_3 \left[
\Psi_{sc},\rho_{{\bf Q}} \right]. \label{e4}
\end{equation}
Here $\mathcal{F}_1$ is an arbitrary functional of $\Psi_{sc}$
containing only terms which are invariant under Eq.~(\ref{e3}),
while $\mathcal{F}_2$ is a function of the $\rho_{{\bf Q}}$
invariant under space group operations like Eq.~(\ref{e2}). These
order parameters are coupled by $\mathcal{F}_3$, and the symmetry
considerations above imply that $\mathcal{F}_3$ only contains
terms which are products of terms already contained in
$\mathcal{F}_1$ and $\mathcal{F}_2$. In other words, only the
energy densities of the two order parameters couple to each other,
and neither order has an information on the local `orientation' of
the other order.  The crucial coupling between the orders is
actually in an oscillatory Berry phase term, but this leaves no
residual contribution in the continuum limit implicit in the LGW
analysis. To leading order in the order parameters, the terms in
Eq.~(\ref{e4}) are
\begin{eqnarray}
\mathcal{F}_1 \left[ \Psi_{sc} \right]  &=& r_1 |\Psi_{sc}|^2 +
\ldots \nonumber \\
 \mathcal{F}_2 \left[ \rho_{{\bf Q}} \right] &=& \sum_{{\bf Q}}
r_{|{\bf Q}|} | \rho_{{\bf Q}} |^2 + \ldots \nonumber \\
\mathcal{F}_3 \left[ \Psi_{sc},\rho_{{\bf Q}} \right] &=&
\sum_{{\bf Q}} v_{|{\bf Q}|} |\Psi_{sc}|^2 | \rho_{{\bf Q}} |^2 +
\ldots . \label{e5}
\end{eqnarray}
The phase diagrams implied by the LGW functionals Eqs.~(\ref{e4})
and (\ref{e5}) were determined by Liu and Fisher \cite{fisherliu},
and are summarized here in Fig~\ref{liu}.
The transition between a state with only superconducting order
($\left\langle \Psi_{sc} \right\rangle \neq 0$, $\left\langle
\rho_{{\bf Q}} \right\rangle = 0$) and a state with only density
wave order ($\left\langle \Psi_{sc} \right\rangle = 0$,
$\left\langle \rho_{{\bf Q}} \right\rangle \neq 0$) can follow one
of 3 possible routes: (a) via a direct first order transition; (b)
via an intermediate supersolid phase with both order parameters
non-zero, which can be bounded by two second order transitions;
and (c) via an intermediate ``disordered'' phase with both order
parameters zero, which can also be bounded by two second order
transitions. An important prediction of LGW theory is that there
is generically no direct second order transition between the
flanking states of Fig~\ref{liu}: such a situation requires
fine-tuning of at least one additional parameter.

While the predictions of LGW theory in Fig~\ref{liu} are
appropriate for classical phase transitions at nonzero
temperatures, they develop a crucial shortcoming upon a na\"ive
extension to quantum phase transitions at zero temperature.
Importantly, the intermediate ``disordered'' phase in
Fig~\ref{liu} has no clear physical interpretation. At the level
of symmetry, LGW theory informs us that this intermediate phase
fully preserves electromagnetic gauge invariance and the space
group symmetry of the lattice. Thinking quantum mechanically, this
is not sufficient to specify the wavefunction of such a state. We
could guess that this ``disordered'' state is a metallic Fermi
liquid, which does preserve the needed symmetries. However, a
Fermi liquid has a large density of states of low energy
excitations associated with the Fermi surface, and these surely
have to be accounted for near the quantum phase transitions to the
other phases; clearly, such excitations are not included in the
LGW framework. At a more sophisticated level, a variety of
insulating states which preserve all symmetries of the Hamiltonian
have been proposed in recent years \cite{topo,lfs,sp}, but these
have a subtle `topological' order which is not contained in the
LGW framework. Such topologically ordered states have global (and
possibly local) low energy excitations and these have to be
properly accounted for near quantum critical points \cite{deccp};
again, such excitations are entirely absent in the LGW framework.
Thus the purported existence of a truly featureless intermediate
``disordered'' state in Fig~\ref{liu}c suggests serious
shortcomings of the LGW description of multiple order parameters
in quantum systems\cite{deccp,rs,lfs,sp}.

Conceptually, it is useful to emphasize that the LGW approach to
the conducting--insulating transition is quite the reverse of the
Mott picture.  Indeed, the above symmetry considerations do not at
all refer to transport, or to localization of the carriers. To
address such properties, a conventional application of the LGW
method would be to introduce the $\rho_{\bf Q}$ and $\Psi_{sc}$
order parameters by a mean-field decoupling of a quantum model of
bosons\cite{BoseMott} or fermions\cite{HertzMillis}.  These order
parameters, once established, appear then as perturbations to the
``quasiparticles'' in the decoupled model.  If $\rho_{\bf Q}$
becomes strong enough, the coincident potential barriers can
localize the carriers.  Thus in the LGW approach the Mott
transition eventually arises out of the competing order, not
vice-versa.

In this paper, we will present an alternative approach to studying
the interplay between the superfluid and density wave ordering in
two spatial dimensions. Our analysis is based upon the recent work
of Senthil {\em et al.}  \cite{deccp} (and its
precedents\cite{rs,lfs,sp}), although we will present a different
perspective. Also, the previous work \cite{deccp} was focused on
systems at half-filling, while our presentation is applicable to
lattice systems at arbitrary densities. One of the primary
benefits of our approach is that the formalism naturally excludes
the featureless `disordered' states that invariably appear in the
more familiar LGW framework. Our theory also makes it evident that
the experimental observations of `phase' fluctuations and pinned
density wave order noted earlier are closely tied to each other,
and, in principle, allows a compact, unified description of these
experiments and also of the pinned density wave order around
vortices in the superconductor \cite{hoffman}.

We will present our results here in the context of boson systems:
specifically, in a general model of bosons hopping on a square lattice
with short-range interactions. However, all the results of this paper
apply essentially unchanged to electronic systems with short-range
pairing.  This is clearly the case for a toy model in which electrons
experience a strong attractive interaction in the $s$-wave (or
$d$-wave, though this is less obvious) channel, causing them to form
tightly bound ``molecular'' Cooper pairs.  Below the pair binding
energy -- i.e. the gap to unbound electron excitations -- a
description of the physics in terms of bosonic Cooper pairs clearly
applies within such a model.  Since we discuss {\sl universal} physics
near a bosonic superfluid to charge ordered transition, the theory
here clearly describes such a transition from a short-range paired
(also called the ``strong pairing'' regime/phase) superconductor to a
density-wave ordered phase of {\sl any} model, provided the electronic
gap is maintained throughout the critical region.  As we discuss
below, a key parameter determining the character of the theories
presented here is the number $f$, which for the lattice boson models
is the average density of bosons per site of the square lattice in the
Mott insulating state. Clearly, for electrons on the lattice with
filling $1-\delta$, i.e. hole density $\delta$ relative to
half-filling, the density of bosonic Cooper pairs is
\begin{equation}
f = \frac{1-\delta}{2}. \label{eq:fdelta}
\end{equation}
We note in passing that the density of carriers in the superfluid
proximate to the Mott insulator is allowed to be different from that
of the Mott insulator, as elaborated in Section~\ref{sec:irrat}.

These features will be explicitly demonstrated in a companion
paper \cite{psgdimers}, hereafter referred to as II. We argue in
II that a convenient phenomenological model for capturing the spin
$S=0$ sector of electrons near a superfluid-insulator transition
is the doped quantum dimer model of the cuprate superconductors
proposed by Fradkin and Kivelson \cite{fradkiv}. At zero hole
concentration, the quantum dimer model \cite{rk} generically has a
Mott insulating ground state with valence bond solid (VBS) order
\cite{ssdimer,rs} (this order constitutes a particular realization
of the generalized density wave order discussed above). Above some
finite hole density, the dimer model has a superconducting ground
state in which neutral observables preserve all lattice
symmetries. (As we will discuss in II, the electronic pairing in
this superconductor could have a $d$- or $s$- (or other) wave
character, depending upon microscopic details not contained in the
dimer model). The evolution of the phase diagram between the
insulating VBS state and the superconductor can occur via many
possible sequences of transitions, all of which will be shown to
be equivalent to those in the simpler boson models discussed in
the present paper.  A duality analysis of the doped dimer model
presented in II directly gives the expected relation,
Eq.~(\ref{eq:fdelta}).  The paper II will also extend the dimer
model to include fermionic $S=1/2$ excitations: this theory will
be shown to have a close connection to other\cite{su2} U(1) or
SU(2) gauge-theoretic ``slave-particle'' approaches to cuprate
physics.  Alternative formulations appropriate to {\sl long-range}
paired superconducting states (i.e. those with gapless nodal
quasiparticle excitations) will also be discussed in II.

Let us turn, then, to a model of bosons hopping on a square lattice,
with short-range interactions,\footnote{Long range Coulomb
  interactions between Cooper pair bosons can be readily included, and
  do not significantly modify the present discussion.}  and at a
density $f$ per site of the square lattice. Our primary technical tool
will be a dual description of this boson system using vortex degrees
of freedom \cite{dh,nelson,fisherlee}. Briefly, the world lines of the
bosons in three-dimensional spacetime are reinterpreted as the
trajectories of vortices in a dual, classical, three-dimensional
`superconductor' with a dual `magnetic' field oriented along the
`time' direction. The dual superconducting order parameter, $\psi$, is
the creation and annihilation operator for vortices in the original
boson variables. The density of the vortices in $\psi$ should equal
the density of the bosons, and hence the dual `magnetic' field acting
on the $\psi$ has a strength of $f$ flux quanta per unit cell of the
dual square lattice.

It is valuable to characterize the phases in Fig~\ref{liu} in
terms of the dual field $\psi$. It is easy to determine the
presence/absence of superfluid order (which is associated with the
presence/absence of a condensate in $\Psi_{sc}$). This has a dual
relationship to the dual `superconducting' order and is therefore
linked to the absence/presence of a condensate in $\psi$:
\begin{eqnarray}
\mbox{Superfluid} &:& \langle \psi \rangle = 0 \nonumber \\
\mbox{Insulator} &:& \langle \psi \rangle \neq 0. \label{e6}
\end{eqnarray}
A condensate in $\psi$ implies a proliferation of vortices in
$\Psi_{sc}$, and hence the loss of superfluid order. Strictly
speaking, the loss of superfluidity does not necessarily require
the appearance of a condensate of elementary vortices. It is
sufficient that at least a composite of $n$ vortices condense,
with $\langle \psi^n \rangle \neq 0$, where $n$ is a positive
integer. The insulating states with $n > 1$ and $\langle \psi
\rangle = 0$ have topological order \cite{topo}. For simplicity,
we will not consider such insulating states here, although it is
not difficult to extend our formalism to include such states and
the associated multi-vortex condensates.

The characterization of the density wave order in terms of the
vortex field, $\psi$, is more subtle, and a key ingredient in our
analysis. We have noted that the field $\psi$ experiences a dual
magnetic field of strength $f$ flux quanta per unit cell.
Consequently, just as is familiar from the Hofstadter problem of
electrons moving in a crystal lattice in the presence of a
magnetic field \cite{zak,hofstadter}, the $\psi$ Hamiltonian
transforms under a {\em projective\/} representation of the space
group \cite{hamermesh,wen} or a PSG. A central defining property
of a projective representation is that the operations $T_x$, $T_y$
of translation by one lattice site along the $x$, $y$ directions
do not commute (as they do in any faithful representation of the
space group), but instead obey
\begin{equation}
T_x T_y = \omega T_y T_x , \label{e7}
\end{equation}
where
\begin{equation}
\omega \equiv e^{2 \pi i f} \label{e8} \:.
\end{equation}
The PSG also contains elements corresponding to all other members of
the square lattice space group. Among these are $R_{\pi/2}^{\rm dual}$
(rotation by angle $\pi/2$ about a site of the dual lattice {\em
  i.e.\/} the sites of the lattice upon which the vortex field $\psi$
resides) and $I_{x}^{\rm dual},I_{y}^{\rm dual}$ (reflections about
the $x,y$ axes of the dual lattice); these obey\cite{arbfoot} group
multiplication laws identical to those in the ordinary space group,
{\em e.g.}
\begin{eqnarray}
 T_x R_{\pi/2}^{\rm dual} &=& R^{\rm dual}_{\pi/2} T_y^{-1}  \nonumber \\
 T_y R_{\pi/2}^{\rm dual}
&=& R_{\pi/2}^{\rm dual} T_x \nonumber \\
\left( R_{\pi/2}^{\rm dual} \right)^4 &=& 1 \;. \label{rtr}
\end{eqnarray}
All the elements of the PSG are obtained by taking arbitrary
products of the elements already specified.

A useful and explicit
description of this projective representation is obtained by
focusing on the low energy and long wavelength fluctuations of
$\psi$ at boson density
\begin{equation}
f = \frac{p}{q} \label{e9}
\end{equation}
where $p$ and $q$ are relatively prime integers. We assume that
the average density in the Mott insulator has the precise
commensurate value in Eq.~(\ref{e9}). In most of the paper we will
also assume that the superfluid has the density value in
Eq.~(\ref{e9}), but it is not difficult to extend our analysis to
allow the density of the superfluid (or supersolid) to stray from
commensurate values to arbitrary incommensurate values: this will
be discussed in Section~\ref{sec:irrat}. As we will review in
Section~\ref{sec:boson}, the spectrum of $\psi$ fluctuations in
the phase with $\langle \psi \rangle=0$ has $q$ distinct minima in
the magnetic Brillouin zone. We can use these minima to define $q$
complex fields, $\varphi_{\ell}$, $\ell = 0, 1, \ldots,q-1$ which
control the low energy physics. These $q$ vortex fields play a
central role in all our analyses, as they allow us to efficiently
characterize the presence or absence of superfluid and/or density
wave order. In a convenient Landau gauge, the $q$ vortex fields
can be chosen to transform under $T_x$ and $T_y$ as
\begin{eqnarray}
T_x &:& \varphi_{\ell} \rightarrow \varphi_{\ell+1} \nonumber \\
T_y &:& \varphi_{\ell} \rightarrow \varphi_\ell \omega^{-\ell}.
\label{txty}
\end{eqnarray}
Here, and henceforth, the arithmetic of all indices of the
$\varphi_\ell$ fields is carried out modulo $q$, {\em e.g.\/}
$\varphi_q \equiv \varphi_0$. The action of all the PSG elements
on the $\varphi_{\ell}$ will be specified in
Section~\ref{sec:field}; here we also note the important
transformation under $R_{\pi/2}^{\rm dual}$
\begin{equation}
 R_{\pi/2}^{\rm dual} : \varphi_{\ell} \rightarrow \frac{1}{\sqrt{q}}
\sum_{m=0}^{q-1} \varphi_m \omega^{-m \ell}, \label{rpi}
\end{equation}
which is a Fourier transform in the space of the $q$ fields. It is
instructive to verify that Eqs.~(\ref{txty}) and (\ref{rpi}) obey
Eqs.~(\ref{e7}) and (\ref{rtr}).

The existence of $q$ degenerate species of vortices in the
superfluid is crucial to all the considerations of this paper. At
first sight, one might imagine that quantum tunnelling between the
different vortex species should lift the degeneracy, and we should
only focus on the lowest energy linear combination of vortices.
However, while such quantum tunnelling may lead to a change in the
preferred basis, it does {\em not\/} lift the $q$-fold degeneracy.
The degeneracy is imposed by the structure of the PSG, in
particular the relation Eq.~(\ref{e7}), and the fact that the
minimum dimension of a PSG representation is $q$.

Using the transformations in Eqs.~(\ref{txty}) and (\ref{rpi}), we
can now easily construct the required density wave order
parameters $\rho_{{\bf Q}}$ as the most general bilinear,
gauge-invariant combinations of the $\varphi_{\ell}$ with the
appropriate transformation properties under the square lattice
space group. The density wave order parameters appear only at the
wavevectors
\begin{equation}
{\bf Q}_{mn} = 2 \pi f (m,n), \label{e10}
\end{equation}
where $m,n$ are integers, and take the form
\begin{equation}
\rho_{mn} \equiv \rho_{{\bf Q}_{mn}} = S\left(|{\bf Q}_{mn}|\right
) \omega^{mn/2} \sum_{\ell = 0}^{q-1} \varphi^{\ast}_\ell
\varphi_{\ell+n} \omega^{\ell m}. \label{e11}
\end{equation}
Here $S(Q)$ is a general `form-factor' which cannot be determined
from symmetry considerations, and has a smooth $Q$ dependence
determined by microscopic details and the precise definition of
the density operator. It is easy to verify that $\rho_{mn}^{\ast}
= \rho_{-m,-n}$, and from Eqs.~(\ref{txty}) and (\ref{rpi}) that
the space group operations act on $\rho_{mn}$ just as expected for
a density wave order parameter
\begin{eqnarray}
T_x &:& \rho_{mn} \rightarrow \omega^{-m} \rho_{mn} \nonumber
 \\
T_y &:& \rho_{mn} \rightarrow \omega^{-n} \rho_{mn} \nonumber
 \\
R_{\pi/2}^{\rm dual} &:& \rho_{mn} \rightarrow \rho_{-n,m}.
\label{e12}
\end{eqnarray}
So by studying the bilinear combinations in Eq.~(\ref{e11}), we
can easily characterize the density wave order in terms of the
$\varphi_{\ell}$. It is worth reiterating here that the
$\rho_{mn}$ order parameters in Eq.~(\ref{e11}) are generic
density wave operators, and could, in principle, be adapted to
obtain the actual density of either the bosons or the vortices.
Indeed, as we noted above Eq.~(\ref{e1}), we can obtain
information on arbitrary observables invariant under spin
rotations and time reversal, including the local density of
states. The different specific observables will differ only in
their values of the form factor $S(Q)$, which we do not specify in
the present paper. Note also that because the $R_{\pi/2}^{\rm
dual}$ rotation is about a dual lattice site, the last relation in
Eq.~(\ref{e12}) implies that the Fourier components $\rho_{{\bf
Q}}$ are defined by taking the origin of the spatial co-ordinates
on a dual lattice site. Upon performing the inverse Fourier
transform from the $\rho_{mn}$ to real space (as in Eq.~(\ref{m1})
below) the resulting density gives a measure (up to a factor of
the unknown $S(Q)$) of the vortex density on the dual lattice
sites, and of the boson density on the direct lattice
sites.\footnote{The vortex degrees of freedom are defined only on
the dual lattice sites, and the reader may wonder how we are able
to obtain information on direct lattice sites in such a facile
manner. The resolution is to note that Eq.~(\ref{e11}) is a
bilinear in the vortices. Therefore, by taking bilinear products
of vortex operators on different sites around a single dual
lattice plaquette, we can construct a measure of ``density'' on
the direct lattice site on its center. We can similarly construct
measures of bond variables such as the exchange or pairing energy.
The advantage of the formalism in the paper is that all of this is
automatically performed simply by studying the PSG properties of
the observables, and then by evaluating Eq.~(\ref{m1}) at the
appropriate spatial position.}

For completeness, we note that the characterizations in
Eq.~(\ref{e6}) of the superfluid order in terms of the vortex
fields has an obvious expression in terms of the $\varphi_{\ell}$:
\begin{eqnarray}
\mbox{Superfluid} &:& \langle \varphi_{\ell} \rangle = 0
\mbox{~for every $\ell$} \nonumber \\
\mbox{Insulator} &:& \langle \varphi_{\ell} \rangle \neq 0
\mbox{~for at least one $\ell$}. \label{e13}
\end{eqnarray}

\begin{figure}
\centering \ifig[width=3in]{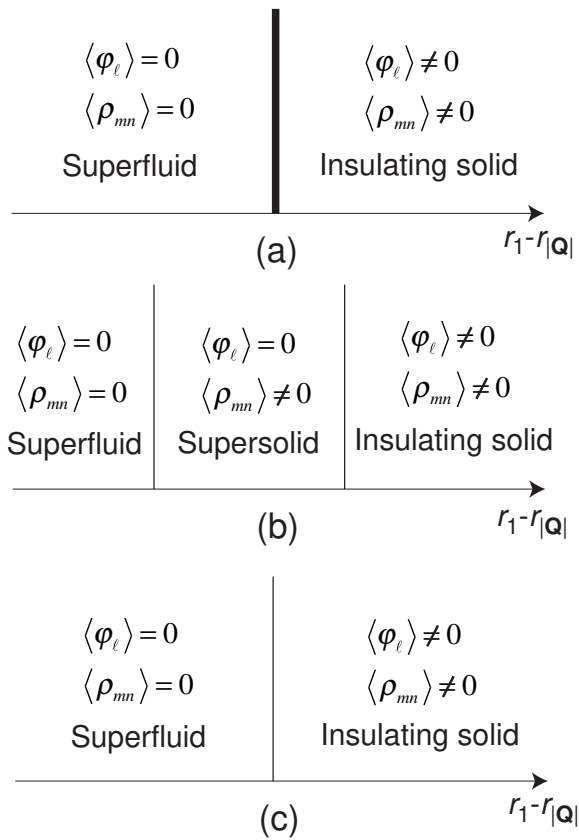} \caption{Possible phase
diagrams in the vortex theory of the interplay of superfluid and
density wave order. Conventions are as in Fig~\ref{liu}. Notice
that the objectionable ``disordered'' phase of the LGW theory has
disappeared, and is replaced by a direct second order transition.
Also, the first order transition in (a) is  ``fluctuation
induced'', {\em i.e.\/} unlike the LGW theory of Fig~\ref{liu},
the mean field vortex theory only predicts a second order
transition, but fluctuations could induce a weak first order
transition.} \label{vortex_pdiag}
\end{figure}

The key relations in Eqs.~(\ref{e11}) and (\ref{e13}) allow us to
obtain an essentially complete characterization of all the phases of
the boson model of interest here: this is a major reason for
expressing the theory in terms of the vortex fields $\varphi_{\ell}$.
Furthermore, the symmetry relations in Eqs.~(\ref{txty}) and
(\ref{rpi}) impose strong constraints on the effective theory for the
$\varphi_{\ell}$ fields whose consequences we will explore in
Section~\ref{sec:cft}. Indeed, even before embarking upon a study of
this effective theory, we can already see that our present theory
overcomes one of the key shortcomings of the LGW approach. We note
from Eq.~(\ref{e13}) that the non-superfluid phase has a condensate of
one or more of the $\varphi_{\ell}$. Inserting these condensates into
Eq.~(\ref{e11}) we observe that $\langle \rho_{mn} \rangle \neq 0$ for
at least one non-zero ${\bf Q}_{mn}$ {\em i.e.} the non-superfluid
insulator has density wave order. Consequently the intermediate
``disordered'' phase in Fig~\ref{liu}c has been automatically excluded
from the present theory. Instead, if the $\varphi_{\ell}$ condensate
appears in a second-order transition (something allowed by the general
structure of the theory), we have the possibility of a generic
second-order transition from a superfluid phase to an insulator with
density wave order, a situation which was forbidden by the LGW
theory.\footnote{We caution that although such continuous transitions
  are allowed in principle, the actual critical theories in question
  are modified by fluctuation effects in $2+1$ dimensions.  We know of
  no reliable analytic means of addressing the properties and
  stability of the strongly-coupled critical theories in question.
  Indeed, both $1/N$ and $\epsilon=3-d$ expansion techniques are known
  to give incorrect results for the few examples of similar theories
  that can be addressed by other means (duality mappings and numerical
  simulation).  These latter methods show that stable critical field
  theories exist for $q=1,2$, at least in a range of parameter space.
  Hence we expect that some but not all of the continuous transitions
  predicted at the mean field (in the $\varphi_\ell$ fields) level
  will remain generically continuous beyond mean field theory, while
  others may be driven first order by fluctuation effects.  This is an
  important -- and vast -- open problem in critical phenomena which we
  will not address.}  We summarize the possible phase diagrams of the
$\varphi_\ell$ theory in Fig~\ref{vortex_pdiag}.

In addition to characterizing modulations in the ``density'' in
terms of the $\rho_{mn}$, the $\varphi_\ell$ fields can also
determine modulations in the local {\em vorticity}. This is a
measure of circulating boson currents around plaquettes of the
direct lattice \cite{ivanov}, and has zero average in all phases
because time-reversal invariance is preserved. However, in the
presence of an applied magnetic field, this constraint no longer
applies. The magnetic field, of course, induces a net circulation
of boson current, and the total number of vortices (or
anti-vortices) is non-zero. However, using a reasoning similar to
that applied above to the density, the PSG of the vortices implies
there are also modulations in the vorticity at the ${\bf Q}_{mn}$
wavevectors. We can define a corresponding set of $V_{mn}$ which
are Fourier components of the vorticity at these wavevectors.
Details of this analysis, and explicit expressions for the
$V_{mn}$ appear in Appendix~\ref{app:vorticity}.

An explicit derivation of the above vortex theory of the Mott
transition of the superfluid appear in Section~\ref{sec:boson}
where we consider a model of bosons on the square lattice. Here we
derive an effective action for the $q$ vortex species in
Eqs.~(\ref{s0}) and (\ref{r1}) which is invariant under the PSG,
and which controls the phases and phase diagrams outlined above.

A complementary perspective is presented in Section~\ref{sec:frac}
where we formulate the physics using the direct boson representation.
In particular, following Senthil {\em et al.}  \cite{deccp}, we
explore the possibility that the LGW forbidden superfluid-insulator
transitions discussed above are associated with the boson
fractionalization. The structure of the theory naturally suggests a
fractionalization of each boson into $q$ components, each with boson
number $1/q$. However, the PSG places strong restrictions on the
fractionalized boson theory. On the rectangular lattice, such a theory
is consistent for all $q$ but only for certain insulating phases
obtained for restricted parameter values. On the square lattice, with
the additional $R_{\pi/2}^{\rm dual}$ element of the PSG, a
fractionalized boson theory is permitted only for a limited set of $q$
values.  Specifically, it is required that the $\varphi_{\ell}$ vortex
fields transform under a {\em permutative\/} representation of the
PSG.  We define a permutative representation as one in which all group
elements can be written as $\Lambda P$, where $\Lambda$ is a unitary
diagonal matrix ({\em i.e.\/} a matrix whose only non-zero elements
are complex numbers of unit magnitude on the diagonal), and P is a
permutation matrix. For the representation defined in
Eqs.~(\ref{txty}) and (\ref{rpi}), $T_x$ and $T_y$ are of this form,
while $R_{\pi/2}^{\rm dual}$ is not.  When a permutative
representation exists, it is possible to globally unitarily transform
the $\varphi_\ell$ fields to a new basis of $\zeta_\ell$ fields, which
realize the permutative representation.  For some range of parameters
-- values of the quantum tunnelling terms between the vortex flavors
-- one can see that the PSG is expected to be elevated at the critical
point to include an emergent $U(1)^q$ (modulo the global $U(1)$ gauge
invariance) symmetry of {\sl independent} rotations of the
$\zeta_\ell$ fields.  This emergent symmetry is the hallmark of
fractionalization, and corresponds to emergence of conserved $U(1)$
gauge fluxes in the direct representation.  Such an emergent symmetry
is clearly not generic, and this turns out to place quite a restrictive
condition on the existence of a fractionalized $1/q$ boson
representation, as we discuss in detail in Section~\ref{sec:frac}, and
in a number of appendices.

Section~\ref{sec:imp} applies our theory to the STM observations
of local density of states modulations in zero and non-zero
magnetic field. A virtue of our approach is that it naturally
connects density wave order with vorticity and allows a unified
description of the experiments in zero and non-zero magnetic
field. As we will discuss, there are several intriguing
consequences of our approach, and we will analyze prospects for
observing these in future STM experiments in
Sections~\ref{sec:imp} and~\ref{sec:conc}.

We briefly mention other recent works which have addressed related
issues. Zaanen {\em et al.}\cite{zaanen} have examined the
superfluid to insulator transition from a complementary
perspective: they focus on the dislocation defects of a particular
insulating solid in the continuum, in contrast to our focus here
on the vortex defects of the superfluid. We do consider the
`melting' of defects in the solid in
Section~\ref{sec:appr-fract-from-1}, but the underlying lattice
plays a crucial role in our considerations. In a work which
appeared while our analysis was substantially complete, Te\v
sanovi\'c \cite{zlatko} has applied the boson-vortex duality to
Cooper pairs and considered the properties of vortices in a dual
`magnetic' field.

\section{Dual vortex theory of bosons on the square lattice}
\label{sec:boson}

We consider an ordinary single-species boson model on the square
lattice. The bosons are represented by rotor operators
$\hat{\phi}_i$ and conjugate number operators $\hat{n}_i$ where
$i$ runs over the sites of the direct square lattice. These
operators obey the commutation relation
\begin{equation}
[\hat{\phi}_i, \hat{n}_j] = i \delta_{ij}.
\end{equation}

A simple boson Hamiltonian in the class of interest has the
structure
\begin{eqnarray}
\mathcal{H} &=& -t \sum_{i\alpha} \cos \left(\Delta_{\alpha}
\hat{\phi}_i - 2 \pi g_{i\alpha} \right) + \sum_i V(\hat{n}_i)
\nonumber \\
&~&~~~~~~~~+ \sum_{i\neq j} \Lambda_{ij} \hat{n}_i \hat{n}_j +
\ldots \label{hubbard}
\end{eqnarray}
where the interaction $V(\hat{n})$ has the on-site terms
$V(\hat{n}) = -\bar \mu \hat{n} + U \hat{n} ( \hat{n} -1)/2$. The
$\Lambda_{ij}$ are repulsive off-site interactions, and can also
include the long-range Coulomb interaction. We will focus on the
case of short-range $\Lambda_{ij}$, and note the minor
modifications necessary for the long-range case. The general
structure of our theory also permits a variety of exchange and
ring-exchange terms, such as those in the studies of Sandvik {\em
et al.} \cite{sandvik}. These off-site or ring exchange couplings
are essential for our analysis, as they are needed to stabilize
insulating phases of the bosons away from integer filling.
Nevertheless, many aspects of our results are independent upon the
particular form of these couplings; their specific form will only
influence the numerical values of the non-linear couplings that
appear in our phenomenological actions (such as those in
Eq.~(\ref{r1})). The index $\alpha$ extends over the spatial
directions $x$, $y$, while we will use indices $\mu$, $\nu$,
$\lambda$ to extend over all three spacetime directions $x$, $y$,
$\tau$. The symbol $\Delta_{\alpha}$ is a discrete lattice
derivative along the $\alpha$ direction: $\Delta_\alpha
\hat{\phi}_i = \hat{\phi}_{i+\alpha} - \hat{\phi}$ (and similarly
for $\Delta_\mu$). We have also included a static external
magnetic field represented by the vector potential $g_{i\alpha}$
for convenience. This is a uniform field which obeys
\begin{equation}
\epsilon_{\mu\nu\lambda} \Delta_{\nu} g_{i\lambda} = h \delta_{\mu
\tau} \label{curlg}
\end{equation}
where $h$ is the strength of the physical magnetic field (which
should be distinguished from the dual ``magnetic'' flux $f$
discussed in Section~\ref{sec:intro}).

\subsection{Dual lattice representation}
\label{sec:dual}

We proceed with a standard duality mapping, following the methods
of Refs.~\onlinecite{dh,nelson,fisherlee} and the notational
conventions of Ref.~\onlinecite{curreac}. We represent the
partition function as Feynman integral over states at a large
number of intermediate time slices, separated by the interval
$\Delta \tau$. The intermediate states use a basis of $\hat{n}_i$
and $\hat{\phi}_i$ at alternate times. The hopping term in
$\mathcal{H}$ acts between $\hat{\phi}_i$ eigenstates, and we
evaluate its matrix elements by using the Villain representation
\begin{eqnarray}
&& \exp\left(t \Delta \tau \cos \left( \Delta_{\alpha}
\hat{\phi}_{i} - 2 \pi g_{i\alpha} \right)\right) \nonumber \\
&& \!\!\!\!\!\!\!\!\!\!\!\!\! \rightarrow \sum_{\{J_{i\alpha}\}}
\exp \left( - \frac{J_{i\alpha}^2}{2 t \Delta \tau} + i J_{i
\alpha} \Delta_{\alpha} \hat{\phi}_{i} - 2\pi i J_{i \alpha} g_{i
\alpha} \right). \label{f4}
\end{eqnarray}
We have dropped an unimportant overall normalization constant, and
will do so below without comment. The $J_{i \alpha}$ are integer
variables residing on the links of the direct lattice,
representing the current of the bosons.

After integrating over the $\phi_i$ on all sites and at all
intermediate times, the partition function becomes
\begin{eqnarray}
\mathcal{Z} &=& \sum_{\{J_{i\mu}\}} \exp \Biggl( - \frac{1}{2e^2}
\sum_{i} \left(J_{i \mu} - H \delta_{\mu\tau} \right)^2 \nonumber
\\ &~&~~-\Delta \tau \sum_{i\neq j} \Lambda_{ij} J_{i\tau} J_{j\tau}
- 2\pi i g_{i \mu} J_{i \mu} \Biggr) \nonumber \\
&~&~~~~~~~~~~~~~~\times\prod_i \delta \left( \Delta_\mu J_{i\mu}
\right) \label{zh}
\end{eqnarray}
where $J_{i\mu} \equiv (n_i, J_{ix}, J_{iy})$ is the
integer-valued boson current in spacetime, $i$ now extends over
the sites of the cubic lattice, and we have chosen $\Delta \tau$
so that $e^2 = t \Delta \tau = 1/U \Delta \tau$, and $H = \bar
\mu/U + 1/2$. We now solve the constraint in Eq.~(\ref{zh}) by
writing
\begin{equation}
J_{i\mu} = \epsilon_{\mu\nu\lambda} \Delta_\nu A_{a \lambda}
\end{equation}
where $a$ labels sites on the dual lattice, and $A_{a \lambda}$ is
an integer-valued gauge field on the links of the dual lattice. We
promote $A_{a\mu}$ from an integer-valued field to a real field by
the Poisson summation method, while ``softening'' the integer
constraint with a fugacity $y_v$. It is convenient to make the
gauge invariance of the dual theory explicit by introducing an
angular field $\vartheta_a$ on the sites of the dual lattice, and
mapping $2 \pi A_{a \mu } \rightarrow 2 \pi A_{a \mu} -
\Delta_{\mu} \vartheta_a$. The operator $e^{i \vartheta_a}$ is
then the creation operator for a vortex in the boson phase
variable $\phi_i$. These transformations yield the dual partition
function
\begin{eqnarray}
 \mathcal{Z}_d = && \prod_a \int d A_{a \mu} \int d \vartheta_a
\exp \Biggl(
\nonumber \\
&& ~~~~~~~~ - \frac{1}{2 e^2} \sum_{\Box} \left(
\epsilon_{\mu\nu\lambda} \Delta_{\nu} A_{a \lambda} -  H
\delta_{\mu\tau} \right)^2
\nonumber \\
&&~~~~~~~~ + y_v \sum_{a \mu} \cos \left( \Delta_{\mu}
\vartheta_{a} - 2 \pi A_{a \mu} \right) \nonumber \\
&&~~~~~~~~- i h \sum_a \left( \Delta_\tau \vartheta_a - 2 \pi A_{a
\tau} \right) \Biggr). \label{zdv}
\end{eqnarray}
We have not explicitly displayed the $\Lambda_{ij}$ term above,
and assumed it has been absorbed into a renormalized value of
$e^2$.\footnote{For the case of long-range Coulomb interactions
with $\Lambda_{ij} \sim 1/|i-j|$, the coupling $e^2$ will acquire
a $\sim |k|$ momentum dependence. This will modify the
renormalization group analysis in Section~\ref{sec:rg}, as
discussed in M.~P.~A.~Fisher and G.~Grinstein, Phys. Rev. Lett.
{\bf 60}, 208 (1988) and J.~Ye, Phys. Rev. B {\bf 58}, 9450
(1998).} This is the theory of a dual vortex boson represented by
the angular rotor variable $\vartheta_a$, coupled to a dual gauge
field $A_{a \mu}$. The Berry phase term proportional to $h$ is
precisely the constraint that the rotor number variable
(canonically conjugate to $\vartheta_a$) takes values which are
integers plus $h$ (see Ref.~\onlinecite{fisherlee}): so there is a
background density of vortices of $h$ per site which is induced by
the magnetic field acting on the direct bosons.

The remainder of Section~\ref{sec:boson} will consider only the
case $h=0$; the analog of the $h \neq 0$ case will appear later in
the dimer model analyses of II.

We obtain the dual theory in its final form on the cubic lattice
by replacing the field $e^{i \vartheta_a}$ by a ``soft-spin'' dual
vortex field $\psi_a$, which yields
\begin{eqnarray}
 \mathcal{Z}_d = && \prod_a \int d A_{a \mu} \int d \psi_a \exp
\Biggl(
\nonumber \\
&&~~~~~~ - \frac{1}{2 e^2} \sum_{\Box} \left(
\epsilon_{\mu\nu\lambda} \Delta_{\nu} A_{a \lambda} - H
\delta_{\mu\tau} \right)^2 \nonumber \\ &&~~~~~~+ \frac{y_v}{2}
\sum_{a \mu} \left[ \psi_{a+\mu}^{\ast} e^{2 \pi i A_{a \mu}}
\psi_a +
\mbox{c.c.} \right] \nonumber \\
&&~~~~~~ -\sum_a \left[ s |\psi_a |^2 + \frac{u}{2} |\psi_a|^4
\right] \Biggr) \label{zdh} \;.
\end{eqnarray}
The last line in Eq.~(\ref{zdh}) is the effective potential for
the complex vortex field $\psi_a$. Increasing the parameter $s$
scans the system from the insulating solid at $s \ll 0$ to the
superfluid at $s \gg 0$ {\i.e.} the system moves from right to
left in the phase diagrams of Fig~\ref{vortex_pdiag}.

\subsection{Symmetries}
\label{sec:field}

We will now present a careful analysis of the symmetries of
Eq.~(\ref{zdh}), with the aim of deducing general constraints that
must be obeyed by the low energy theory near the
superfluid-to-insulator transition.

We begin in the superfluid regime with $s$ large, so that $\langle
\psi_a \rangle = 0$. The direct boson density is the $\tau$
component of the dual `magnetic' flux $\epsilon_{\mu\nu\lambda}
\Delta_{\nu} A_\lambda$, and for $s$ large in the action in
Eq.~(\ref{zdh}) it is clear that the saddle point of the $A_{\mu}$
fluctuations occurs at $A_{a\mu} = \overline{A}_{a\mu}$ with
$\epsilon_{\mu\nu\lambda} \Delta_{\nu} \overline{A}_{a \lambda} =
H \delta_{\mu\tau}$. We want this boson density to be the value
$f$ in Eq.~(\ref{e9}), and so we should choose $H=f$.

We now wish to examine the structure of $\psi_a$ fluctuations
about this saddle point.  Let us choose the Landau gauge with
$\overline{A}_{a\tau} = \overline{A}_{ax} = 0$ and
\begin{equation}
\overline{A}_{ay} = f ~a_x ; \label{landau}
\end{equation}
here $a_x$ is the $x$ co-ordinate of the dual lattice $a$. We now
need to determine the $\psi_a$ spectrum in a background
$\overline{A}_{a\mu}$ field.

The basic symmetry operations are $T_{x}$, $T_{y}$, and
$R_{\pi/2}^{\rm dual}$, introduced in Section~\ref{sec:intro}. The
action of these operators on $\psi_a \equiv \psi (a_x, a_y)$
required to keep the Hamiltonian invariant is
\begin{eqnarray}
T_y &:& \psi(a_x, a_y)  \rightarrow \psi (a_x, a_y - 1) \nonumber
\\
T_x &:& \psi(a_x, a_y)  \rightarrow \psi (a_x - 1, a_y) \omega^{
a_y} \nonumber \\
R_{\pi/2}^{\rm dual} &:& \psi(a_x, a_y)  \rightarrow \psi (a_y,
-a_x ) \omega^{a_x a_y}  \label{atrans}
\end{eqnarray}
Notice that Eqs.~(\ref{e7}) and (\ref{rtr}) are obeyed by the
above.

It is also useful to collect the representation of
Eq.~(\ref{atrans}) in momentum space. Implying that all momenta
are reduced back to the extended Brillouin zone with momenta $-\pi
< k_x, k_y < \pi$ we find
\begin{eqnarray}
T_y &:& \psi(k_x, k_y)  \rightarrow \psi (k_x, k_y) e^{-i k_y}
\nonumber
\\
T_x &:& \psi(k_x, k_y)  \rightarrow \psi (k_x, k_y - 2 \pi f)
e^{-i k_x} \nonumber
\\
R_{\pi/2}^{\rm dual} &:& \psi(k_x, k_y)  \rightarrow \label{ktrans} \\
&~& \frac{1}{q} \sum_{m,n=0}^{q-1}
 \psi (k_y + 2 \pi n f , -k_x - 2 \pi m f ) \omega^{-m n}
 \nonumber
\end{eqnarray}

The most important $\psi_a$ fluctuations will be at momenta at
which the spectrum has minima. It is not difficult to show from
the above symmetry relations that any such minimum is at least
$q$-fold degenerate: Let $|\Lambda \rangle$ be a state at a
minimum of the spectrum. Because, the operator $T_y$ commutes with
the Hamiltonian, this state can always be chosen to be an
eigenstate of $T_y$, with eigenvalue $e^{-i k_y^{\ast}}$. Now the
above relations imply immediately that $T_x |\Lambda \rangle$ is
also an eigenstate of the Hamiltonian with $T_y$ eigenvalue $e^{-i
k_y^{\ast}} \omega^{-1}$. Also, because the $T_y$ eigenvalue is
distinct from that of $|\Lambda \rangle$, this state is an
orthogonal eigenstate of the Hamiltonian. By repeated application
of this argument, we obtain $q$ orthogonal eigenstates of the
Hamiltonian whose $T_y$ eigenvalues are $e^{-i k_y^{\ast}}$ times
integer powers of $\omega^{-1}$. Because we are working with the gauge
of Eq.~(\ref{landau}), our Hamiltonian couples momenta $(k_x,
k_y)$ to momenta $(k_x \pm 2 \pi f, k_y)$.  It is therefore
advantageous to look at
the spectrum in the reduced Brillouin zone with 
$-\pi/q < k_x < \pi/q$ and $-\pi < k_y < \pi$. For the
nearest-neighbor model under consideration here, the minima of the
spectrum are then at the $q$ wavevectors $(0, 2 \pi \ell p/q)$
with $\ell = 0, \ldots q-1$. Let us label the eigenmodes at these
wavevectors $\varphi_\ell$. We therefore have to write down the
field theory in terms of these $q$ complex fields $\varphi_\ell$.

It is useful, especially when analyzing the influence of
$R_{\pi/2}^{\rm dual}$, to make the above symmetry considerations
explicit. In the extended Brillouin zone, for the nearest-neighbor
model under consideration here, let us label the $q^2$ states at
the wavevectors $(2 \pi m p/q,2 \pi n p/q)$ by $|m,n \rangle$.
Then one of the minima of the spectrum corresponds to the state
\begin{equation}
| \varphi_0 \rangle = \sum_{m=0}^{q-1} c_m |m, 0 \rangle
\end{equation}
where the $c_m$ are some complex numbers. Then, by operation of
$T_x$ on $|\varphi_0 \rangle$ we obtain the $q$ degenerate
eigenstates as
\begin{equation}
| \varphi_\ell \rangle = \sum_{m=0}^{q-1} c_m\, \omega^{- \ell m }
|m, \ell \rangle \label{defphil}
\end{equation}
Now let us consider the action of $R_{\pi/2}^{\rm dual}$ on the
states in Eq.~(\ref{defphil}). Using Eq.~(\ref{ktrans}), and after
some simple changes of variables we obtain
\begin{equation}
R_{\pi/2}^{\rm dual} |\varphi_\ell \rangle = \frac{1}{q}
\sum_{m,m',\ell'=0}^{q-1} c_{m}\, \omega^{-(m' \ell' + \ell \ell' - m
m')} |m', \ell' \rangle \label{rphi}
\end{equation}
Now, because $R_{\pi/2}^{\rm dual}$ commutes with $H$, the
right-hand-side of Eq.~(\ref{rphi}) must be a linear combination
of the $| \varphi_\ell \rangle$ states in Eq.~(\ref{defphil}). The
matrix elements of the rotation operator are then given by
$\langle \varphi_{\ell'}| R_{\pi/2}^{\rm dual} | \varphi_{\ell}
\rangle = c\, \omega^{-\ell \ell'}$, where $c=\frac{1}{q} \sum
c_{m'}^{\ast} c_m \omega^{m m'}$ is independent of $\ell$ and
$\ell'$.
For the nearest-neighbor model under consideration here,
it can be easily
checked that the $c_m$'s are invariant under a Fourier transform
such that $c=1/\sqrt{q}$. Hence we find
\begin{equation}
R_{\pi/2}^{\rm dual} |\varphi_\ell \rangle = \frac{1}{\sqrt{q}}
\sum_{\ell'=0}^{q-1} \omega^{- \ell \ell'} |\varphi_{\ell'}
\rangle . \label{rphi2}
\end{equation}

Our discussion above has now established that the low energy
vortex fields must have an action invariant under the
transformations in Eqs.~(\ref{txty}) and (\ref{rpi}). In a similar
manner we can also determine the transformations associated with
the remaining elements of the square lattice space group. These
involve the operations $I_x^{\rm dual}$ and $I_y^{\rm dual}$ which
are reflections about the $x$ and $y$ axes of the dual lattice.
Under these operations we find
\begin{eqnarray}
I_x^{\rm dual} &:& \varphi_\ell \rightarrow \varphi_{\ell}^{\ast}
\nonumber
\\
I_y^{\rm dual} &:& \varphi_\ell \rightarrow \varphi_{-\ell}^{\ast} \;.
\label{ixiy}
\end{eqnarray}
Finally it is interesting to consider the point inversion operator
$I_p^{\rm dual} \equiv (R_{\pi/2}^{\rm dual})^2$, with
\begin{equation}
  I_p^{\rm dual} \, : \, \varphi_\ell \rightarrow \varphi_{-\ell} \;.
\end{equation}
As in the ordinary space group we have
\begin{equation}
  I_p^{\rm dual} = I_x^{\rm dual} I_y^{\rm dual} = I_y^{\rm dual} I_x^{\rm dual} \;.
\end{equation}

\subsection{Continuum field theories}
\label{sec:cft}

We have established that fluctuations of the vortex fields about
the saddle point of Eq.~(\ref{zdh}) in Eq.~(\ref{landau})
transform under a projective representation of the square lattice
space group which is defined by Eqs.~(\ref{txty}), (\ref{rpi}),
and (\ref{ixiy}). In this section we will write down the most
general continuum theory of the $\varphi_\ell$ fields which is
invariant under these projective transformations. The action
should include fluctuations in $A_{a \mu}$ about $\overline{A}_{a
\mu}$ -- this will be represented by the continuum non-compact
U(1) gauge field $a A_\mu /(2 \pi)$, where $a$ is the lattice
spacing.

First, we consider quadratic order terms about the saddle point of
Eq.~(\ref{zdh}). The most general action has the familiar terms of
scalar electrodynamics
\begin{eqnarray}
\mathcal{S}_0 &=& \int d^2 r d \tau \left( \sum_{\ell = 0}^{q-1}
\left[ |(\partial_\mu - i A_{\mu} ) \varphi_\ell |^2 + s
|\varphi_\ell |^2 \right] \right. \nonumber \\
&+& \left. \frac{1}{2e^2} \left( \epsilon_{\mu\nu\lambda}
\partial_\nu A_\lambda \right)^2 \right).
\label{s0}
\end{eqnarray}
We have rescaled the coupling $e$ here by a factor of $2 \pi$ from
Eq.~(\ref{zdh}).

Next, we consider terms which are quartic in the $\varphi_\ell$,
but which contain no spatial or temporal derivatives. These will
be contained in the action $\mathcal{S}_1$. We discuss two
approaches to obtaining the most general quartic invariants. The
first is the most physically transparent, but turns out to be
eventually inconvenient for explicit computations. In this
approach we use density operators defined in Eq.~(\ref{e11}), and
their simple transformation properties in Eq.~(\ref{e12}), to
build up quartic invariants. In particular, the quartic invariants
are only quadratic in the $\rho_{mn}$, and we need only the most
general quadratic term invariant under Eq.~(\ref{e12}). This has
the form
\begin{eqnarray}
\mathcal{S}_1 &=& \int d^2 r d\tau \Biggl( \sum_{n=0}^{q/2}
\sum_{m=0}^n \lambda_{nm} \left[ |\rho_{nm}|^2 + |\rho_{n,-m}|^2
\right. \nonumber \\ &~&~~~~~~~~~~~\left. + |\rho_{mn}|^2 +
|\rho_{m,-n}|^2 \right] \Biggr) \label{s1}
\end{eqnarray}
However, not all the invariants above are independent, and there
are often linear relations between them - this reduces the number
of independent coupling constants $\lambda_{nm}$. Determining the
linear relations between the couplings turns out to be
inconvenient, and we found it easier to proceed by the second
method described below.

In the second approach, we first impose only the constraints
imposed by the translation operations $T_x$, $T_y$. By inspection,
it is easy to see that the most general quartic term invariant
under these operations has the structure
\begin{equation}
\mathcal{S}_1 = \frac{1}{4} \int d^2 r d\tau \sum_{\ell m n}
\gamma_{mn} \varphi_{\ell}^\ast \varphi_{\ell + m}^{\ast}
\varphi_{\ell+n} \varphi_{\ell+m-n}. \label{r1}
\end{equation}
Here the integers $\ell,m,n,\ldots$ range implicitly from 0 to
$q-1$ and all additions over these integers are taken modulo $q$.
Imposing in addition the reflection operations in
Eq.~(\ref{ixiy}), and accounting for the internal symmetries in
Eq.~(\ref{r1}), it is easy to show that the couplings
$\gamma_{mn}$ can always be taken to be real and to obey the
relations
\begin{eqnarray}
\gamma_{mn} &=& \gamma_{-m,-n} \nonumber \\
\gamma_{mn} &=& \gamma_{m,m-n} \nonumber \\
\gamma_{mn} &=& \gamma_{m-2n,-n}. \label{r3}
\end{eqnarray}

We have so far not yet imposed the constraints implied by the
lattice rotation $R_{\pi/2}^{\rm dual}$. Consequently,
Eqs.~(\ref{r1}) and (\ref{r3}) define the most general quartic
terms for a system with {\em rectangular\/} symmetry, which may be
appropriate in some physical situations. By explicit solution of
the constraints defined by Eq.~(\ref{r3}), we found that the
quartic terms are determined by $N_{\rm rect}$ independent real
coupling constants, where
\begin{equation}
N_{\rm rect} = \frac{(n+1)(n+2)}{2}~\mbox{for $q=2n,2n+1$},
\label{rect}
\end{equation}
with $n$ a positive integer.  Note that the number of couplings grows
quite rapidly with increasing $q$: $N_{\rm rect} \sim q^2/8$.

Finally, let us consider the consequences of the $R_{\pi/2}^{\rm
dual}$ symmetry. By applying Eq.~(\ref{rpi}) to Eq.~(\ref{r1}) we
find that $\mathcal{S}_1$ remains invariant provided the
$\gamma_{mn}$ obey the relations
\begin{equation}
\gamma_{\bar{m}\bar{n}}= \frac{1}{q} \sum_{mn} \gamma_{mn}
\omega^{-\left[ n(\bar{m}-\bar{n}) + \bar{n} (m-n) \right]}
\label{r4}
\end{equation}
Na\"ively, there are $q^2$ relations implied by Eq.~(\ref{r4}),
but they are not all independent of each other. By explicitly
solving the relations in Eq.~(\ref{r4}) for a range of $q$ values
we found that the number of independent quartic coupling constants
for a system with full square lattice symmetry is
\begin{equation}
N_{\rm square} = \left\{
\begin{array}{cc} (n+1)^2 & \mbox{for $q=4n,4n+1$} \\
(n+1)(n+2) & \mbox{for $q=4n+2,4n+3$}
\end{array} \right. \label{square}
\end{equation}
with $n$ a positive integer.  The additional restrictions of square
 relative to rectangular symmetry reduce the number of independent coupling
constants by roughly half at large $q$: $N_{\rm square}\sim q^2/16$.

\subsection{Mean field theory}
\label{sec:mft}

This section will examine the mean-field phase diagrams of the
general theory $\mathcal{S}_0+\mathcal{S}_1$ proposed in
Section~\ref{sec:cft}. From the discussion in
Section~\ref{sec:intro}, it is clear that such a procedure can yield
a direct second order transition from a superfluid state (with
$\langle \varphi_\ell \rangle = 0$) to an insulating state with
density wave order (with $\langle \varphi_\ell \rangle \neq 0$).
We are interested in determining the possible configurations of
values of the $\varphi_{\ell}$ and the associated patterns of
density wave order for a range of $q$ values (in this subsection,
we will simply write $\langle \varphi_{\ell} \rangle$ as
$\varphi_\ell$ because there is no distinction between the two
quantities in mean field theory). After determining the
$\varphi_\ell$, we determined the $\rho_{mn}$ by Eq.~(\ref{e11})
(using a Lorentzian for the form factor $S(Q)= 1/(1 + Q^2)$), and
then computed the density wave order using
\begin{equation}
\delta \rho ({\bf r}) = \sum_{m,n=-q}^{q-1} \rho_{mn} e^{2 \pi i f (m
r_x + n r_y)} \label{m1}
\end{equation}
We evaluated Eq.~(\ref{m1}) at ${\bf r}$ values corresponding to
the sites, bonds, and plaquettes of the direct lattice, and
plotted the results in the square lattice figures that appear
below. In the formalism of Section~\ref{sec:boson}, ${\bf r}$
values with integer co-ordinates correspond to the sites of the
dual lattice, and hence to plaquettes of the direct lattice. The
value of $\delta \rho ({\bf r})$ on such plaquette co-ordinates
can be considered a measure of the ring-exchange amplitude of
bosons around the plaquette. By a similar reasoning, ${\bf r}$
values with half-odd-integer co-ordinates represent sites of the
direct lattice, and the values of $\delta \rho ({\bf r})$ on such
sites measure the boson density on these sites. Finally, ${\bf r}$
values with $r_x$ integer and $r_y$ half-odd-integer correspond to
horizontal links of the square lattice (and vice versa for
vertical links), and the values of $\delta \rho ({\bf r})$ on
the links is a measure of the mean boson kinetic energy; if the
bosons represent a spin system, this is a measure of the spin
exchange energy.

Our results appear for a range of $q$ values in the following
subsections.

\subsubsection{$q=2$}
\label{sec:q2}

After imposing the symmetry relations in Eqs.~(\ref{r3}) and
(\ref{r4}), the quartic potential in Eq.~(\ref{r1}) (defined by
$\mathcal{S}_1 = \int d^2 r d \tau \mathcal{L}_4$) is
\begin{eqnarray}
\mathcal{L}_4 = \frac{\gamma_{00}}{4} \left( |\varphi_0|^2 +
|\varphi_1|^2 \right)^2 + \frac{\gamma_{01}}{4} \left( \varphi_0
\varphi_1^\ast - \varphi_0^\ast \varphi_1 \right)^2. \label{m2}
\end{eqnarray}
We now make the change of variables
\begin{eqnarray}
\varphi_0 &=& \frac{\zeta_0 + \zeta_1}{\sqrt{2}} \nonumber \\
\varphi_1 &=& -i \frac{\zeta_0 - \zeta_1}{\sqrt{2}}. \label{m3}
\end{eqnarray}
As we will discuss in detail in Section~\ref{sec:fract-vort-pict},
the new variables have the advantage of realizing a {\em
permutative\/} representation of the PSG, {\em i.e.\/} all
representation matrices of the PSG have only one non-zero
unimodular element in every row and column. From
Eqs.~(\ref{txty}), (\ref{rpi}), and (\ref{m3}) it is easy to
deduce that in the $\zeta_\ell$ variables,
\begin{eqnarray}
T_x = \left( \begin{array}{cc} 0 & -i \\ i & 0 \end{array}
\right)~&;&~ T_y = \left( \begin{array}{cc} 0 & 1 \\ 1 & 0
\end{array} \right) \nonumber \\
R_{\pi/2}^{\rm dual} &=& \left( \begin{array}{cc} 0 & e^{-i \pi/4}
\\ e^{i\pi/4} & 0 \end{array} \right). \label{m3a}
\end{eqnarray}
The action in Eq.~(\ref{m2}) reduces to
\begin{eqnarray}
\mathcal{L}_4 = \frac{\gamma_{00}}{4} \left( |\zeta_0|^2 +
|\zeta_1|^2 \right)^2 - \frac{\gamma_{01}}{4} \left( |\zeta_0|^2 -
| \zeta_1|^2 \right)^2. \label{m4}
\end{eqnarray}
The result in Eq.~(\ref{m4}) is identical to that found in earlier
studies \cite{lfs,sp} of the $q=2$ case.

Minimizing the action implied by Eq.~(\ref{m4}), it is evident
that for $\gamma_{01} < 0$ there is a one parameter family of
gauge-invariant solutions in which the relative phase of $\zeta_0$
and $\zeta_1$ remains undetermined. Computing the values of
$\delta \rho ({\bf r})$ establishes that this phase is physically
significant, because it leads to distinguishable density wave
modulations. This indicates that the phase will be determined by
higher order terms in the effective potential. As shown in earlier
work \cite{lfs}, one needs to go to a term which is eighth order
in the $\varphi_\ell$ before the value of $\arg (\zeta_0/\zeta_1)$
is pinned at specific values. In the present formalism, the
required eighth order term is $\rho_{0,1}^2 \rho_{1,0}^2$, which
is invariant under all the transformations in Eq.~(\ref{e12});
this contains a term $\sim (\zeta_0 \zeta_1^\ast )^4 +
\mbox{c.c}$.

\begin{figure}
\centering
\ifig[width=3in]{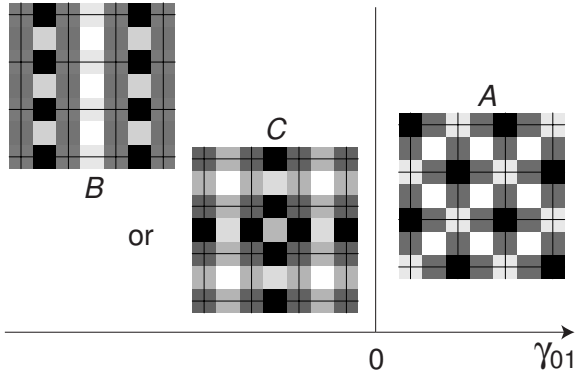}
\caption{Mean field phase diagram of $\mathcal{S}_0 +
\mathcal{S}_1$ defined in Eqs.~(\ref{s0}) and (\ref{r1}). The
couplings obey Eqs.~(\ref{r3}) and (\ref{r4}). We assumed $s=-1$
to obtain an insulating state, and $\gamma_{00}=1$. Plotted are
the values of $\delta \rho ({\bf r})$, defined as in
Eq.~(\ref{m1}), on the sites, links and plaquettes of the direct
lattice: the lines represent links of the direct lattice. As
discussed in the text, these values represent the boson density,
kinetic energy, and ring-exchange amplitudes respectively. The
saddle point values of the fields in the states are shown in
Eq.~(\protect\ref{m5}). The choice between the states $B$ and $C$
is made by a eighth order term in the action.} \label{figq2}
\end{figure}

Our specific results from the mean field theory for $q=2$ are
summarized in Fig~\ref{figq2}.
The state (A) has an ordinary charge density wave (CDW) at
wavevector ($\pi,\pi$). The other two states are VBS states in
which all the sites of the direct lattice remain equivalent, and
the VBS order appears in the (B) columnar dimer or (C) plaquette
pattern. The saddle point values of the fields associated with
these states are:
\begin{eqnarray}
(A)~&:&~ \zeta_0 \neq 0~,~\zeta_1 = 0~\mbox{or}~\zeta_0
= 0~,~\zeta_1 \neq 0. \nonumber \\
(B)~&:&~ \zeta_0 = e^{i n \pi/2} \zeta_1 \neq 0. \nonumber \\
(C)~&:&~ \zeta_0 = e^{i (n+1/2) \pi/2} \zeta_1 \neq 0, \label{m5}
\end{eqnarray}
where $n$ is any integer. At quartic order, the states $B$ and $C$
are degenerate with all states with $\zeta_0 = e^{i \theta}
\varphi_1$, with $\theta$ arbitrary; the value of $\theta$ is
selected only by the eighth order term.

\subsubsection{$q=3$}
\label{sec:q3}

\begin{figure}
\centering
\ifig[width=2.7in]{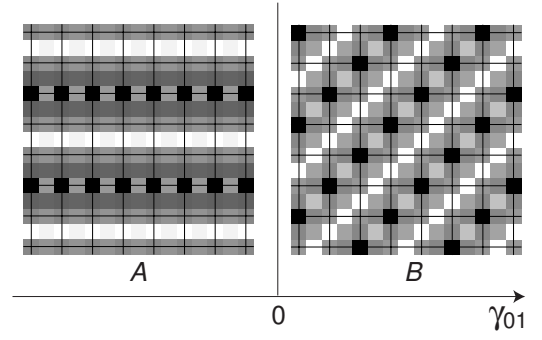}
\caption{As in Fig~\protect\ref{figq2} but for $q=3$ }
\label{figq3}
\end{figure}

Now there are 3 $\varphi_\ell$ fields, and the quartic potential
in Eq.~(\ref{m2}) is replaced by
\begin{eqnarray}
\mathcal{L}_4 &=& \frac{\gamma_{00}}{4} \left( |\varphi_0|^2 +
|\varphi_1|^2  + |\varphi_2|^2 \right)^2 \nonumber \\ &+&
\frac{\gamma_{01}}{2} \left( \varphi_0^\ast \varphi_1^\ast
\varphi_2^2 + \varphi_1^\ast \varphi_2^\ast \varphi_0^2 +
\varphi_2^\ast \varphi_0^\ast
\varphi_1^2 + \mbox{c.c.} \right. \nonumber \\
&-& \left. 2 |\varphi_0|^2 |\varphi_1|^2 - 2 |\varphi_1|^2
|\varphi_2|^2- 2 |\varphi_2|^2 |\varphi_0|^2\right). \label{m6}
\end{eqnarray}
We show in Appendix~\ref{sec:noq3} that there is now no
transformation of the $\varphi_\ell$ variables which realize a
permutative representation of the PSG, and in which Eq.~(\ref{m6})
may take a more transparent form. The results of the minimization
of $\mathcal{S}_0+\mathcal{S}_1$ for the insulating phases are
shown in Fig~\ref{figq3}.
The states have stripe order, one along the diagonals, and the
other along the principle axes of the square lattice. Both states
are 6-fold degenerate, and the characteristic saddle point values
of the fields are
\begin{eqnarray}
(A)~&:&~ \varphi_0 \neq 0~,~\varphi_1 = \varphi_2 =
0~\mbox{or}\nonumber \\
&~&~~~~~~~~~~~~e^{i 4n \pi/3}\varphi_2 = e^{i 2 n \pi/3} \varphi_1
=
\varphi_0 \neq 0 \nonumber \\
(B)~&:&~ \varphi_0=\varphi_1 = e^{\pm 2 i \pi/3}
\varphi_2~\nonumber \\
&~&~~~~~~~~~~~~\mbox{and permutations}. \label{m7}
\end{eqnarray}

\subsubsection{$q=4$}
\label{sec:q4}

\begin{figure}
\centering \ifig[width=3.3in]{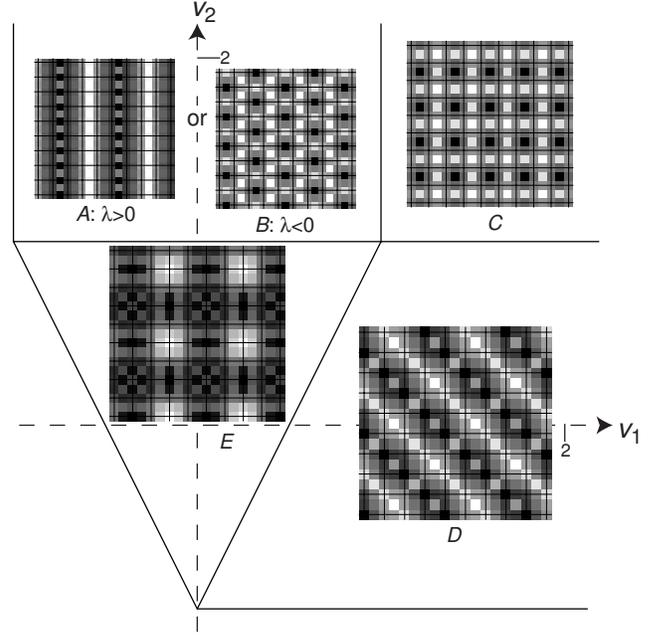} \caption{As in
Fig~\protect\ref{figq2}, but for $q=4$ with the quartic potential
as in Eq.~(\protect\ref{m10}). We have set $u=1$, and then
determined the phase diagram as function of $v_1$ and $v_2$ for
infinitesimal values of $\lambda$. The dashed lines represent the
$v_1$ and $v_2$ axes, while the full lines are phase boundaries.}
\label{figq4a}
\end{figure}

\begin{figure}
\centering
\ifig[width=3in]{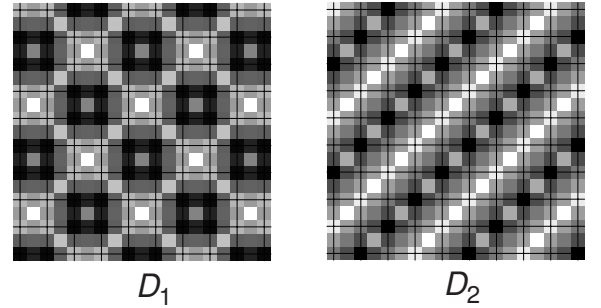}
\caption{Higher symmetry cases of the state $D$ in
Fig.~\protect\ref{figq4a}. With the angle $\theta$ defined as in
the text, state $D_1$ corresponds to $\theta=n \pi/2$, and the
state $D_2$ to $\theta = (n+1/2)\pi/2$ ($n$ integer).}
\label{figq4b}
\end{figure}

For $q=4$, there are 4 independent quartic invariants in Eq.
(\ref{r1}), and they can be combined into the following forms.
\begin{eqnarray}
&& I_1 = \rho_{0,0}^2 = \left( |\varphi_0|^2 + |\varphi_1|^2 +
|\varphi_2|^2 +
|\varphi_3 |^2 \right)^2 \nonumber \\
&& I_2 = \rho_{2,2}^2 = \left( \varphi_0^{\ast} \varphi_2 +
\varphi_2^{\ast} \varphi_0 - \varphi_1^{\ast} \varphi_3 -
\varphi_3^{\ast}
\varphi_1 \right)^2 \nonumber \\
&& I_3 = |\varphi_0|^2 |\varphi_1|^2
 + |\varphi_1|^2 |\varphi_2|^2 + |\varphi_2|^2 |\varphi_3|^2
 + |\varphi_3|^2 |\varphi_0|^2
\nonumber \\ &&- \varphi_0^\ast \varphi_1^\ast \varphi_2 \varphi_3
- \varphi_1^\ast \varphi_2^\ast \varphi_3 \varphi_0-
\varphi_2^\ast \varphi_3^\ast \varphi_0 \varphi_1 - \varphi_3^\ast
\varphi_0^\ast \varphi_1 \varphi_2 \nonumber \\
&& I_4 = \varphi_0^\ast \varphi_2^\ast \left( \varphi_1^2 +
\varphi_3^2 \right) + \mbox {c.c.} + \varphi_1^\ast \varphi_3^\ast
\left( \varphi_0^2 + \varphi_2^2 \right) + \mbox {c.c.} \nonumber
\\ &&~~~~~~~~~~~~ - 4 |\varphi_0|^2 |\varphi_2 |^2 - 4 |\varphi_1 |^2
|\varphi_3|^2 \label{m8}
\end{eqnarray}
Now as for $q=2$, the invariants for $q=4$ do simplify by
redefinitions of the fields analogous to Eq.~(\ref{m3}), which
realize a permutative representation of the PSG. We define
\begin{eqnarray}
\varphi_0 & = & (\zeta_0 + \zeta_1)/\sqrt{2} \nonumber \\
\varphi_2 & = & (\zeta_0 - \zeta_1)/\sqrt{2} \nonumber \\
\varphi_1 & = & (\zeta_3 + \zeta_2)/\sqrt{2} \nonumber \\
\varphi_3 & = & (\zeta_3 - \zeta_2)/\sqrt{2}. \label{m9}
\end{eqnarray}
The PSG operations on the $\zeta_\ell$ can then easily be shown to
take the permutative form specified in Eq.~(\ref{eq:3.5}) (in a
slightly different notation). In the new variables, the quartic
potential can be written in the following form
\begin{eqnarray}
\mathcal{L}_4 &=& \frac{u}{2} \left(|\zeta_0|^4 + |\zeta_1|^4 +
|\zeta_2|^4
+ |\zeta_3 |^4 \right) \nonumber \\
&+& v_1 \left( |\zeta_0|^2 + |\zeta_2 |^2 \right)
\left( |\zeta_1|^2 + |\zeta_3 |^2 \right) \nonumber \\
&+& v_2 \left( |\zeta_0|^2 |\zeta_2 |^2 +
 |\zeta_1|^2 |\zeta_3 |^2 \right)  \nonumber \\
&-& \frac{\lambda}{2} \left( \zeta_0^{\ast 2} \zeta_1^2 +
\zeta_1^{\ast 2} \zeta_0^2  - \zeta_1^{\ast 2} \zeta_2^2 -
\zeta_2^{\ast 2} \zeta_1^2 \right. \nonumber \\ &+& \left.
\zeta_2^{\ast 2} \zeta_3^2 + \zeta_3^{\ast 2} \zeta_2^2  +
\zeta_3^{\ast 2} \zeta_0^2 + \zeta_0^{\ast 2} \zeta_3^2 \right)
\label{m10}
\end{eqnarray}
where $u$, $v_{1,2}$ and $\lambda$ are coupling constants. Notice
that at $\lambda =0$ the action is independent of the relative
phases of the $\zeta_\ell$, and so there is a 3 parameter family
of degenerate states. It is therefore convenient to first
determine the minima at $\lambda=0$ (which can be done
analytically), and to then determine the fate of the minima so
found at small $\lambda$. The results of such a procedure are
summarized in Fig~\ref{figq4a}.
The states in Fig~\ref{figq4a} are characterized by the following
parameter values
\begin{eqnarray}
(\mbox{$A$ and $B$})~&:&\mbox{One of $\zeta_0$ or $\zeta_2$
non-zero, one of $\zeta_1$ or $\zeta_3$} \nonumber \\
&~&\mbox{non-zero, magnitudes of non-zero $\zeta_\ell$ equal}
\nonumber \\
C~&:&\mbox{Only one of $|\zeta_\ell|$ non-zero} \nonumber \\
D~&:&\mbox{Either $\zeta_0=\zeta_2=0$ and
$|\zeta_1|=|\zeta_3| \neq 0$,} \nonumber \\
&~&\mbox{or $\zeta_1=\zeta_3=0$ and $|\zeta_0|=|\zeta_2|
\neq 0$} \nonumber \\
E~&:&\mbox{All $|\zeta_\ell|$ equal and non-zero}
 \label{m11}
\end{eqnarray}
By inserting the saddle point values of state $D$ in
Eq.~(\ref{m10}) we observe that this state has a continuous
degeneracy with $\zeta_1 = e^{i \theta} \zeta_3$ (for the first
choice of the saddle point) and $\theta$ arbitrary. We expect that
higher order terms in the action will select the value of
$\theta$, and we show the higher symmetry states at special values
of $\theta$ in Fig~\ref{figq4b}.

The phases in Fig~\ref{figq4a} all survive for a finite range of
$\lambda$. However, for large enough $|\lambda|$ a number of
additional phases, with distinct patterns of lattice symmetry
breaking, appear. We have not determined the full phase diagram in
the three-dimensional $v_1$, $v_2$, $\lambda$ space, but here
merely list the additional phases that were found in our numerical
analysis. The new large $|\lambda|$ phases are shown in
Fig~\ref{figq4c}, and the typical saddle points had the following
characteristics:
\begin{figure}
\centering
\ifig[width=3in]{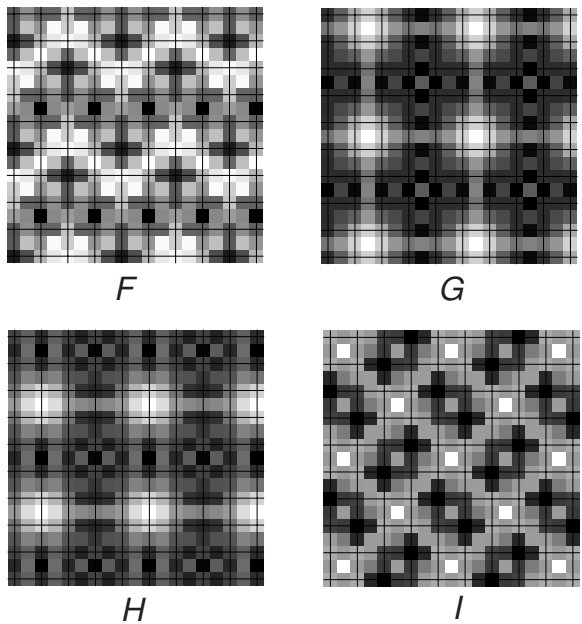}
\caption{Additional insulating mean-field states for $q=4$ with
the quartic potential as in Eq.~(\protect\ref{m10}). These states
appear for larger values of $|\lambda|$ and the field
configurations are described in the text.} \label{figq4c}
\end{figure}
\begin{eqnarray}
F~&:&\mbox{$\zeta_0 = \zeta_3$, $\zeta_1 = \zeta_2$, $|\zeta_0|
\neq |\zeta_1 |$}
\nonumber \\
G~&:&\mbox{$\zeta_0 = i\zeta_2$, $|\zeta_1| \neq |\zeta_0
|$, $\arg(\zeta_0) = \arg(\zeta_1) + \pi$}\nonumber \\
&~&~~\mbox{and $\zeta_3 = 0$}
\nonumber \\
H~&:&\mbox{$\zeta_3 = i\zeta_0$, $\zeta_1 = i\zeta_2$, $|\zeta_0|
\neq |\zeta_1 |$ }
\nonumber \\
I~&:&\mbox{$\zeta_1 = i\zeta_3$, $ |\zeta_1| \neq
|\zeta_2|$, $\arg(\zeta_2) = \arg(\zeta_1) $}\nonumber \\
&~&~~\mbox{and $\zeta_0 = 0$}. \label{m12}
\end{eqnarray}

\subsubsection{General $q$}
\label{sec:q16}

The insulating states obtained above clearly have the dimensions
of their unit cells constrained by the value of $q$, and we will
discuss these constraints more explicitly here with special
attention to the value $q=16$. From Eq.~(\ref{eq:fdelta}) and the
analysis of II, this value of $q$ is of particular relevance to
the cuprates at the hole density of 1/8. From Eq.~(\ref{square}),
we note that the $q=16$ case has 25 independent quartic coupling
constants. The full parameter space of insulating states is
therefore of immense complexity, and we will not attempt to map
out any phase diagram. Instead we shall examine the spatial
structure of a few simple configurations of the $\varphi_\ell$.

Let us assume that the insulating state has a unit cell of size $a
\times b$ lattice sites ($a$, $b$ integers). We now discuss the
general constraints on the values of $a$ and $b$. From the allowed
values of the wavevectors in Eq.~(\ref{e10}) we see that
generically we will have $a=q$ and $b=q$. However, it could be
that the $\varphi_\ell$ take special values so that some of the
$\rho_{mn}$ vanish. In this case, the periods $a$ and $b$ could be
any integer divisor of $q$. However, an interesting property of
Eq.~(\ref{e11}) is that it is not possible to simultaneously
reduce the values of both $a$ and $b$. The non-commutativity of
$T_x$ and $T_y$ in Eq.~(\ref{e7}) effectively imposes an
``uncertainty relation'' between their values so that the product
$ab$ must be a multiple of $q$: intuitively, we see this as a
requirement that the unit cell must contain an integer number of
bosons. Summarizing, the values of $a$ and $b$ are expected to be
constrained by the requirements that
\begin{equation}
\frac{q}{a},~\frac{q}{b},~\frac{ab}{q}~\mbox{are all integers.}
\label{int1}
\end{equation}

To illustrate the origin of the constraints in Eq.~(\ref{int1}),
let us consider the $q=16$ case, and attempt to construct
insulating states with $a$ and $b$ as small as possible. Motivated
by the application to the cuprates at hole density $\delta=1/8$,
let us look for solutions in which $b$ is 4 or smaller. From
Eq.~(\ref{e11}) it is clear that a simple way to achieve this is
to take $\varphi_\ell=0$ unless $\ell$ is a multiple of 4, so that
$\rho_{mn}$ is zero unless $n$ is a multiple of 4. A generic
solution with this property has $\rho_{mn}$ non-zero for all $m$,
and hence a period $a=16$. To obtain a smaller value of $a$ we
must make the $\rho_{mn}$ vanish for as large a set of $m$ values
as possible. However, with this particular choice of the
$\varphi_\ell$ notice that $\rho_{mn}$ is proportional to
$\rho_{m+4,n}$. So the best we can do is to have $\rho_{mn}$
non-zero only for $m$ a multiple of 4. In this case we obtain
$a=4$, which results in a unit cell consistent with
Eq.~(\ref{int1}).

\begin{figure}
\centering
\ifig[width=3in]{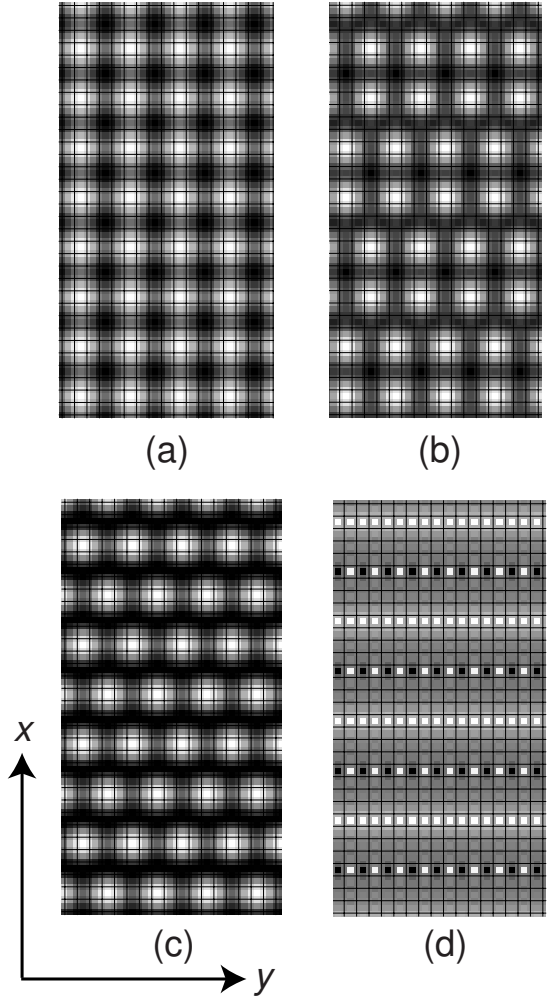}
\caption{A few $\varphi_\ell$ configurations for $q=16$ with
higher symmetry and unit cells smaller than the maximal $16 \times
16$. All figures have $\varphi_\ell =0$ unless $\ell$ is a
multiple of 4. ({\em a\/}) $\varphi_0 = 1$, $\varphi_4=1$,
$\varphi_8=1$, $\varphi_{12} = 1$, $4 \times 4$ unit cell.  ({\em
b\/}) $\varphi_0 = 1$, $\varphi_4=-1$, $\varphi_8=1$,
$\varphi_{12} = 1$, $16 \times 4$ unit cell.  ({\em c\/})
$\varphi_0 = 1$, $\varphi_4=1$, $\varphi_8=-1$, $\varphi_{12} =
-1$, $8 \times 4$ unit cell.  ({\em d\/}) $\varphi_0 = 1$,
$\varphi_4=0$, $\varphi_8=1$, $\varphi_{12} = 0$, $8 \times 2$
unit cell.} \label{figq16}
\end{figure}

We used the above strategy to identify a number of $\varphi_\ell$
configurations with highly symmetric, small unit cell density
modulations. A few sample results are shown in Fig~\ref{figq16}.
All unit cells are clearly consistent with Eq.~(\ref{int1}).

\subsection{Renormalization group analysis}
\label{sec:rg}

We now briefly address fluctuation effects across the mean field
transitions found in Section~\ref{sec:mft}. The analysis follows
previous work \cite{hlm,chen} on scalar quantum electrodynamics in
three dimensions who examined the theory $\mathcal{S}_0 +
\mathcal{S}_1$ for the case where the quartic interactions had a
full U($q$) invariance. Here we will extend the earlier one loop
results to the particular quartic couplings in Eq.~(\ref{r1}).
Unfortunately, we find runaway flows to strong coupling for all
values of $q$ we have studied, and no stable fixed points which
are accessible in a one loop analysis.

For the renormalization group (RG) analysis it is useful to write
the quartic terms in $\mathcal{S}_1$ in the more general form
\begin{equation}
 \mathcal{S}_1 = \frac{1}{4} \int d^2 r d \tau  \sum_{\ell mni}  u_{\ell m;ni}
\varphi_{\ell}^{\ast} \varphi_{m}^\ast \varphi_n \varphi_i
\label{s1u}
\end{equation}
where
\begin{equation}
u_{\ell m;ni} \equiv \delta_{\ell+m,n+i} \gamma_{m-\ell,n-\ell}.
\label{ugam}
\end{equation}
These couplings obey the symmetry relations
\begin{eqnarray}
u_{\ell m;ni} &=& u_{m \ell ;ni} \nonumber \\
u_{\ell m;ni} &=& u_{\ell m;in} \nonumber \\
u_{\ell m;ni} &=& u_{ni; \ell m} \nonumber \\
u_{\ell m;ni} &=& u_{\ell+j,m+j;n+j,i+j},
\end{eqnarray}
(as always, all index arithmetic is modulo $q$) and the trace
identity\cite{bgz}
\begin{equation}
\sum_\ell u_{\ell m; \ell i} = U \delta_{mi}
\end{equation}
with $U \equiv \sum_\ell \gamma_{\ell 0}$.

The RG equations can be derived by a simple generalization of the
methods in Refs.~\onlinecite{chen,bgz}. At one loop order the
results are
\begin{eqnarray}
\frac{d e^2}{d l} &=& \epsilon e^2 - \frac{C q}{3} e^4
\nonumber \\
\frac{d u_{\ell m;nk}}{d l} &=& (\epsilon - 2 \eta) u_{\ell m; nk}
- C \sum_{ij} \Biggl( \frac{1}{2} u_{\ell m;ij} u_{ij;nk}
\nonumber \\ &+& u_{\ell i; n j} u_{m j; k i} + u_{\ell i; k j}
u_{m j; n i}
\Biggr)\nonumber \\
&-& a_2 e^4 C ( \delta_{\ell n} \delta_{mk} + \delta_{\ell k}
\delta_{mn} ) \nonumber \\
\eta &=& - a_1 e^2 C. \label{rgflow1}
\end{eqnarray}
The equations have been derived in $3-\epsilon$ spatial dimensions
and $\eta$ is the anomalous dimension of the $\varphi_{\ell}$
field. The numerical constants $C$, $a_{1,2}$ can be obtained both
in the $\epsilon$ expansion and in the fixed dimension RG
performed directly in two spatial dimensions. For the $\epsilon$
expansion we have $C^{-1} = 8 \pi^2$, $a_1 = 3$, $a_2 = 6$. For
the fixed dimension expansion we have $C^{-1} = 8 \pi$, $a_1 =
8/3$, $a_2=4$. Using Eq.~(\ref{ugam}), the equation for the
quartic couplings becomes
\begin{eqnarray}
\frac{d \gamma_{mn}}{d l} &=& (\epsilon - 2 \eta) \gamma_{mn} -
C_d \sum_{\ell} \Biggl( \frac{1}{2} \gamma_{m\ell} \gamma_{2n-m,
n-\ell} \nonumber \\ &+&
\gamma_{\ell n } \gamma_{m+n-\ell,n} + \gamma_{\ell,m-n} \gamma_{ \ell-n,m-n} \Biggr) \nonumber \\
&-& a_2 e^4 C_d ( \delta_{n0} + \delta_{mn} ) \label{rgflow2}
\end{eqnarray}
We have verified that this flow equation for the couplings
$\gamma_{mn}$ preserves the symmetry constraints in Eq.~(\ref{r3})
(for rectangular symmetry) and Eq.~(\ref{r4}) (for square
symmetry).

We searched for fixed points of Eqs.~(\ref{rgflow1}) and
Eqs.~(\ref{rgflow2}) for small values of $q$, and found that all
were unstable (as expected from Refs.~\onlinecite{hlm,chen}). The
analysis becomes rapidly more complicated with increasing values
of $q$, and we do not have general understanding of the flow
structure of these equations.

\subsection{Deviations from density $p/q$}
\label{sec:irrat}

Our formalism has so far been explicitly designed for rational
boson densities $f$ obeying Eq.~(\ref{e9}). In principle, it would
appear then we can model essentially any density simply by taking
$q$ large enough. However, this is not an attractive way to
proceed for reasons discussed below.

First, recall that there are $q$ minima in the Brillouin zone, and
so these minima become more closely spaced with increasing $q$.
So, to consider these minima independent degrees of freedom (as is
implicitly done in our continuum field theory), we can only work
at momenta which are significantly smaller than the spacing
between the minima. Hence the continuum theory of
Section~\ref{sec:cft} only applies beyond a large length scale
which increases linearly with $q$.

Second, the $q$ minima are very shallow for large $q$.  In other
words, the bandwidth of the lowest lowest energy Hofstadter band
containing these minima is very narrow.  When integrating out the
vortex modes away from the minima to arrive at the continuum field
theory, one encounters energy denominators determined by this
bandwidth.  In fact, this is likely the most severe restriction on the
na\"ive application of the $q$-vortex method.  In particular, one may
argue that for $p=1$ (and likely for any $p$ by some refinement of
this argument) the bandwidth of the lowest Hofstadter band is {\sl
  exponentially small}, $O(e^{-c q})$ with some $O(1)$ constant $c$,
for large $q$.\footnote{We were unable to find a similar result in the
  literature, though it or a more precise refinement of it likely
  exists.  Therefore we give a brief argument here.  For $f=1/q$, the
  magnetic field is very small, and for the lowest energy states the
  continuum limit is a good zeroth order approximation.  The lowest
  Hofstadter band should then be regarded as the lowest Landau level
  (LLL) of the continuum problem, weakly broadened by the periodic
  potential of the lattice.  The states in the LLL correspond to
  electrons in Landau orbits that can be taken as Gaussian (e.g. in
  symmetric gauge) packets with a width of order the magnetic length,
  $\ell_B \sim \sqrt{q}$.  The broadening of this LLL is due to the
  non-zero harmonics of the periodic potential, which oscillates on
  the lattice scale.  Its matrix elements in the LLL are thus of order
  the Fourier transform of the aforementioned Gaussian packet at a
  reciprocal lattice vector $Q=(2\pi,2\pi)$.  Hence one expects the
  bandwidth $e^{-{\rm const} Q^2\ell_B^2 } \sim e^{-c q}$.  }  This is
a much smaller energy scale than the splitting between Landau
bands, which is also small but $O(1/q)$ for large $q$.  A
quantitative physical consequence of the narrow bandwidth is a
reduction of the vortices' tendency to condense, i.e. an
enhancement of the domain of the boson superfluid state.

As a consequence of these difficulties, it is preferable to take the
continuum limit appropriate to the low energy fluctuations of some
nearby insulating commensurate state with a moderate value of $q$. The
actual density of the boson system may not exactly equal $p/q$.
However, it is easy to modify the theory for the system at density
$p/q$ (presented in Section~\ref{sec:cft}), to allow for this density
deviation. As the boson density is the average $A_{\mu}$ `magnetic'
flux, we simply have to replace the action $\mathcal{S}_0$ in
Eq.~(\ref{s0}) by
\begin{eqnarray}
\mathcal{S}_0 &=& \int d^2 r d \tau \left( \sum_{\ell = 0}^{q-1}
\left[ |(\partial_\mu - i A_{\mu} ) \varphi_\ell |^2 + s
|\varphi_\ell |^2 \right] \right. \nonumber \\
&+& \left. \frac{1}{2e^2} \left( \epsilon_{\mu\nu\lambda}
\partial_\nu A_\lambda - \frac{2 \pi \delta f}{a^2} \delta_{\mu\tau} \right)^2
\right),
\label{s0p}
\end{eqnarray}
where $\delta f$ is the difference between $p/q$ and the actual
boson density, and $a$ is the lattice spacing. All the higher
order terms in the action remain as discussed in
Section~\ref{sec:cft}.

The theory in Eq.~(\ref{s0p}) has a structure similar to the
Ginzburg-Landau model for a `superconductor' in a `magnetic'
field, and has corresponding possibilities for its phases. It can
behave like a type I `superconductor' and expel the flux $\delta
f$, while condensing the $\varphi_\ell$: this clearly yields the
commensurate Mott insulator with density $p/q$. In the `normal'
phase of this `superconductor' the $\varphi_\ell$ are uncondensed
and flux $\delta f$ pierces the system: this is the superfluid
with density $p/q + \delta f$. This `normal' phase can also have a
condensate of vortex-anti-vortex composites with some of $\langle
\rho_{mn} \rangle$ non-zero, and this is a supersolid with density
$p/q + \delta f$, but with density wave order similar to that in
the commensurate Mott insulator. Finally, the model
Eq.~(\ref{s0p}) can also behave like a type II `superconductor'
and allow the `flux' to penetrate the system in an Abrikosov
lattice: this is an incommensurate insulator (or floating Wigner
solid) in which the particle density is $p/q+\delta f$. Partial
`flux' penetration is also possible, and this yields Mott
insulators at other densities.

\section{Boson fractionalization}
\label{sec:frac}

The discussion of the previous sections generalizes the dual
effective action of Refs.~\onlinecite{deccp} (and its
predecessors) for $f=1/2$.  In that work, it was pointed out that
for $f=1/2$ the theory with $\gamma_{01}<0$ is itself equivalent
to a {\sl fractionalized} representation in which the elementary
boson is split into two ``half'' bosons, carrying equal
($\frac{1}{2}$ the elementary value) physical charge but opposite
values of an emergent $U(1)$ gauge charge.  These fractional
particles are {\sl deconfined} at the critical point in the sense
that the gauge magnetic flux is conserved by the fixed point
theory.  In the language of the present paper, this conservation
law is nothing but the conservation of the {\sl
  difference} of the two ($\zeta_0$ and $\zeta_1$) vorticities.

Two important features are critical to this conclusion. First, we
used the equivalence of the dual $2-$vortex action -- in a
particular regime ($\gamma_{01}<0$) -- to a fractionalized
$1/2$-boson representation.  This mapping is non-trivial and
indeed does not generalize to all $f$, as we will discuss below.
Second, even when such a representation does obtain, deconfinement
at the QCP is contingent upon the fixed point theory possessing
the appropriate emergent conservation law(s).  This is true for
$f=1/2$ roughly because the most relevant operator to
violate the conservation law ($(\zeta_0^*\zeta_1^{\vphantom*})^4 +
{\rm h.c.}$) is sufficiently high order that it can be
persuasively argued\cite{deccp} to be irrelevant.  In general, the
relevance or irrelevance of such operators is far from obvious, as
the interacting $q$-vortex gauge theory in $2+1$ dimensions is
strongly coupled, so that the critical indices needed to answer
this question are not calculable analytically.  For this reason,
we will have little to say in this paper about this second issue,
which is best resolved by numerical (Monte Carlo) studies.

In this section we will however discuss the first requirement for a
fractionalized boson interpretation of these Mott QCPs.  Perhaps
surprisingly, we will see that this can be satisfied in a theory with
full square lattice symmetry only for a rather special set of values
of rational $f$.  We do not at present have a general criteria to
determine these values of $f$.  By explicit example, we find that a
fractionalized formulation is possible for $f=1/2,1/4,1/8,1/9$, and
have proven (see below and Appendix~\ref{sec:noq3}) that none exists
for $f=1/3$.  We speculate that the sequences of $f=p/q$ with
$q=n^2,2n^2$ and integer $n$ all admit a fractionalized formulation.
If the microscopic symmetry is relaxed from square to rectangular
(i.e. invariance under only twofold rather than fourfold rotations),
however, the fractionalization requirements can be satisfied for any
rational $f$. We show how these conclusions can be obtained both from
an analysis of the dual $q$-vortex action of this paper, and from a
complementary direct analysis of the original boson problem.  The
agreement between these two completely distinct methods provides an
important check on the results, as well as useful physical
interpretation.  The direct method is a generalization of the $U(1)$
gauge theory approach employed in Ref.~\onlinecite{deccp}.

Although this analysis provides a formal means of understanding
fractionalization at such Mott QCPs, a more physical picture is
also desirable.  In Ref.~\onlinecite{LevinSenthil}, Levin and
Senthil provided such a simple real-space picture of charge-$1/2$
excitations for $f=1/2$ coming from the Mott side of the
transition.  They showed that these particles can be understood as
$Z_4$ vortices -- points of connection of four elementary domain
walls -- in the density wave (valence bond solid) order parameter.
For the specific case of $f=1/4$, we discuss, in
Section~\ref{sec:appr-fract-from-1}, a similar picture -- as $Z_4
\times Z_4$ vortices -- for the charge $1/4$ excitations that
(may) become deconfined at the Mott transition.

\subsection{Fractionalization in the vortex picture}
\label{sec:fract-vort-pict}

Let us consider the general $q$-vortex action at filling $f=p/q$.
By blind analogy with half filling ($f=1/2$), one might expect
this to describe a QCP in which the elementary boson has been
fractionalized into $q$ bosons with charge $1/q$.  If we examine
this proposition in a little detail, it becomes clear that this is
not quite generally correct.

For half-filling,\cite{deccp} this conclusion was based on a
duality mapping of the 2-vortex theory back to a half-boson
theory.  The physical content of this mapping is that a `vortex'
in a single vortex field $\varphi_\ell$ (this is a sometimes
confusing double duality which unfortunately cannot be avoided in
this particular method of analysis) is a point particle whose
creation or annihilation operator becomes the fundamental field in
a new representation.  Here and in the following we use single
quotes to distinguish this `vortex' from a true physical vortex,
which is created by a linear combination of $\varphi_\ell$ fields.
As in any other Ginzburg-Landau-like theory, such a `vortex' will
bind a quantized amount of `flux' $\int\! d^2r\,
\epsilon_{ij}\partial_i A_j$. The `flux' however is $2\pi$ times
the physical {\sl charge} (boson number), so this `vortex' is
actually a particle carrying some amount of boson charge, which we
will see is fractional under favorable circumstances.

The above argument has one crucial un-stated assumption: that
vortex-like topological defects in the $\varphi_\ell$ fields are
well-defined.  This is true only if the magnitudes
$|\varphi_\ell|$ are constrained to be approximately constant,
i.e. that the dominant terms in the action are minimized by this
condition.  Moreover, phases of these fields should be
unconstrained.  Formally, this requirement is necessary to pass
from the soft-spin Ginzburg-Landau theory of the $\varphi_\ell$
fields to a hard-spin description in terms of phase-only
variables.  Duality transformations are formulated in a hard-spin
description.

Actually, we can relax this requirement slightly, as already hinted by
the mean-field analysis of the previous section.  In particular, since
the quadratic action ${\cal S}_0$ is invariant under a global $U(q)$
unitary rotation of the $\varphi_\ell$ fields, it is sufficient that
some unitarily transformed fields $\zeta_\ell = \sum_{\ell'}
U_{\ell\ell'} \varphi_{\ell'}$ (with $U^\dagger
U^{\vphantom\dagger}=1$) can be regarded as having approximately
constant magnitude, $|\zeta_\ell|=\overline\varphi$, a constant, but
arbitrary phases.  One may think of this as follows.  Consider the
evolution of the system from the superconducting to insulating state,
by reducing $s$ starting from $s>0$.  Due to fluctuations, the actual
location of the critical point $s=s_c$ at which vortex condensation
occurs is {\sl reduced} to $s_c<0$.  Thus one may expect that locally,
for $s_c \lesssim s <0$, the $\zeta_\ell$ fields may develop some
non-zero magnitude.  This occurs because of a balance between the
negative quadratic term $-|s|\sum_\ell |\zeta_\ell|^2$ and the quartic
terms in ${\cal L}_4$.  From the form of the negative quadratic term,
this balance clearly favors states with a non-zero generalized
``radius'' $\sqrt{\sum_\ell|\zeta_\ell|^2}$ in $\zeta_\ell$ space.
The preferred directions in this $2q$-dimensional (due to real and
imaginary parts of each $\zeta_\ell$) space, however, are determined
by the terms in ${\cal L}_4$.  To obtain a phase-only description, one
needs that the dominant terms in ${\cal L}_4$ favor states with
$|\zeta_\ell|=\overline\varphi$.

This requirement places strong constraints on the form of ${\cal
  L}_4$. In particular, one must have that, for fixed
$\sum_\ell|\zeta_\ell|^2= q \overline\varphi^2$, the minima of
${\cal
  L}_4$ satisfy $|\zeta_\ell|=\overline\varphi$ for all $\ell$.
Explicit examples with rotational invariance can be seen for $q=2,4$
from Sec.~\ref{sec:q2},\ref{sec:q4}.  For $q=2$, this is true if
$\gamma_{01}<0$.  For $q=4$ this is true in region E when $\lambda=0$.
Thus if $\lambda$ is irrelevant in region E at the QCP, the
requirement is satisfied.  On the other hand, for $q=3$, it is clearly
not satisfied.  For $\gamma_{01}=0$, there is a full $2q-1=5$-sphere
of degenerate states, clearly of different topology than the $3$-torus
of states of three angular variables.  For $\gamma_{01}> 0$, the
degeneracy of minima of ${\cal L}_4$ is just a discrete multiple of
the overall $U(1)$ (singular angular degree of freedom) gauge
symmetry.

We do not have a general analysis of when the constant amplitude
condition can be satisfied.  One can, however, place a necessary
condition on the existence of any Lagrangian invariant under the
symmetries of the problem for which all minimum action configurations
satisfy $|\zeta_\ell|=\overline\varphi$ and conversely all such
configurations have minimal action.  In particular, for any such
configuration, any new configuration equivalent to it by symmetry must
have equal action.  Thus if these are the unique configurations with
minimal action, all symmetry operations must preserve this condition.
That is, the space of states with $|\zeta_\ell|=\overline\varphi$
should be closed under all symmetry operations.  Unfortunately, this
is not trivial to check, since we still have a large freedom to choose
the unitary transformation $U_{\ell\ell'}$ defining the $\zeta_\ell$
fields.

For the simpler case in which the requirement of rotational invariance
is removed, i.e. for a system with rectangular rather than square
symmetry, it is straightforward to see that this condition can be
achieved for all $f$.  This is easily seen from
Eqs.~(\ref{r1})~(\ref{r3}).  In particular, a quartic term of the form
${\cal L}_4= \frac{u}{2}\sum_\ell |\varphi_\ell|^4$ is allowed for all
$q$, and favors equal amplitude states.  It is clearly invariant under
all symmetry operations save rotations.  Hence no unitary rotation is
required, and we can simply take $\zeta_\ell=\varphi_\ell$.

The case with fourfold rotational symmetry remains an interesting
open problem.  While we have not yet obtained a general solution,
we formulate the problem mathematically.  In particular, the
requirement that the constant magnitude configurations be closed
under the action of the PSG is quite strong.  In particular, an
arbitrary such configuration, $\zeta_\ell =\overline\varphi
e^{i\vartheta_\ell}$, must map to another such configuration.
Since the transformations are linear, and the $\vartheta_\ell$ are
arbitrary, this requires that the representation matrices of the
PSG elements in this basis must take the form $G=\Lambda P$,
where, as we noted in Section~\ref{sec:intro}, $\Lambda$ is a
unitary diagonal matrix (i.e. a matrix whose only non-zero
elements are complex numbers of unit magnitude on the diagonal),
and $P$ is a permutation matrix. Thus it is sufficient to
determine whether the algebra of the generators of the PSG (unit
$x$ and $y$ translations, and $\pi/2$ rotations) can be faithfully
represented by such matrices.

We believe that such {\em permutative representations\/} of the full
(square symmetry) PSG exist only for specific $q$.  As indicated
above, we have constructed explicit examples in detail in the text for
$q=2,4$.  The ability to do so in these two cases is not obvious in
the Landau gauge construction of the previous Sections. In
Appendix~\ref{sec:symmetric}, we show that it can be clarified for
$q=4$ (for $q=2$ it is rather straightforward) by choosing a different
``symmetric'' gauge which more closely respects the fourfold
rotational symmetry of the lattice.  The dual vortex theory
constructed in this gauge immediately yields a permutative
representation of the PSG.  Generally, a permutative representation
exists only for specially chosen $q$. In Appendix~\ref{sec:noq3}, we
prove that no such representation exists for $q=3$, and also give
explicit examples of such representations for $q=8,9$.  The latter
examples are obtained from a construction which can be generalized to
$q=n^2,2n^2$ with any integer $n$, but which we cannot guarantee
yields a solution.  Nonetheless, we speculate that all members of this
family support a permutative representation.

For those cases for which the phase-only description can be
achieved, we return to the question of charge fractionalization by
considering the `flux' (charge) of the dual `vortices'.  Supposing
the constant amplitude condition, for slowly varying
$\vartheta_\ell$ the effective phase-only action can be written
\begin{equation}
  \label{eq:rhos}
  S_{eff} = \int\! d^2r d\tau\, \sum_\ell \frac{\rho'_s}{2}
  |\partial_\mu \vartheta_\ell - A_\mu|^2 + \frac{1}{2e^2}
  (\epsilon_{\mu\nu\lambda}\partial_\nu A_\lambda)^2,
\end{equation}
with $\rho'_s = 2 \overline\varphi^2$.  Consider a fixed `vortex'
(independent of $\tau$) in one -- say $\vartheta_0$ -- of the $q$
phase fields, centered at the origin $r=0$.  One has the spatial
gradient $\vec\nabla\vartheta_0 = \hat\phi/r$, all other
$\vec\vartheta_\ell=0$ for $\ell\neq 0$ (here $\hat\phi =
(-y,x)/r$ is the tangential unit vector) .  Clearly, the action is
minimized for tangential $\vec{A} = A \hat\phi$.  Far from the
`vortex' core, the Maxwell term
$(\epsilon_{\mu\nu\lambda}\partial_\nu A_\lambda)^2$ is
negligible, so one need minimize only the first term in
Eq.~\ref{eq:rhos}.  The corresponding Lagrange density at a
distance $r$ from the origin is thus
\begin{equation}
  \label{eq:vortexlag}
  {\cal L}'_v = \frac{\rho'_s}{2} \left[(1/r-A)^2 + (q-1)A^2\right].
\end{equation}
Minimizing this over $A$, one finds $A=1/(qr)$.  Integrating this
to find the flux gives
\begin{equation}
  \label{eq:fluxprime}
  \oint \vec{A}\cdot d\vec{r} = \frac{2\pi}{q}.
\end{equation}
Since the physical charge is just this dual flux divided by
$2\pi$, the `vortex' in $\varphi_0$ indeed carries fractional
boson charge $1/q$.

\begin{figure}
\centering \ifig[width=3in]{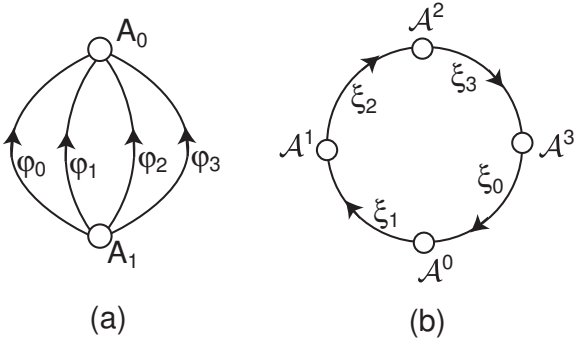} \caption{Graphical
representation of the field theories discussed towards the end of
Section~\protect\ref{sec:fract-vort-pict} in ``moose'' diagrams
for $q=4$. The circles represent non-compact U(1) gauge fields,
while the directed lines represent complex scalar fields which are
charged +1/-1 under the gauge field site at their beginning/end.
The $\varphi_\ell$ vortex theory in (a) is specified in and above
Eq.~(\protect\ref{moose1}) with $A_\mu = A_{0 \mu} - A_{1 \mu}$.
Its dual theory of charge $1/q$ bosons $\xi_\ell$ (which are
`vortices' in $\varphi_\ell$) in (b) is specified in
Eqs.~(\protect\ref{moose2}) and (\protect\ref{moose3}). In both
diagrams the center-of-mass gauge field, $A_{0\mu} + A_{1\mu}$ or
$\sum_{\ell} \mathcal{A}^{\ell}_{\mu}$, decouples from all matter
fields, and is therefore immaterial. If we visualize the
$A_0$/$A_1$ gauge fields on the north/south pole of a sphere, and
the ring in (b) along the equator, then the geometric
interpretation of the duality becomes evident. It is also amusing
to note that the two diagrams are identical only for $q=2$, and
hence only this theory is self-dual, as noted by Motrunich and
Vishwanath. \protect\cite{mv}} \label{figmoose}
\end{figure}

This simple analysis can be confirmed by a straightforward but
more formal duality calculation on a lattice-regularized hard-spin
$\vartheta_\ell$ model. In the continuum, the theory of the $q$
$\varphi_\ell$ vortex fields coupled to the single non-compact
U(1) gauge field $A_\mu$ is described by the action $\mathcal{S}_0
+ \int d^2 x d \tau \mathcal{L}_4$ where $\mathcal{S}_0$ is as in
Eq.~(\ref{s0}) and $\mathcal{L}_4$ has the structure (in an
appropriate basis which realizes the permutative PSG)
\begin{equation} \mathcal{L}_4 =
\sum_{m,n} v_{mn} |\varphi_m|^2 |\varphi_n|^2, \label{moose1}
\end{equation}
where the co-efficients $v_{mn}$ are constrained by the
permutative PSG; specifically, the matrix $v$ must be invariant
under all the permutative transformations $P$ of the PSG:
$v=P^{-1}vP$. As we have discussed above, this theory has $q-1$
globally conserved charges associated with the symmetries of phase
rotations of the differences of the $\vartheta_\ell$ fields. The
na\"ive dual form of this continuum theory obtained
\cite{ssmott,sudbo} from a lattice-regularized hard-spin
$\vartheta_\ell$ model has the following continuum representation
in a theory of $q$ bosons $\xi_\ell$ (these are the `vortices' in
the vortex field $\varphi_\ell$) each carrying boson charge $1/q$,
and $q$ non-compact U(1) gauge fields
$\widetilde{\mathcal{A}}^{\ell}_{\mu}$ with the action
\begin{eqnarray}
\mathcal{S}_\xi &=& \int d^2 x d \tau \Biggl [ \sum_{\ell=0}^{q-1}
\left[\left| \left(
\partial_{\mu} - i \widetilde{\mathcal{A}}^{\ell}_{\mu}  \right)
\xi_{\ell} \right|^2 + \widetilde{s} |\xi_\ell |^2 \right]  \nonumber \\
&+& \sum_{m,n} K_{mn} \left( \epsilon_{\mu\nu\lambda}
\partial_{\nu} \widetilde{\mathcal{A}}^{m}_{\lambda} \right)  \left( \epsilon_{\mu\rho\sigma}
\partial_{\rho} \widetilde{\mathcal{A}}^{n}_{\sigma} \right) \nonumber \\
&+& \widetilde{K} \left( \sum_{\ell=0}^{q-1}
\widetilde{\mathcal{A}}^{\ell}_{\mu} \right)^2 + \sum_{m,n}
\widetilde{v}_{mn} |\xi_m|^2 |\xi_n|^2 \Biggr] .
~~~~~~~~~\label{moose2}
\end{eqnarray}
As below Eq.~(\ref{moose1}), the matrices $K$ and $\widetilde{v}$
are also invariant under the permutative transformations $P$ of
the PSG. Note that the `center-of-mass' gauge field $\sum_{\ell}
\widetilde{\mathcal{A}}^{\ell}_{\mu}$ is `Higgsed' out into a
massive mode by the coupling $\widetilde{K}$. It is therefore
convenient to make a new choice of variables in which this
`center-of-mass' mode has been explicitly set to zero
\begin{equation}
\widetilde{\mathcal{A}}^{\ell}_{\mu} = \mathcal{A}^{\ell}_{\mu} -
\mathcal{A}^{\ell-1}_{\mu}\;. \label{moose3}
\end{equation}
In these new variables, the $\sum_{\ell} \mathcal{A}^{\ell}_{\mu}$
gauge field is decoupled from all matter fields, and so is a
redundant degree of freedom which drops out. The conserved gauge
fluxes of the remaining $(q-1)$ $\mathcal{A}^{\ell}_{\mu}$ gauge
fields represent the $(q-1)$ conserved currents of the global
symmetries of the $\varphi_\ell$ theory noted above. The field
theories discussed in this paragraph have a convenient
representation in `moose' or `quiver' diagrams \cite{nima}, as
illustrated in Fig~\ref{figmoose}.

The following subsections will present a careful analysis of the
reverse of the above duality, with special attention being paid to
lattice Berry phases and symmetries.

\subsection{Fractionalization in the direct picture: $U(1)^{q-1}$ slave bosons}
\label{sec:fract-direct-pict}

Having seen that the dual $q$-vortex theory is under particular
circumstances itself dual to a theory of $q$ charge $1/q$ bosons,
it is natural to ask whether this fractionalized representation
can be directly obtained from the original boson problem, without
recourse to duality.  We show how this is done here, and further
how the $q$-vortex theory can be recovered from this route by
dualizing the fractional boson theory directly.

\subsubsection{Construction of an effective gauge theory Hamiltonian}

Starting from a microscopic Hamiltonian like Eq.~(\ref{hubbard}),
one may introduce charge $1/\tilde{q}$ slave-rotor variables,
$\hat{\phi}_{i\ell},\hat{n}_{i\ell}$, with
$[\hat{\phi}_{i\ell},\hat{n}_{j\ell'}]=i\delta_{ij}
\delta_{\ell\ell'}$:
\begin{eqnarray}
  \label{eq:slaverotor}
  e^{i\hat{\phi}_i} & = & \prod_{\ell=0}^{\tilde{q}-1}
  e^{i\hat{\phi}_{i\ell}}, \\
  \hat{n}_i & = & \frac{1}{\tilde{q}}\sum_{\ell=0}^{\tilde{q}-1}
  \hat{n}_{i\ell} .
\end{eqnarray}
Clearly, the operator $\hat{n}_{i\ell}$ counts the number of
flavor $\ell$ bosons, each of which has physical charge
$1/\tilde{q}$. This is a microscopically faithful representation
of the original rotor states if we, e.g. impose the constraints
\begin{equation}
  \label{eq:hardconstraint}
  \hat{n}_{i\ell} -\hat{n}_{i\ell'}=0
\end{equation}
for all $\ell,\ell'$.  Thus, the physical state with
$\hat{n}_i=n_i$ is represented by the state with all slave rotors
identical $\hat{n}_{i\ell}=n_i$.  In this subsubsection we will
keep $\tilde{q}$ arbitrary, with no a priori connection to the
filling $f=p/q$.  This connection is made in the following
subsubsection.

Eq.~(\ref{eq:hardconstraint}) expresses $\tilde{q}-1$ independent
constraints, which generate the same number of independent
unphysical gauge rotations of the $\hat{\phi}_{i\ell}$.  That is,
any rotation $\hat{\phi}_{i\ell} \rightarrow
\hat{\phi}_{i\ell}+\chi_{i\ell}$ with $\sum_\ell \chi_{i\ell}=0$
leaves the physical boson creation/annihation operators invariant.
It should be noted that the representation above has another set
of discrete ($S_{\tilde{q}}$) gauge symmetries: the $\tilde{q}$
slave rotor variables can be permuted arbitrarily and
independently on each site without changing $\hat{n}_i$ or
$\hat{\phi}_i$.  If the constraint in
Eq.~(\ref{eq:hardconstraint}) is imposed exactly on every lattice
site, however, the $S_{\tilde{q}}$ gauge symmetry does not lead to
any additional constraints, and it leaves the $U(1)^{\tilde{q}-1}$
constrained states invariant (i.e. these states form a trivial
identity representation of $S_{\tilde{q}}$).

Such a microscopic slave-rotor rewriting of the problem has,
unfortunately, no physical content.  It becomes physical only if,
on long scales, a set of renormalized effective degrees of freedom
``descended'' from the slave rotors become physically meaningful
quasiparticles, which locally violate these hard constraints.  We
would like, therefore, to consider a renormalized effective theory
in which the fractional rotors have some local independence.  In
principle such a theory can be constructed by implementing the
hard constraints in a path integral formulation, and studying
fluctuations around a mean-field solution of this lattice field
theory.  For the universal phenomenological purposes of this
paper, however, this is unnecessary, and we instead prefer to
proceed on conceptual grounds.  Importantly, it is essential that
globally, in any finite system, all quantum numbers be physical,
i.e. the total number of bosons must be integer.  This goal is
satisfied along with local fluctuations of the fractional rotors
by introducing dynamical compact $U(1)$ gauge fields:
\begin{equation}
  \label{eq:gauge}
  [{\cal E}_{i\alpha}^\ell,{\cal A}_{j\beta}^{\ell'}]=
  i\delta_{ij}\delta_{\alpha\beta}\delta^{\ell\ell'},
\end{equation}
with $\{\alpha,\beta\}\in \{x,y\}$.  Here ${\cal
E}_{i\alpha}^{\ell}$ has integer eigenvalues, and ${\cal
A}_{i\beta}^{\ell}$ is a $2\pi$-periodic angular variable.  To
keep the formalism as symmetric as possible, we have introduced
$\tilde{q}$ ($\ell,\ell' =0\ldots \tilde{q}-1$) such fields
despite there being only $\tilde{q}-1$ independent constraints.
The redundancy is eliminated by requiring
\begin{equation}
  \label{eq:redundant}
  \sum_{\ell=0}^{\tilde{q}-1} \vec{\cal E}_i^\ell = 0.
\end{equation}
The vector overline indicates the usual two-dimensional vector
($x,y$) notation.  With these
fields, we replace Eq.~(\ref{eq:hardconstraint}) by a set of
Gauss' law constraints
\begin{equation}
  \label{eq:gauss}
  \left[\vec{\Delta} \cdot \vec{\cal E}^\ell\right]_i = \hat{n}_{i,\ell}
  -\hat{n}_{i,\ell+1},
\end{equation}
where, as usual, $\hat{n}_{i,\tilde{q}}=\hat{n}_{i{\scriptscriptstyle
    0}}$, and the $\vec\Delta\cdot$ symbol indicates a lattice
divergence.  If we do not allow the gauge electric flux to exit the
system boundaries (or there are no boundaries), Gauss' theorem applied
to Eq.~(\ref{eq:gauss}) implies as desired that the total boson charge
in the system ($\frac{1}{\tilde{q}}\sum_{i\ell} \hat{n}_{i\ell}$) is
integral.  The associated effective Hamiltonian is then na\"ively
\begin{eqnarray}
  \label{eq:heff1}
  {\cal H}_{\tilde{q}}^0 & = & -t \sum_{i\ell} \cos( \Delta_\alpha
    \hat{\phi}_{i\ell}- {\cal A}_{i\alpha}^\ell + {\cal
      A}_{i\alpha}^{\ell-1} ) +u (\hat{n}_{i\ell}-f)^2\nonumber \\
  & & + \frac{v}{2} \sum_{i\ell} |\vec{\cal E}_i^\ell|^2 -
  K \sum_{a\ell} \textrm{``cos''}(\vec\Delta \times \vec{\cal
    A}^\ell)_a.
\end{eqnarray}
In the second line above we have added quotes around the cosine
term to indicate that care must actually be taken here to ensure
that the unphysical center of mass of the gauge fields ($\sum_\ell
\vec{\cal
  A}^\ell$) is decoupled from the other physical orthogonal linear
combinations.  This will be achieved in the path integral
formulation to which we will soon turn through the use of a
Villain potential.

We now come to a subtle and crucial point in the gauge theory
formulation.  Eqs.~(\ref{eq:gauss}),~(\ref{eq:heff1}) incorporate
on long scales the gauge constraints necessary to remove the
unphysical aspects of the $U(1)^{\tilde{q}-1}$ gauge redundancy.
Further, the Gauss' law relations as formulated in
Eq.~(\ref{eq:gauss}) largely remove the $S_{\tilde{q}}$
permutation gauge symmetry.  What remains of the latter is an
invariance under some group of {\sl global} permutations, under which
\begin{eqnarray}
  \label{eq:globperm}
  \hat{n}_\ell & \rightarrow & \hat{n}_{P(\ell)}, \\
  \hat{\phi}_\ell & \rightarrow & \hat{\phi}_{P(\ell)}, \nonumber
\end{eqnarray}
where $P(\ell)$ is a permutation on $0\ldots \tilde{q}-1$.  These
permutations $P(\ell)$ are not arbitrary: only for particular choices
can linear transformations of $\vec{\cal A}^\ell$ and $\vec{\cal
  E}^\ell$ be chosen along with Eqs.~(\ref{eq:globperm}) to keep
Eq.~(\ref{eq:heff1}) invariant.  This global permutation invariance in
general depends upon the particular value of $\tilde{q}$.  At a
minimum, it includes all cyclic permutations (generated by the
elementary cyclic permutation $P(\ell)=\ell+1$, for which $\vec{\cal A}^\ell
\rightarrow \vec{\cal A}^{\ell+1}$ and $\vec{\cal E}^\ell \rightarrow
\vec{\cal E}^{\ell+1}$ leaves Eqs.~(\ref{eq:heff1}) invariant), but it
is larger for special values of $\tilde{q}$.  The permutation symmetry
has been reduced to a global one since the gauge fields couple
different lattice sites, so bosons can no longer be permuted locally.
This permutation invariance, if the microscopic constraints,
Eq.~(\ref{eq:hardconstraint}), are strictly enforced, has no
consequence.  In our effective theory, however, for which states
violating these constraints are included, this global symmetry has
meaning.  Since the original microscopic boson model has no
corresponding symmetry, this will lead to unphysical aspects of the
effective theory if it is not corrected somehow.

An appealing and simple way to remove the permutation invariance
is to add a term to the Hamiltonian which removes this symmetry.
We will construct such a term, guided by a few requirements.
First, it should reduce to a trivial constant if the hard
constraints in Eq.~(\ref{eq:hardconstraint}) are enforced exactly.
Second, it should not affect the average density of the slave
bosons.  Third, for reasons that will become apparent in the next
subsubsection, it should enlarge the na\"ive unit cell of the
fractional boson kinetic terms.  Fourth, it should {\sl allow for
some projective implementation of all the symmetry
  operations of the lattice}.  The last requirement is the most
non-trivial, and moreover seems somewhat at odds with the third
one.  As we will see, it precludes us from constructing an
appropriate slave-boson theory at an arbitrary filling factor.
Our tentative choice is to include a term of the form
\begin{eqnarray}
  \label{eq:staggpot}
  {\cal H}_{\tilde{q}}^1 = - \mu_s \sum_{i\ell} \left(m_{i\ell}-
  \frac{1}{\tilde{q}}\right)\hat{n}_{i\ell}.
\end{eqnarray}
Here $m_{i\ell}$ is an integer field defined by introducing
$\tilde{q}$ sublattices (the choice of sublattices will be made
later), with
\begin{equation}
  \label{eq:mdef}
  m_{i\ell} = \left\{
\begin{array}{cl} 1 & \mbox{for $i\in$ sublattice $\ell$} \\
0 & \mbox{otherwise}
\end{array} \right. .
\end{equation}
This term clearly satisfies the first three requirements, but the
implementation of symmetry operations is problematic.  In
particular, as we have actually tried to do (requirement 3),
Eq.~(\ref{eq:staggpot}) breaks the na\"ive translational and
rotational invariances of the original model.  Indeed, a
non-trivial implementation of these symmetries is possible only
for special values of $\tilde{q}$ and/or lattice symmetries.

A relatively simple case is that of rectangular symmetry.  In this
case, it is natural to choose the set of $\tilde{q}$ sublattices
by defining the $\ell^{th}$ sublattice as the set of sites $i$ for
which $i_x=\ell (\mbox{mod $\tilde{q}$})$ -- just labelling
sequentially columns by $\ell=0,1,\ldots\tilde{q}-1,0,\ldots$.
Then translations along $y$ are trivially preserved, while an $x$
translation is implemented by translating by one lattice spacing
{\sl and} simultaneously cyclically permuting $\ell\rightarrow
\ell+1$.

Unfortunately, this prescription fails to provide a means of
implementing the $\pi/2$ rotation operation of the square symmetry
group.  In general, we have been unable to find {\sl any} choice
of sublattices which does so for arbitrary $\tilde{q}$.  In fact,
the only two cases for which we have succeeded in finding a
sublattice choice that fully implements the square symmetry group
are $\tilde{q}=2,4$. For $\tilde{q}=2$, it is achieved by taking
the two checkerboard sublattices of the square lattice.  Then both
$x$ and $y$ translations and $\pi/2$ rotations about a dual
lattice site must be accompanied by a (there is only one!) cyclic
permutation, while a $\pi/2$ rotation about a direct lattice site
requires no permutation.  For $\tilde{q}=4$, one defines the four
sublattices according to
\begin{equation}
  \label{eq:foursubs}
  i \in \mbox{sublattice}  \left\{
    \begin{array}{cl}
      0 & \mbox{for $(i_x,i_y)=(0,0)\, ({\rm mod} 2)$}, \\
      1 & \mbox{for $(i_x,i_y)=(1,0)\, ({\rm mod} 2)$}, \\
      2 & \mbox{for $(i_x,i_y)=(1,1)\, ({\rm mod} 2)$}, \\
      3 & \mbox{for $(i_x,i_y)=(0,1)\, ({\rm mod} 2)$}.
\end{array} \right. .
\end{equation}
These are constructed by choosing the sites of the $2\times 2$
square including the origin at its lower-left corner, labeling
them counter-clockwise $0,1,2,3$, and translating this square
vertically and horizontally to cover the plane.  This can also be
seen (less trivially) to preserve all symmetry operations.  For
instance, a $\pi/2$ rotation about a dual lattice site should be
combined with a cyclic permutation, and a unit translation in the
$x$ direction is effected by
\begin{eqnarray}
  \label{eq:txfour}
  \hat{n}_{i0} & \leftrightarrow & \hat{n}_{i+\hat{x},1},\nonumber \\
  \hat{n}_{i2} & \leftrightarrow & \hat{n}_{i+\hat{x},3},\nonumber \\
  \vec{\cal E}_i^0 & \rightarrow & -\vec{\cal E}_{i+\hat{x}}^0,
  \nonumber  \\
  \vec{\cal E}_i^1 & \rightarrow & -\vec{\cal E}_{i+\hat{x}}^3,
  \nonumber  \\
  \vec{\cal E}_i^2 & \rightarrow & -\vec{\cal E}_{i+\hat{x}}^2,
  \nonumber  \\
  \vec{\cal E}_i^3 & \rightarrow & -\vec{\cal E}_{i+\hat{x}}^1,
\end{eqnarray}
and similar relations for $\hat{\phi}_{i\ell},\vec{\cal
A}^\ell_i$. One can catalogue and verify the remaining symmetries
for $\tilde{q}=4$.

Crucially, we see that it is not even possible in principle to
choose a set of sublattices that is preserved under rotations for
most values of $\tilde{q}$, which is a minimal requirement for
successfully implementing rotational symmetry.

\subsubsection{Analysis of the effective gauge theory}
\label{sec:analys-effect-gauge}

Having discussed the construction of the effective gauge theory
${\cal
  H}_{\tilde{q}} = {\cal H}^0_{\tilde{q}} + {\cal H}^1_{\tilde{q}}$ in
Eqs.~(\ref{eq:heff1}),~(\ref{eq:staggpot}), with the constraint of
Eq.~(\ref{eq:gauss}), we now turn to its analysis.  Notably, up to
this point, we have rather arbitrarily chosen to split the
original boson into $\tilde{q}$ parts, without any direct
reference to the filling factor.  From the dual analysis of
Sec.~\ref{sec:fract-vort-pict}, we expect however that the
physically appropriate fraction corresponds to $\tilde{q}=q$.  The
connection is as follows.  Due to the presence of the staggered
potential $\mu_s$, each of the fractional bosons moves in an
environment of reduced translational symmetry, with a
$\tilde{q}$-site enlarged Bravais lattice unit cell.  At a filling
$f=p/q$, the average number of each fractional boson per enarged
unit cell is then $f \tilde{q} = p \tilde{q}/q$.  The smallest
amount of fractionalization then consistent with an integer number
of fractional bosons per unit cell is then $\tilde{q}=q$.
Intuitively, with this condition, it is possible for the
fractional bosons to form a featureless Mott insulator, apart from
the effect of gauge fluctuations.  We therefore now focus on this
case, $\tilde{q}=q$.  The problem as described by ${\cal H}_q =
{\cal H}_q^0 + {\cal H}_q^1$
(Eqs.~(\ref{eq:heff1}),~(\ref{eq:staggpot})), is amenable to a
simple direct analysis.

Consider first the Mott phase.  In Eq.~(\ref{eq:heff1}), this
occurs for large $u \gg t$, for which $\hat{n}_{i\ell}$ will be
approximately good quantum numbers with only only weak local
fluctuations in the ground state.  For a properly chosen range
(fine-tuning is not necessary) of $\tilde\mu_s$, one expects
$\hat{n}_{i\ell}=p m_{i\ell}$ to be a qualitatively good
approximation to the ground state.\footnote{Note that other Mott
  states are possible for $p>1$, e.g. ones in which $p$ of the $q$
  sites of the sublattice unit cell are occupied by one fractional
  boson, rather than one site being occupied by $p$ fractional bosons
  as assumed here for simplicity.  Different such choices correspond
  to different PSGs.}  Having determined the state of the fractional
bosons, one may then study just the pure $U(1)^{q-1}$ gauge theory
\begin{equation}
  \label{eq:puregauge}
  {\cal H}_{U(1)} = \frac{v}{2} \sum_{i\ell} |\vec{\cal E}_i^\ell|^2 -
  K \sum_{a\ell} \textrm{``cos''}(\vec\Delta \times
  \vec{\cal A}^\ell)_a,
\end{equation}
with the simplified Gauss' law constraint
\begin{equation}
  \label{eq:puregauss}
  \left[\vec{\Delta} \cdot \vec{\cal E}^\ell\right]_i = \epsilon_{i\ell},
\end{equation}
where $\epsilon_{i\ell} = p(m_{i,\ell}-m_{i,\ell+1})$.

In arriving at Eqs.~(\ref{eq:puregauge}),~(\ref{eq:puregauss}), we
have, to a first approximation, treated the Mott insulator as a
deconfined $U(1)^{q-1}$ ``spin liquid'' with only gapped
gauge-charged excitations.  The particular liquid in question is
characterized by the presence of $q$ gapped boson excitations and
the PSG of the previous subsubsection specifying the
implementation of physical symmetries.  As is well-known for
$q=2$, however, the deconfined (or more precisely Coulomb) phase
of a pure compact $U(1)$ gauge theory in $2+1$ dimensions is
unstable to confinement via the proliferation of monopole
instantons, as first understood by Polyakov.  This is true here as
well, and the monopole unbinding will lead to the various possible
density-wave orders associated with the Mott state.

To see this explicitly, one can perform a simple duality
transformation for the gauge theory.  In its Hamiltonian
version\footnote{The reader unfamiliar with $U(1)$ dualities may
find
  it educational to compare this to the Villain version of duality
  carried out in the partition function, used elsewhere in the
  paper.}, this is simply the change of variables
\begin{eqnarray}
  \label{eq:hdualtransf}
  {\cal E}^\ell_\alpha & = & \frac{\epsilon_{\alpha\beta} \Delta_\beta
    \chi_\ell}{2\pi} + \overline{\cal E}^\ell_\alpha, \\
  \frac{\vec\Delta \times \vec{\cal A}^\ell}{2\pi} & = & B^\ell,
\end{eqnarray}
where $\overline{\cal E}_\alpha^\ell$ is an integer-valued
classical $c$-number field satisfying
\begin{equation}
  \label{eq:puregaussbg}
  \left[\Delta_\alpha \overline{\cal E}_\alpha^\ell\right]_i =
  \epsilon_{i\ell},
\end{equation}
and $\chi_\ell$ is an integer scalar field.  Due to the redundant
extra gauge field that has been introduced, we need to enforce
$\sum_\ell \chi_\ell=\sum_\ell\overline{\cal E}^\ell = 0$.  The
former can be implemented by including a large mass term, $m^2
(\sum_\ell\chi_\ell)^2$ term in ${\cal H}$, but for simplicity of
presentation we will just keep this constraint implicit.  One can
verify that the new variables are canonical
\begin{equation}
  \label{eq:canon}
  \left[B^\ell_a, \chi^{\ell'}_b\right] = i \delta_{ab}\delta_{\ell\ell'}.
\end{equation}
This reformulates the gauge theory in terms of gauge-invariant
variables.  The Hamiltonian becomes
\begin{equation}
  \label{eq:puregauge1}
  {\cal H}_{U(1)} =  \sum_{\ell}\left[
  \frac{v}{8\pi^2}|\Delta_\alpha\chi_{\ell} - 2\pi
  \epsilon_{\alpha\beta}\overline{\cal E}_{\beta}^\ell|^2 -
  K  \textrm{``cos''}(2\pi B^\ell)\right].
\end{equation}
To treat the theory analytically is simplest for large $K\gg v$,
where fluctuations of $B^\ell$ are typically small.  One may the
na\"ively expand the ``cos'' term.  This approximation, however,
spoils the periodicity $B^\ell \leftrightarrow B^\ell + 1$, or
equivalently the $2\pi\times$integer constraint on $\chi_{\ell}$.
Hence, when making this expansion, one should include an extra
term in the Hamiltonian to favor integer values of $\chi_\ell$.
One obtains then
\begin{eqnarray}
  \label{eq:pg2}
    {\cal H}_{U(1)} & = &  \sum_{\ell}\bigg[
  \frac{\tilde{v}}{2}|\Delta_\alpha\chi_{\ell} - 2\pi
  \epsilon_{\alpha\beta}\overline{\cal E}_{\beta}^\ell|^2
  +\frac{\tilde{K}}{2} (B^\ell)^2  \nonumber
  \\ && - \lambda \cos \chi_{\ell}\bigg],
\end{eqnarray}
with $\tilde{K}= (2\pi)^2 K, \tilde{v}=v/(2\pi)^2$. It is useful
now to decompose $\overline{\cal E}^\ell$ into transverse and
longitudinal parts,
\begin{equation}
  \label{eq:ebardecomp}
  \overline{\cal E}_{\alpha}^\ell = -\frac{\epsilon_{\alpha\beta} \Delta_\beta
    \eta_\ell}{2\pi} + \Delta_\alpha \gamma_\ell.
\end{equation}
Again, it is preferable to take $\sum_\ell \eta_\ell=0$. Provided
one chooses $\eta_\ell,\gamma_\ell$ bounded at infinity, the
latter is decoupled from $\chi_\ell$ and contributes only an
unimportant constant to ${\cal H}$.  Since $\chi_\ell$ has been
promoted now to a continuous field, one can then make the shift
$\chi_\ell \rightarrow \chi_\ell + \eta_\ell$.  One obtains
finally
\begin{eqnarray}
  \label{eq:sinegordon}
    {\cal H}_{U(1)} & = &  \sum_{a\ell}\bigg[
  \frac{\tilde v}{2}|\Delta_\alpha\chi_{a\ell}|^2
  +\frac{\tilde{K}}{2} (B_a^\ell)^2  \nonumber
  \\ && - \lambda \cos (\chi_{a\ell}+ \eta_{a\ell})\bigg].
\end{eqnarray}
This is just a set of $q$ $2+1$-dimensional lattice sine-Gordon
theories (with the implicit constraint that the center of mass
does not fluctuate, $\sum_\ell \chi_\ell=0$).  To interpret the
sine-Gordon terms, recall that $B^\ell$ and $\chi_{\ell}$ are
canonically conjugate.  Hence, the operator $e^{i\chi_\ell}$
generates a unit shift of $B^\ell$.  It thus creates a single
quantum of the $\ell^{th}$ gauge flux.  Based on the corresponding
space-time configurations, such operators are called ``monopole''
(instanton) operators.

It is clear from this discussion that such monopole operators
transform non-trivially under space group operations.  In
particular, if for a particular space group element $g$, the site
$a \rightarrow g(a)$ and under the PSG $\ell \rightarrow G(\ell)$,
one has
\begin{equation}
  \label{eq:montransf}
  e^{i\chi_{a\ell}} \rightarrow e^{i\chi_{a\ell}}
  e^{i(\eta_{g(a),G(\ell)}-\eta_{a,\ell})}.
\end{equation}

For large $\tilde{K}\gg\tilde{v}$, the fluctuations of
$\chi_{a\ell}$ are large on short scales, and one can integrate
out these short scale fluctuations (``coarse grain'')
perturbatively in $\lambda$.  This gives a continuum theory in
which the sine-Gordon terms are spatially averaged in a cumulant
expansion.  One expects on general grounds that all those and only
those multi-monopole terms that combine to form scalars under the
space group survive in the continuum theory:
\begin{eqnarray}
   \label{eq:pg3}
    {\cal H}_{U(1)} & = &  \int\!d^2r\bigg[
  \frac{\tilde v}{2}|\Delta_\alpha\chi_{\ell}|^2
  +\frac{\tilde{K}}{2} (B^\ell)^2  +V(\{\chi_\ell\})\bigg], \nonumber \\
\end{eqnarray}
where
\begin{equation}
  \label{eq:chipot}
  \nonumber
  V(\{\chi_\ell\}) = - \sum_n \frac{\lambda^n}{n!} \left\langle
    \left(\sum_\ell \cos (\chi_{\ell}+
    \eta_{a\ell}) \right)^n \right\rangle_{a,C}.
\end{equation}
Here the angular brackets indicate (cumulant) spatial averaging of
the oscillating $\eta_{a\ell}$ factors over sites $a$ of the dual
lattice (it is sufficient in practice to average over the $2q$
sites in which $\eta_{a\ell}$ takes independent values).

In general, this averaging leaves some non-vanishing $2$-monopole
and/or $4$-monopole sine-Gordon terms $\lambda_2,\lambda_4$, which
depend upon the form of $\eta_{a\ell}$.  As Polyakov pointed out,
although the fluctuations of $\chi_{a\ell}$ are large on small
scales, they are always bounded in $2+1$-dimensions, so that
ultimately these non-vanishing sine-Gordon terms ``pin'' the
$\chi_\ell$ fields.  This has the consequence of gapping out the
``photons'' of the putative Coulomb phase, and furthermore, giving
a non-zero average to the single-monopole operators
$e^{i\chi_\ell}$.  According to Eq.~(\ref{eq:montransf}), such
averages break space group symmetries, so the resulting state
necessarily has some kind of density-wave order.
\begin{figure}
\centering
\ifig[width=2.5in]{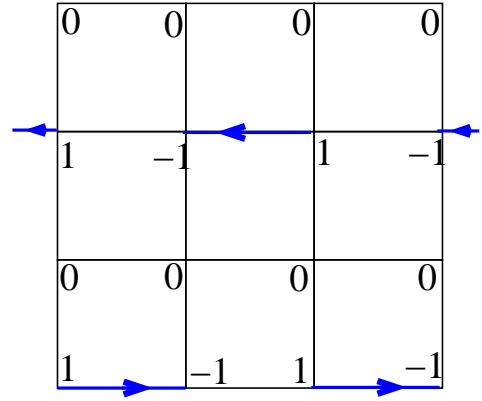}
\caption{Background electric field ${\cal E}^0_i$ configuration
for
  $q=4$.  The other three background fields can be obtained by
  rotation.} \label{fig:ebar0}
\end{figure}
\begin{figure}
\centering
\ifig[width=3in]{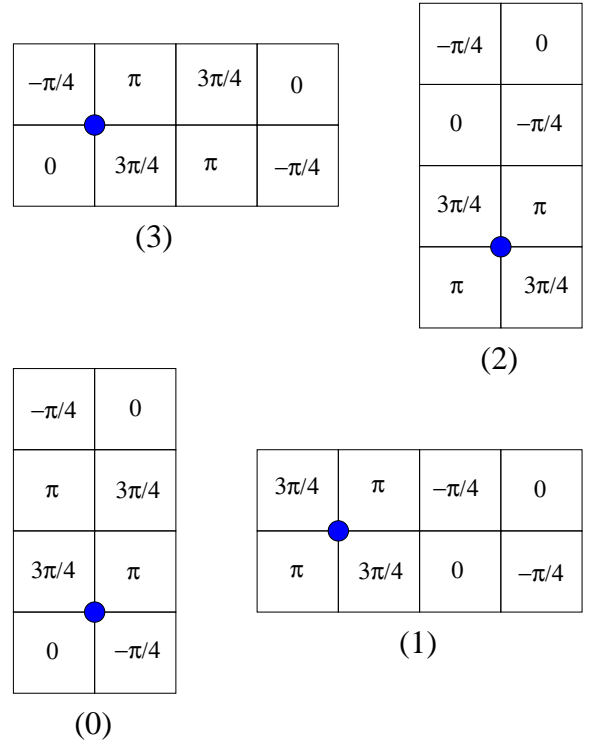}
\caption{Background scalar fields $\eta_\ell$ for $q=4$.
Rectangular
  unit cells are shown for $\ell=0,1,2,3$.  The filled circle
  indicates the origin of the direct lattice. } \label{fig:eta}
\end{figure}
To make these manipulations concrete, we give the example of
$f=1/4$, with the sublattice choice of Eq.~(\ref{eq:foursubs}).
First, one must solve the Gauss' law constraint,
Eq.~(\ref{eq:puregaussbg}). Then, the fields $\eta_{a\ell}$ are
determined from Eqs.~(\ref{eq:ebardecomp}) by the Poisson
equations,
\begin{equation}
  \label{eq:etapoisson}
  \Delta^2 \eta_{a\ell} = 2\pi \left(\epsilon_{\alpha\beta} \Delta_\alpha
  \overline{\cal E}^\ell_\beta\right)_a,
\end{equation}
and the requirement that they be bounded and satisfy $\sum_\ell
\eta_{a\ell}=0\, \textrm{mod $2\pi$}$.  The results are shown in
Figs.~\ref{fig:ebar0},\ref{fig:eta}. For this set of
$\eta_{a\ell}$, one can readily calculate the spatial averages in
Eq.~(\ref{eq:pg3}) up to $O(\lambda^4)$.  One finds the effective
potential takes the form
\begin{eqnarray}
\label{eq:potq4}
  V(\{\chi_\ell\}) & = & -\sum_\ell \left[\lambda_2 \cos (2\chi_\ell-\frac{\pi}{4}) + \lambda_4
  \sin 4\chi_\ell\right]  \nonumber \\ && - \lambda_{22}
\sum_{\ell<\ell'} \cos 2(\chi_\ell-\chi_{\ell'}),
\end{eqnarray}
with $\lambda_2 \sim \lambda^2, \lambda_4, \lambda_{22} \sim
\lambda^4$.  This indicates that for $q=4$, individual monopoles occur
{\sl doubled} rather than quadrupled as for $q=2$, in the partition
function.  They are, however, created with the non-trivial phase
factor above ($\pi/4$ phase shift inside the cosine proportional to
$\lambda_2$).

Let us now turn to the superfluid phase, occurring for $t \gg u$
in Eq.~(\ref{eq:heff1}).  In this case, we expect the
fractionalized bosons to condense, i.e. neglecting gauge
fluctuations, $\langle e^{i\hat\phi_\ell}\rangle \neq 0$.  By the
usual Anderson-Higgs mechanism, the $U(1)^{q-1}$ gauge fields will
thereby acquire a gap. With all $q$ fractional bosons condensed,
of course the physical boson is also condensed, and one has a
superfluid.

What are the elementary excitations in this phase?  One expects
these to be $q$ flavors of vortices and anti-vortices,
corresponding to phase windings in each of the $q$ boson fields.
Despite the presence of the gauge fields, these will have a
logarithmic energy cost, since due to the constraint $\sum_\ell
\vec{\cal A}_\ell = 0\, ({\rm mod}\, 2\pi)$, the gauge fields
cannot screen a single vortex.  Thus from this (and indeed any
other) universal point of view the superfluid is completely
conventional.

One can imagine, however, constructing ``vortex excitons'', i.e. a
vortex in the $\ell^{th}$ boson field and an anti-vortex in
$\ell^{\prime th}$ field.  These excitations carry zero physical
vorticity, and indeed can be screened by the gauge fields keeping the
constraint satisfied.  Most simply, a vortex in the $\ell^{th}$ and an
anti-vortex in the $(\ell+1)^{th}$ field is screened by a $2\pi$
(elementary) flux in the field $\vec{\cal A}^\ell$.  Thus each vortex
exciton binds an elementary gauge flux.  Conversely, an elementary flux
of this type will necessarily bind a vortex exciton in the superfluid
phase, as the ``bare'' gauge flux can only penetrate the superfluid in
combination with a vortex exciton (at finite energy) by the Meissner
effect.

Moreover, there is no microscopic conservation law for such gauge
fluxes/vortex excitons, as they carry no net vorticity.  Thus
there will be some amplitude for tunneling between states with
different numbers of such objects.  A space-time event at which
such a tunneling occurs is nothing but a monopole of the type
discussed in the pure gauge theory.  We have already described how
to calculate the amplitudes for such processes in the preceding
discussion of the insulating problem.  In the superfluid state,
due to this binding of monopoles to vortex excitons, one may
replace the notion of a flux-changing event by the creation of a
vortex exciton. Equivalently, if one thinks of the action of such
a term on a configuration with one vortex present, this may also
be considered as a vortex flavor-changing event.

Regardless of the language used, one may deduce from this
discussion the form of the dual vortex action.  In particular, it
should be written in terms of $q$ vortex fields $\zeta_q$,
describing vortices in each of the $q$ fractional bosons.  By the
usual considerations applicable in such a vortex description, the
vortex fields should be coupled to a single non-compact $U(1)$
gauge field, $A_\mu$, whose $3$-curl represents the physical
conserved boson current.  As usual, in zero applied magnetic
field, time-reversal symmetry implies symmetry between vortices
and anti-vortices, so that the theory must be ``particle-hole
symmetric'', i.e. relativistic in these variables.  Provided the
PSG contains enough symmetry between the different bose fields (as
in all cases we have considered), the quadratic part of the vortex
action is therefore fixed to have the form in Eq.~(\ref{s0}).  The
full action ${\cal S} = {\cal S}_0+\int\! d^2 r d\tau\,
V(\zeta_\ell)$ has an additional ``potential'' part.  One
expects
\begin{equation}
  \label{eq:vpot}
  V(\zeta_\ell)=V_0(|\zeta_\ell|)+
  V_\lambda(\zeta_\ell).
\end{equation}
Here the first term is a vortex-conserving potential deriving from
terms not involving monopole events.  It can be chosen as, e.g.
$V_0=u\sum_\ell |\zeta_\ell|^4$, simply to favor equally
all vortex flavors.  The second term is deduced from the monopole
terms already derived above by the rule
\begin{equation}
  \label{eq:vlambda}
  V_\lambda(\zeta_\ell) =
  \left. V(\{\chi_\ell\})\right|_{e^{i\chi_\ell} \rightarrow
    c\zeta^*_\ell \zeta^{\vphantom{*}}_{\ell+1}},
\end{equation}
the latter replacement encoding the coincidence of monopole with
vortex flavor-changing events.  Here $c$ is a constant associated
with amplitude of $\zeta_\ell$.  In the example of $q=4$,
one can readily see that the $\lambda_2$ term in
Eq.~(\ref{eq:potq4}) leads to a term in $V_\lambda$
gauge-equivalent to the $\lambda$ term in Eq.~(\ref{m10}).

Thus we have recovered the form of the vortex action studied in
the remainder of the paper by this reasoning.  Of course, rather
than the physical arguments used above, we could proceed rather
more mechanically by performing duality transformations both for
the $U(1)$ gauge fields {\sl and} for the fractional boson fields.
This procedure recovers identical results, and is sketched in
Appendix~\ref{sec:duality-q-slave}.

\subsection{Approach to fractionalization from the density-wave state}
\label{sec:appr-fract-from-1}

At half-filling, a very intuitive picture of fractionalization at the
deconfined QCP could be developed based upon topological defects in
the valence bond ordered state.  In this subsection, we consider the
generalization of this notion to a potential deconfined QCP at
$f=1/4$.  We will see that indeed there are intersections of
particular ``elementary'' domain walls in the charge ordered state
that can be regarded as discrete vortices.  Under not very restrictive
conditions, such vortices trap a fractional charge in units of $1/4$.
Though we frame the discussion in this rather specific context, it is
clear that the arguments in this subsection are rather general in
nature.  Indeed, we expect that it is a common (not universal but true
in a sizeable fraction of cases) feature of charge ordered states
breaking discrete lattice symmetries, that intersections of domain
walls carry fractional charge.  Of course, this does not imply in and
of itself any exotic physics.   Such discrete vortex configurations
are certainly ``confined'' in the charge ordered state, having an
energy linear in system size due to their trailing domain walls.

\begin{figure}
\centering
\ifig[width=2.8in]{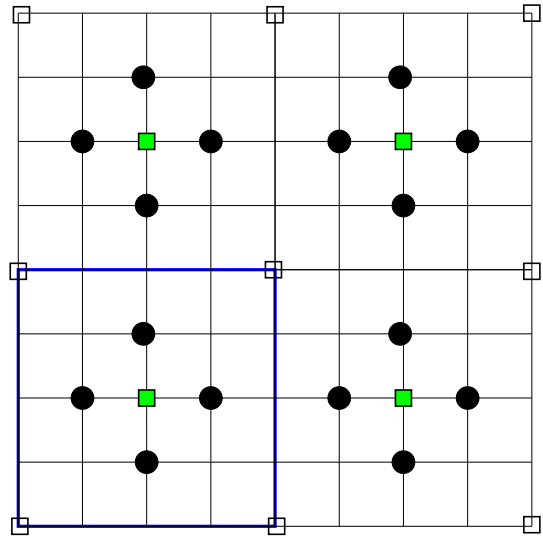}
\caption{Caricature of the ground state for the CDW phase at
  $f=1/4$. Here the black circles are occupied sites.  A unit cell is
  outlined in the lower left corner.  Sites on the boundary
  are shared amongst 2 or 4 unit cells.  The squares (open and filled)
  indicate the two inequivalent centers of $\pi/2$ rotations within
  each unit cell.} \label{fig:trivunit}
\end{figure}

\begin{figure}[t]
\ifig[width=0.3\textwidth]{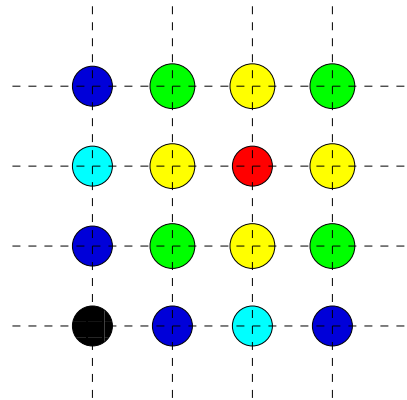}
\caption{A less extreme cartoon of a unit cell of the 16-fold
  degenerate CDW ground state.  Different shades of gray represent different
  boson densities.}
\label{cdw16}
\end{figure}

Turning now back to the specific problem of $f=1/4$, we first need to
understand the order parameter space of the density wave state.  As
discussed above, the appropriate phase in question corresponds to that
in region E of Fig.~\ref{figq4a}.  From the figure, one can
immediately see that the density wave has a $16$ site unit cell with a
period of $4$ lattice spacings in both the $x$ and $y$ directions.
For reference, these have the following symmetry properties (see
Sec.~\ref{sec:some-ground-states}):
\begin{enumerate}
\item{All 16 states are invariant under $T_x^4$ and $T_y^4$, i.e.  the
    charge density pattern in the ground state has a period of 4
    lattice constants in both $x$- and $y$-directions.} \item{The unit
    cell contains two inequivalent centers of $\pi/2$ rotation,
    separated by two lattice spacings in the $x$ and $y$ directions
    from one another.}
\item{Reflection planes along the $x$ and $y$ axes pass through
the centers of rotation.} \item{All 16 states are related to each
other by successive $T_x$ or $T_y$ operations.}
\end{enumerate}
All these symmetry properties, and the proper particle filling,
can be reproduced trivially by a wavefunction taken as an exact
boson number eigenstate on every site, of the form shown in
Fig.~\ref{fig:trivunit}.  It is then clear that this phase may be
interpreted as a charge density wave (CDW).  Another caricature
which is also useful to consider is a state which is a direct
product of states defined on the $4\times 4$ unit cells, with
exactly $4$ bosons per unit cell, but with charge fluctuations
within each unit cell (see Fig.~\ref{cdw16}).  More generally, the
ground state of course contains further local charge fluctuations.
Nevertheless, we expect that qualitative properties within the CDW
phase can be determined by adiabatic continuity from the simpler
caricatured states.

\begin{figure}[t]
\ifig[width=0.3\textwidth]{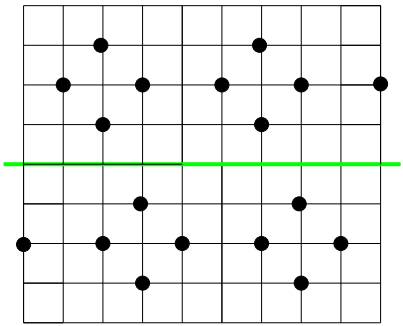}
\caption{Cartoon of an elementary domain wall oriented along the
$x$
  axis.  Upon crossing from top to bottom, the state is shifted by one
  unit in the $x$ direction.}
\label{cdw16d}
\end{figure}

Clearly, the order parameter space can be specified by giving the
lattice co\"ordinates of one of the two inversion centers, i.e. an
ordered pair of integers $(m,n)$ {\sl modulo $4$}.  Let us now
turn to topological defects.  First consider elementary domain
walls, which connect ``neighboring'' states in this order
parameter space, i.e. where the two co\"ordinates $(m',n')$ and
$(m,n)$ differ by a unit translation in $x$ or $y$.  Clearly, any
one state can be separated from four others in this manner.  A
cartoon of such an elementary domain wall, oriented for simplicity
along one of the principle axes, is indicated in
Fig.~\ref{cdw16d}. An important question is the nature of the
local excitations in the vicinity of the domain wall, assuming for
simplicity that it is flat (i.e. not in a rough phase) and pinned
at the boundaries, so that it has no translational degrees of
freedom.. Theoretically, one may imagine that such a domain wall
may be either insulating or conducting.  In the simplest models,
as in the cartoon of Fig.~\ref{cdw16d}, the charge configuration
in the vicinity of the wall is fully rigid, and one expects it to
be insulating. More precisely, the excitation spectrum remains
gapped in the presence of the domain wall. Moreover, one can count
the excess charge (above $1/4$-filling) per unit length associated
with the domain wall. In the caricature of Fig.~\ref{cdw16d}, the
excess charge is clearly vanishing. Due to the gap in the
excitation spectrum, one expects this zero excess charge to be
preserved as one perturbs around the caricatured state.

\begin{figure}[t]
\ifig[width=0.4\textwidth]{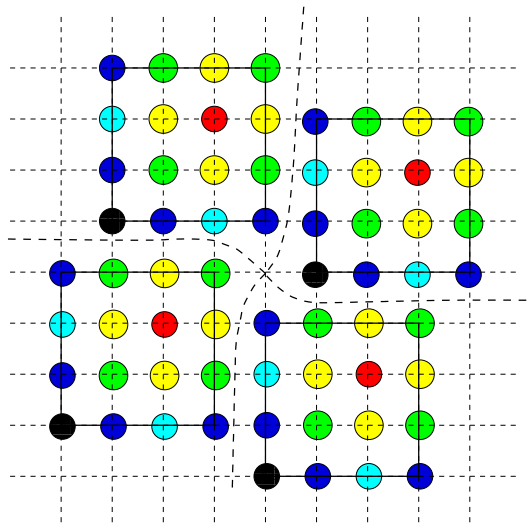}
\caption{A $Z_4\times Z_4$ ``vortex'' defect CDW phase.
  The site where 4 domain walls intersect does not belong to any unit
  cell and thus corresponds to an excitation of charge $-1/4$ (if
  empty as drawn).}
\label{vortex}
\end{figure}

Henceforth we will assume that the elementary domain walls are
insulating and uncharged.  Now consider the nature of vortices in the
CDW order parameter. Some inspiration can be obtained from the point
of view of the preceding sections.  We have seen that charge $\pm 1/q$
particles correspond to $\pm 2\pi$ vortices in each of the
$\zeta_\ell$ fields.  Thus one expects to find a total of $8$ distinct
such vortex-like excitations in the density wave order parameter.
Indeed, by using the explicit transformations of the $\zeta_\ell$
fields under translations in Eqs.~(\ref{eq:3.5}), one can
straightforwardly see that each such vortex corresponds to the
intersection of four elementary domain walls.  Across each such domain
wall, the CDW pattern is translated to that of another of the 16
ground states.  For instance, one finds that a positive vorticity
vortex in $\zeta_0$ corresponds to the sequence of operations:
\begin{equation}
  \label{eq:vort0}
  T_y^{-1} T_x^{-1}  T_y^{\vphantom{-1}} T_x^{\vphantom{-1}}.
\end{equation}
More generally, upon moving clockwise around the center of the
``vortex'' in real space, one walks either clockwise or
counter-clockwise in unit steps in the order parameter space.  We
denote these $Z_4\times Z_4$ vortices.  A schematic illustration is
shown in Fig.~\ref{vortex}.

Since both the domain walls extending away from the $Z_4\times Z_4$
vortex and the bulk CDW states are gapped Mott insulators, one expects
this state to have a definite and quantized charge.  Again, our
cartoon wavefunction gives some insight.  Both the bulk and domain
wall wavefunctions can be written in a form with a definite $4$ bosons
per $4\times 4$ unit cell.  In the $Z_4\times Z_4$ state of
Fig.~\ref{vortex}, all sites are covered by such unit cells {\sl
  except one site in the center of the ``vortex''}.  This fact is
inevitable by symmetry: four domains related by unit translations that
can be covered by non-overlapping unit cells are related by rotation
around a site of the direct lattice.  Thus one can write two such
caricatured wavefunctions, having either zero or one boson on this
central site. As the occupancy per site in the bulk is $f=1/4$, such
states have, relative to the ground state, a definite relative charge
of $-1/4$ (unoccupied central site) or $+3/4$ (occupied central site).
Because both the bulk and domain walls have been assumed insulating,
we expect those quantized charges to be a robust feature of the
$Z_4\times Z_4$ vortex, even for more realistic wavefunctions.  This
is because, although some charge may ``leak'' away from the central
site, it cannot escape through the insulating environs, so the total
excess charge summed over the region around the vortex will not
change.

It is notable that one does not obtain in this way two
symmetry-related states with charges $Q=\pm 1/4$.  The difference
from the case of $f=1/2$ is that there is no particle-hole
symmetric limit corresponding to $f=1/4$.  Indeed, there is no
symmetry relating the charge $Q=-1/4$ and $q=+3/4$ states, and
generally these will have different energy.  Of course, such a
strongly localized form of the wavefunctions can be even
approximately appropriate only deep in the CDW state, where any
particle-hole symmetry is very strongly broken. Thus there is no
intrinsic contradiction with the expectations of the preceding
considerations of this section.  Of course, deep in the CDW phase
such a state has an energy linearly diverging with system size,
due to the four domain walls extending to the system boundaries.
Its emergence as a deconfined ``particle''\footnote{Of course,
even in the deconfined scenario, there is no sharp sense in which
this excitation is particle-like at the QCP.} in the spectrum near
the critical point is not at all obvious from this perspective.

Apparently if a deconfined QCP obtains in this case, particle-hole
symmetry must be at least approximately restored in its vicinity,
which is remarkable from this point of view.  Its absence deep in
the CDW phase, while not intrinsically troubling, makes an attempt
to formulate a theory of the QCP directly in terms of such
microscopic vortex defects not very promising.  We therefore leave
it an open subject for future work.

\section{Impurity pinning}
\label{sec:imp}

This section will show how our approach offers a natural framework
for describing recent STM observations
\cite{krmp,fang,ali,mcelroy,hanaguri,hoffman} of modulations in
the local density of states of the cuprate superconductors. As we
noted in Section~\ref{sec:intro} and demonstrated in II, the
bosonic theories presented so far apply also to paired electron
systems, with the PSG determined in Eq.~(\ref{eq:fdelta}) by the
density of Cooper pairs. We will interpret the observed density
STM modulations as representing density wave order associated with
a proximate superfluid-to-insulator transition. Indeed, as we have
discussed in Section~\ref{sec:intro}, the PSG requires density
wave order to be present near all such transitions in which the
elementary vortices do not form composites.

As we will demonstrate below, an important advantage of our focus
on the $\varphi_\ell$ vortex degrees of freedom, and their
connection to density wave modulations, is that we are able to use
the same theory to compute the impurity-induced pinning of density
wave order in zero magnetic field, and also the vortex-induced
density wave order in a finite magnetic field. At zero magnetic
field, there are vacuum fluctuations of vortex-anti-vortex pairs
which, when pinned by impurities, lead to modulations observable
in STM. At finite magnetic field, there is a net excess of
vortices over anti-vortices (say) but precisely the same pinning
potentials again lead to density wave order. This section will
consider a simple toy model of the pinning process, with the aim
of exposing the very general relationship between the zero field
and finite field STM experiments.

The perspective offered here differs considerably from previous
analyses of the STM experiments
\cite{zds,tolyastm,podolsky,chenting,zhu,franz,zhang,ghosal,andersen02,machida}
(although VBS order associated with vortices was suggested prior
to the experiments in Ref.~\onlinecite{ps}). In
Refs.~\onlinecite{zds,tolyastm,podolsky} the spin and charge order
were treated as the primary fluctuating degrees of freedom which
were pinned by localized impurities. However, the connection of
this pinning to the location of the vortices was indirect, and in
a sense, had to be put in by hand. It was assumed that a large
coupling $v_{|{\bf Q}|}$ in Eq.~(\ref{e5}) caused each vortex to
induce a large pinning potential for the fluctuating density wave
order. Our present approach dispenses with this ad hoc procedure,
and shows how the connection between vorticity and density wave
order emerges directly from the underlying quantum physics.

We will work within the continuum field theory of low energy
vortex fluctuations in the superfluid discussed in
Section~\ref{sec:cft}. As our purpose is mainly illustrative, we
will only consider this theory within a leading Gaussian
approximation; our approach can, in principle, be extended to
include the non-linear terms discussed in Section~\ref{sec:cft}
using the formalism presented below. However, the gauge potential
$A_\mu$ has to be considered with some care, and cannot na\"ively
be dropped: we have to include an average value of $A_\tau$ which
accounts both for the vortex ``chemical potential'' (proportional
to the applied physical magnetic field) and for the interaction
between vortices in the vortex lattice. We therefore consider the
following action for the $\varphi_\ell$ vortices
\begin{eqnarray}
  \mathcal{S}_0 &=& \int d^2 r \, d \tau \sum_{\ell = 0}^{q-1}
  \Bigl[ \left|\left( \partial_\tau - i \overline{A}_\tau ({\bf r}) \right) \varphi_\ell \right|^2 +
c^2 \left| \partial_\alpha \varphi_\ell \right|^2
   \nonumber \\
   &~&~~~~~~~~~~~~~~~~~~~~~+ m_v^2 c^4
  |\varphi_\ell |^2 \Bigr] \;,
  \label{eq:S0:A=0}
\end{eqnarray}
where the index $\alpha$ extends over the two spatial directions.
This is clearly derived from the action in Eq.~(\ref{s0}), written
in a notation convenient for our purposes here: the energy gap
towards creation of a vortex or anti-vortex is $m_v c^2$, and $c$
is a velocity. Translational invariance is broken by adding a
pinning potential: we assume a static impurity potential which
couples linearly to the density operator:
\begin{equation}
\mathcal{S}_V = \int d^2 r \, d \tau \sum_{m n} V_{mn}({\bf r})\,
\rho_{mn}^{\ast} ({\bf r},\tau) \;, \label{vp3}
\end{equation}
where $\rho_{mn}$ is specified as in Eq.~(\ref{e11}). Since
$\rho_{mn}^{\ast}({\bf r},\tau) = \rho_{-m,-n}({\bf r},\tau)$ we
must have $V_{mn}^{\ast}({\bf r}) = V_{-m,-n}({\bf r})$. We will
now consider the properties of the action $\mathcal{S}_0 +
\mathcal{S}_V$ in zero and non-zero magnetic field in turn in the
following subsections.

\subsection{Pinning in zero magnetic field}
\label{sec:pinzero}

In this case we can set $\overline{A}_\tau = 0$, and perform a
perturbative expansion in powers of $V_{mn}$. A one-loop
computation of the $\varphi_\ell$ susceptibility shows
\begin{equation}
  \langle \rho_{mn}({\bf r}) \rangle = S(|{\bf Q}_{mn}|)^2 \int
  d^2 r' \,  \Pi_0({\bf r} - {\bf r'}) \, V_{mn}({\bf r'}) \;,
  \label{eq:SV}
\end{equation}
where the susceptibility is
\begin{eqnarray}
  \Pi_0 ({\bf r}) &=& - q \int \frac{d^2 k}{4 \pi^2} \frac{d^2 p}{4 \pi^2} \frac{d \omega}{2 \pi}
\frac{e^{i {\bf k} \cdot {\bf r}}}{( \omega^2 + p^2 c^2 + m_v^2
c^4)} \nonumber \\
&~&~~~~~~~~~~~\times \frac{1}{(\omega^2 + ({\bf k} + {\bf p})^2 c^2 + m_v^2 c^4)} \nonumber \\
&=& -q \int \frac{d^2 k}{4 \pi^2} \frac{e^{i {\bf k} \cdot {\bf
r}}}{4 \pi c^3 k} \tan^{-1} \left( \frac{k}{2 m_v c} \right)
\nonumber \\
&=& -\frac{q}{8 \pi^2 c^3} \int_0^{\infty} dk J_0 (kr) \tan^{-1}
\left( \frac{k}{2 m_v c} \right) \nonumber \\
&=& - \frac{q m_v}{4 \pi^2 c^2} \int_1^{\infty} dy \int_0^{\infty}
\frac{ k dk J_0 (2 m_v c r k )}{(y^2 + k^2)} \nonumber
\\ &=& - \frac{q m_v}{4 \pi^2 c^2} \int_1^{\infty} dy K_0 (2 m_v r c
y) \nonumber \\
& \approx & - \frac{q m_v}{4 \pi^2 c^2} \sqrt{\frac{\pi}{2}}
\frac{ e^{- 2 m_v c r}}{(2 m_v c r)^{3/2}} ~\mbox{as $r
\rightarrow \infty$.}\label{pz1}
\end{eqnarray}
So the pinned order decays exponentially with $r$ over the length
scale $1/(2 m_v c)$.

\subsection{Pinning in a non-zero magnetic field}
\label{sec:pinnonzero}

A non-zero magnetic field induces a finite density of vortices,
and consequently we have to consider the theory in the presence of
a non-zero vortex ``chemical potential''. Here we wish to focus on
the behavior of a single vortex localized near the origin,
surrounded by an infinite vortex lattice. In this case, the
quantity $\overline{A}_\tau$ has two distinct contributions: ({\em
i\/}) a spatially independent vortex chemical potential, $\mu_v$,
proportional to the applied magnetic field, and ({\em ii\/}) a
${\bf r}$ dependent potential created by the surrounding vortices
of the vortex lattice. The latter potential arises from the
superflow around each vortex: in the present dual theory it is the
logarithmic ``Coulomb'' interaction energy associated with the
$A_\mu$ gauge field created by the total vortex ``charge''
density. For the vortex under consideration, this ``Coulomb''
potential is assumed to have a minimum near ${\bf r}=0$, and we
can safely perform a quadratic expansion of the potential around
its minimum. Collecting these contributions, we write
\begin{equation}
i \overline{A}_\tau ({\bf r}) = -\mu_v + \frac{1}{2} m_v
\omega_v^2 r^2 .\label{harmonic}
\end{equation}
The second contribution above represents the interaction energy
between the vortices, and the frequency $\omega_v$ is determined
by the superfluid stiffness of the bosons, and the spacing of the
vortex lattice. At a more sophisticated level, we really have to
consider the collective oscillations of the entire vortex lattice,
and quantize the motion of all the vortices together \cite{bbs};
however, we will be satisfied here with the simple `Einstein
model' of the vortex lattice oscillation spectrum which is
implicit in Eq.~(\ref{harmonic}).

A finite density of vortices requires that the vortex chemical
potential be larger than the energy gap to create a vortex. Hence
we have $\mu_v > m_v c^2 $. In the analytic computations of this
subsection, we will consider here only the additional density
modulation created by the vortices, while neglecting the `vacuum'
quantum fluctuations of the vortex-anti-vortex pairs; provided $|i
\overline{A}_\tau - m_v c^2| \ll m_v c^2$, these `vacuum'
polarization contributions can be assumed to yield an additive
contribution to the density modulation equal to that already
computed in Section~\ref{sec:pinzero}. Technically, we can ignore
the vortex-anti-vortex pairs simply by performing a low frequency
expansion of Eq.~(\ref{eq:S0:A=0}) in the presence of a non-zero
$\mu_v$. To this end we define
\begin{equation}
\varphi_\ell = \frac{1}{\sqrt{2 i \overline{A}_\tau}} \Psi_\ell
\approx \frac{1}{\sqrt{2 m_v c^2}} \Psi_\ell, \label{vp1}
\end{equation}
and expand Eq.~(\ref{eq:S0:A=0}) to leading order in temporal
derivatives and in $|i \overline{A}_\tau - m_v c^2|$. This yields
the following familiar action for a bosonic vortex ``particle''
with $q$ `flavors' $\Psi_\ell$
\begin{eqnarray}
\mathcal{S}_0 &=& \int d^2 r \, d \tau \sum_{\ell = 0}^{q-1}
  \Biggl[ \Psi_\ell^{\ast} \frac{\partial \Psi_\ell}{\partial \tau}
   + \frac{|\partial_\alpha \Psi_\ell|^2}{2
  m_v}  \nonumber \\ &~&~~+ \left( m_v c^2 - \mu_v + \frac{1}{2} m_v \omega_v^2 r^2
  \right) |\Psi_\ell|^2\Biggr] \label{vp2}
\end{eqnarray}
We also have to insert Eq.~(\ref{vp1}) into Eq.~(\ref{vp3}) and so
obtain the pinning potential acting on the vortices: this
potential has a non-trivial structure in the `flavor' space of the
$q$ vortices, and this will naturally be crucial in determining
the induced density wave modulation.

Just as in Section~\ref{sec:pinzero}, we proceed in a perturbative
expansion in powers of $V_{mn}$. However, unlike the previous zero
magnetic field case, we now find that the perturbation theory is
degenerate, and remarkably the leading order result turns out to
be {\em independent\/} of the overall scale of the $|V_{mn}|$. To
zeroth order in the $V_{mn}$, notice that the vortex particle
described by Eq.~(\ref{vp2}) has a $q$-fold degenerate ground
state, with its spatial wavefunction given by the ground state of
the harmonic oscillator
\begin{equation}
\Psi_\ell ({\bf r}) = \mathcal{U}_{\ell} \left( \frac{m_v
\omega_v}{\pi} \right)^{1/2} \exp \left( - \frac{m_v \omega_v
r^2}{2} \right) \label{vp4}
\end{equation}
where $\mathcal{U}_{\ell}$ is, for now, an arbitrary unit vector
in the internal vortex space. The $\mathcal{U}_\ell$ vector is
determined by diagonalizing the pinning potential projected into
this degenerate ground state manifold. From Eqs.~(\ref{vp3}) and
(\ref{e11}) we deduce that $\mathcal{U}_\ell$ is the ground state
eigenvector of the matrix $\mathcal{M}_{\ell m}$,
\begin{equation}
\sum_m \mathcal{M}_{\ell m} \mathcal{U}_m = \epsilon_0
\mathcal{U}_\ell ,\label{vp5}
\end{equation}
where
\begin{eqnarray}
&& \mathcal{M}_{\ell m} = \frac{\omega_v}{2 \pi c^2}  \int d^2 r
\exp \left( - m_v \omega_v r^2 \right) \nonumber \\ &&\times
\sum_n \Biggl[ S(|{\bf Q}_{n,\ell-m}|) V_{n,\ell-m} ({\bf r})
\omega^{-n (\ell+m)/2} \Biggr].~~~~~~~~ \label{vp6}
\end{eqnarray}
Note that the overall scale of $\mathcal{M}_{\ell m}$ is
immaterial at this order in perturbation theory because it has no
influence on $\mathcal{U}_\ell$. Finally, the density wave order
associated with this vortex is determined by inserting
Eqs.~(\ref{vp1}) and (\ref{vp4}) into Eq.~(\ref{e11}):
\begin{eqnarray}
\langle \rho_{mn} ({\bf r}) \rangle &=& \frac{S\left(|{\bf
Q}_{mn}|\right ) \omega_v }{2 \pi c^2}  \exp \left( - m_v \omega_v
r^2 \right) \nonumber
\\ &~&~~\times \omega^{mn/2} \sum_{\ell} \mathcal{U}^{\ast}_\ell
\mathcal{U}_{\ell+n} \omega^{\ell m}. \label{vp7}
\end{eqnarray}
As promised, this result has no prefactor of $V_{mn}$: the
dependence on the pinning potential is entirely in
Eqs.~(\ref{vp5}) and (\ref{vp6}) where the relative values of
$V_{mn}$ determine the orientation of the unit vector
$\mathcal{U}_\ell$ in the complex $q$ dimensional vortex space.

We also note here that the above calculation can be extended to
allow for small deviations from the commensurate density of $p/q$.
We do this by extending Eq.~(\ref{eq:S0:A=0}) as in
Eq.~(\ref{s0p}), and include an average dual `magnetic' field of
strength $2 \pi \delta f/a^2$ that acts on the vortices. However,
it cannot be assumed that the dual `magnetic' field is spatially
uniform, and its space dependence must be computed
self-consistently. We expect that the dual `magnetic' flux will be
partially expelled from the vicinity of the vortices (where the
dual `matter' resides). In the direct boson language, this means
that the density of bosons is not spatially uniform in the
presence of the vortex lattice, and that the boson density is
closer to $p/q$ than average at the positions of the vortex cores.
This effect has been discussed in Ref.~\onlinecite{congjun} for
the case of Mott insulators with integer filling. Let us assume
that the actual boson density deviation from commensurability at
the vortex lattice positions is $\delta \widetilde{f}$, where
$|\delta \widetilde{f}| < | \delta f |$. We now need to determine
the vortex wavevfunction by solving the Schr\"odinger equation
associated with  Eq.~(\ref{vp2}) and a `magnetic' field of
strength $\delta \widetilde{f}$. This is equivalent to the problem
of a particle in a parabolic potential and a magnetic field; the
wavefunction of the vortex retains the Gaussian form in
Eq.~(\ref{vp4}), and all subsequent expressions remain as in the
paragraphs above apart from the shift in the frequency $\omega_v
\rightarrow \left(\omega_v^2 + (2 \pi \delta \widetilde{f}/(2m_v
a^2))^2 \right)^{1/2}$.

\subsection{Discussion}
\label{sec:discuss}

The above computations of pinned density wave order in zero and
non-zero magnetic field have a number of unusual features which
deserve further comment. It would be useful to have tests of these
features in future STM experiments.

The pinned order in zero field in Eq.~(\ref{pz1}) has a
conventional form similar to that obtained in other approaches.
The induced order decays exponentially from the pinning site, on a
scale $1/(2 m_v c)$ related to the distance to the
superfluid-to-insulator transition. This decay length equals the
intrinsic correlation length for density wave order fluctuations
in the superfluid, as might na\"ively be expected. Also, for weak
pinning, the amplitude of the pinned order is proportional to the
strength of the pinning.

In contrast, the finite magnetic field result in Eq.~(\ref{vp7})
has some remarkable features. Quite generally, for weak pinning,
the envelope of the density wave order has the functional form of
a Gaussian in the small $r$ region where the harmonic
approximation holds. The width of this Gaussian is {\em not\/}
determined by the correlation length of the density wave order,
but by the scale over which the vortex undergoes quantum zero
point motion. The latter distance is determined by the inverse
square root of the mass of the vortex, $m_v$, and the vortex
lattice oscillation frequency, $\omega_v$. It should be possible
to compute the magnetic field dependence of $\omega_v$, and thus
quantitatively compare Eq.~(\ref{vp7}) with STM experiments in a
varying magnetic field. Finally, the prefactor of Eq.~(\ref{vp7})
is independent of the strength of the pinning field, but has an
interesting dependence on $\omega_v$, which places further
constraints on possible experimental comparisons; the independence
on the pinning field strength can readily tested in numerical
studies such as those in
Refs.~\onlinecite{sandvik},~\onlinecite{sandvik2}, and
~\onlinecite{nikolai}.

\section{Conclusions}
\label{sec:conc}

This paper has presented a general framework for describing the
superfluid-to-insulator transition on regular lattices. We
examined simple model boson systems on the square lattice, but we
expect our results to be quite general, and applicable also to
paired electron systems exhibiting transitions between
superconducting and Mott insulating states. The latter connection
is established in more detail in a companion paper
\cite{psgdimers} II.

There were three important themes underlying our approach. The
first is a feature common to several recent studies of quantum
phase transitions with multiple order parameters
\cite{rs,lfs,deccp,LevinSenthil}: the quantum mechanics of the
defects of the order parameter in one phase induce correlations of
the second phase. In our case, we focused on the vortex defects of
the superfluid state. These were found to have remarkable symmetry
properties which linked them intricately to density wave order.
The condensation of such vortices, apart from leading to the
destruction of superfluid order, also led to the simultaneous
appearance of density wave order in the Mott insulator. This
constitutes an important mechanism for the breakdown of the
Landau-Ginzburg-Wilson theory of the phase transition.

The second theme was the crucial role played by projective
representations of the space group of the lattice (the PSGs). PSGs
have been highlighted in the recent work of Wen \cite{wen} in his
study of ground states insulating spin systems (equivalently, of
bosons at half-filling). We argued that the vortices of the
superfluid in the vicinity of a commensurate Mott insulator state
transform under a PSG defined by Eqs.~(\ref{e7}) and (\ref{rtr}).
The key defining parameter of this PSG is the unimodular complex
number $\omega$, and for a Mott insulator with average density of
$p/q$ bosons per lattice site $\omega$ was defined in
Eqs.~(\ref{e8}) and (\ref{e9}). Note that the superfluid (or a
supersolid phase, if present) need {\em not\/} be exactly at this
density, and our formalism allows small variations in density away
from that of the commensurate Mott insulator (see
Section~\ref{sec:irrat}). The existence of such a PSG was shown to
imply that there are $q$ degenerate species of vortices in the
superfluid. Moreover, we argued that any linear superposition of
such vortices was necessarily associated with density wave order
at wavelengths which are integer divisors of $q$ lattice spacing:
consequently, as noted above, vortex condensation produces a Mott
insulator with long-range density wave order.

Section~\ref{sec:imp} presented an analysis of the consequences of
impurities in the superfluid. Any impurity breaks the lattice
symmetry, and hence it also lifts the degeneracy between the $q$
vortex species. It then follows that any localized vortex has an
associated halo of density wave order, and we offered this as the
fundamental explanation of the STM observations of Hoffman {\em et
al.} \cite{hoffman} on Bi$_2$Sr$_2$CaCu$_2$O$_{8+\delta}$.

The third theme of our paper was one of particle number
fractionalization. This arose in an attempt to interpret the dual
vortex theory of the LGW-forbidden superfluid-insulator transition
in the language of the direct lattice bosons. Following the
`deconfined criticality' proposal of Senthil {\em et al.}
\cite{deccp}, we hypothesized a theory in which the boson
fractionalized into $q$ constituent particles each with particle
number $1/q$. By performing a duality analysis upon such a
fractionalized boson theory, we were able to reproduce the dual
theory of $q$ vortices, but only for special values of $q$. The
selected values of $q$ were those for which it was possible to
obtain a {\em permutative\/} representation of the PSG.

Moving beyond simple boson systems, our analysis makes clear that
the key ingredient needed for other superfluids is the value of
the parameter $\omega$ controlling the PSG of its vortices. With
the knowledge of $\omega$, the general structure of the field
theory in Section~\ref{sec:cft} is fully determined, and
subsequent results follow. We invite the reader to now proceed to
II, where we consider electron systems with short range pairing,
and show that $\omega$ is determined by the density of Cooper
pairs.

\begin{acknowledgments}

We thank M.~P.~A.~Fisher, T.~Senthil, Z.~Te\v sanovi\'c, and
J.~Zaanen for valuable discussions. This research was supported by
the National Science Foundation under grants DMR-9985255 (L.
Balents), DMR-0098226 (S.S.), and DMR-0210790, PHY-9907949 at the
Kavli Institute for Theoretical Physics (S.S.), the Packard
Foundation (L. Balents), the Deutsche Forschungsgemeinschaft under
grant BA 2263/1-1 (L. Bartosch), and the John Simon Guggenheim
Memorial Foundation (S.S.). S.S. thanks the Aspen Center of
Physics for hospitality. K.S. thanks S.~M.~Girvin for support
through ARO grant 1015164.2.J00113.627012.

\end{acknowledgments}

\appendix

\section{Vorticity modulations}
\label{app:vorticity}

We have shown in the body of the paper that the vortex fields
$\varphi_\ell$ are naturally connected to modulations in the
generalized density (which could be any observable invariant under
time reversal and spin rotations); explicitly, the two were
connected by the formula in Eq.~(\ref{e11}). In this appendix we
will examine the issue of the modulations in the {\em vorticity}.

Following Ivanov {\em et al.}\cite{ivanov}, we define the
vorticity as a observable associated with each plaquette of the
direct lattice (or each site of the dual lattice), equal to the
sum of the particle (boson) currents flowing anti-clockwise around
the plaquette. The vorticity is odd under time-reversal, and
because all the states considered in the body of the paper are
time-reversal invariant, its expectation value vanishes. However,
when a magnetic field is applied, as in
Section~\ref{sec:pinnonzero}, then the vorticity can have quite a
rich spatial structure. In the superfluid phase, we know that the
magnetic field induces a lattice of vortices. We also argued in
Section~\ref{sec:pinnonzero} that there is a density modulation
superimposed on this vortex lattice described by Eq.~(\ref{vp7}).
Here we shall show that there is also a corresponding pattern of
vorticity modulations at the wavevectors ${\bf Q}_{mn}$. Thus the
appearance of ``staggered vorticity'' is quite a general feature
of superfluids near a Mott transition, and not a special
characteristic of certain spin liquids as implied by Ivanov {\em
et al.} \cite{ivanov}.

We can obtain an expression for the vorticity either by explicit
derivation or by symmetry considerations alone. In the first
approach, we notice that the vorticity is conjugate to a magnetic
field applied on the plaquette of a dual lattice, and the latter
couples to the operator $\psi_a^\ast \partial_\tau \psi_a  -
\partial_\tau \psi_a^\ast \psi_a$. In the second approach, we need
the action of the time-reversal operation, $\mathcal{T}$, on the
varies fields in the paper; a simple analysis shows that under
$\mathcal{T}$
\begin{eqnarray}
&& \tau \rightarrow -\tau~~;~~x \rightarrow x~~;~~y \rightarrow
y~~;~~ \hat{\phi} \rightarrow -\hat{\phi}~~; \nonumber \\
&& J_\tau \rightarrow J_\tau~~;~~J_x \rightarrow - J_x~~;~~J_y
\rightarrow -J_y~~; \nonumber \\
&& A_\tau \rightarrow -A_\tau~~;~~A_x \rightarrow A_x~~;~~A_y
\rightarrow A_y~~; \nonumber \\
&&~~~~~~~~~~~~~\psi \rightarrow \psi~~;~~\varphi_\ell \rightarrow
\varphi_\ell.
\end{eqnarray}
From these mappings we can construct the simplest observable which
is odd under $\mathcal{T}$, $I_x^{\rm dual}$, $I_y^{\rm dual}$,
and transforms like Eq.~(\ref{e12}) under translations and lattice
rotations. We define $V_{mn}$ to be the Fourier component of the
vorticity at the wavevector ${\bf Q}_{mn}$ defined in
Eq.~(\ref{e10}), and obtain (compare Eq.~(\ref{e11}))
\begin{eqnarray}
V_{mn} &=& S_V (|{\bf Q}_{mn}|) \omega^{mn/2} \sum_{\ell=0}^{q-1}
\omega^{\ell m} \nonumber \\
&~&~~~~~~~~~\times\left( \varphi_\ell^\ast \frac{\partial
\varphi_{\ell+n}}{\partial \tau} - \frac{\partial
\varphi_\ell^\ast}{\partial\tau} \varphi_{\ell+n} \right).
\end{eqnarray}
Here $S_V (Q)$ is an unknown structure factor which depends
smoothly on $Q$.

We can now easily extend the discussion in
Section~\ref{sec:pinnonzero} to obtain the mean vorticity
modulation in the vortex lattice. The analog of the result in
Eq.~(\ref{vp7}) is
\begin{eqnarray}
\langle V_{mn} ({\bf r}) \rangle &=& \frac{S_V \left(|{\bf
Q}_{mn}|\right ) m_v \omega_v }{\pi}  \exp \left( - m_v \omega_v
r^2 \right) \nonumber
\\ &~&~~\times \omega^{mn/2} \sum_{\ell} \mathcal{U}^{\ast}_\ell
\mathcal{U}_{\ell+n} \omega^{\ell m}. \label{vp7a}
\end{eqnarray}

\section{Other lattices}

%

We illustrate the extension of the PSG to other lattices by
considering the case of bosons on a honeycomb lattice. We consider
boson density $f$ per site, and as in the case of a square lattice
we write $f=p/q$, with $p$ and $q$ relatively prime integers and
define $\omega = e^{2\pi i f}$. The vortices reside on the dual
triangular lattice, and so we have to consider the PSG of motion
on a triangular lattice in the presence of a `magnetic' field.

The triangular lattice is spanned by the two basis vectors
\begin{equation}
  {\bf a}_1 =  {\bf e}_x \;,\qquad
  {\bf a}_2 =  -\frac{1}{2} {\bf e}_x + \frac{\sqrt 3}{2} {\bf e}_y \;.
\label{eq:triangularbasisvectors}
\end{equation}
Here ${\bf e}_x$ and ${\bf e}_y$ are orthogonal unit vectors. Any
position vector of the lattice can be written as ${\bf a} = a_1
{\bf a}_1 + a_2 {\bf a}_2$ with $a_1$ and $a_2$ integers. It is
also convenient to define
\begin{equation}
  {\bf a}_d = {\bf a}_1 + {\bf a}_2 = \frac{1}{2} {\bf e}_x + \frac{\sqrt 3}{2} {\bf e}_y \;.
\end{equation}
Here we will use the Landau gauge $\overline{A}_{a\tau} =
\overline{A}_{a1} = 0$ and
\begin{equation}
\overline{A}_{a2} = 2 f \, a_1 \;,\quad \overline{A}_{ad} = f (2
a_1 +1) \;. \label{eq:triangularlandau}
\end{equation}
For the translation and rotation operators we obtain
\begin{align}
T_1 &: \psi(a_1, a_2)  \rightarrow \psi (a_1 - 1, a_2) \, \omega^{
2 a_2} \nonumber \\
T_2 &: \psi(a_1, a_2)  \rightarrow \psi (a_1, a_2 - 1) \,
\nonumber
\\
T_d &: \psi(a_1, a_2)  \rightarrow \psi (a_1 - 1, a_2 - 1) \,
\omega^{
2 a_2 -1} \nonumber \\
R_{\pi/3}^{\rm dual} &: \psi(a_1, a_2)  \rightarrow \psi (a_2,a_2
-a_1 ) \, \omega^{2 a_1 a_2 - a_2^2}  \;.
\label{eq:triangularatrans}
\end{align}
As in the case of a square lattice, the translation operators do
not commute, and in addition to $T_1 T_2 = \omega^2 T_2 T_1$  we
have
\begin{equation}
  T_1\, T_2 = \omega \, T_d \;.
    \label{eq:T1T2Td}
\end{equation}
As expected, the following relations of the triangular lattice are
satisfied,
\begin{eqnarray}
 T_1 R_{\pi/3}^{\rm dual} &=& R_{\pi/3}^{\rm dual} T_2^{-1}  \nonumber \\
 T_2 R_{\pi/3}^{\rm dual} &=& R_{\pi/3}^{\rm dual} T_d  \nonumber \\
 T_d R_{\pi/3}^{\rm dual} &=& R_{\pi/3}^{\rm dual} T_1 \nonumber \\
 \left( R_{\pi/3}^{\rm dual} \right)^6 &=& 1 \;. \label{eq:triangularrtr}
 \label{eq:TR3=R3T}
\end{eqnarray}

To discuss the translation and rotation operations in momentum
space it is convenient to introduce the reciprocal lattice to our
triangular lattice.  The reciprocal lattice is also a triangular
lattice which, with respect to the original (dual) triangular
lattice, is rotated by $30$ degrees and spanned by the basis
vectors
\begin{equation}
  {\bf b}_1 = {\bf e}_x + \frac{1}{\sqrt 3} {\bf e}_y \;,\quad {\bf b}_2 = \frac{2}{\sqrt 3} {\bf e}_y \;.
\end{equation}
We can now write any wave-vector ${\bf k}$ as ${\bf k} = k_1 {\bf
b}_1 + k_2 {\bf b}_2$. Using ${\bf a}_i \cdot {\bf b}_j =
\delta_{ij}$ we have ${\bf k} \cdot {\bf a} =  k_1 a_1 +  k_2 a_2$
and obtain for the translation and rotation operations in momentum
space
\begin{eqnarray}
T_1 &:& \psi(k_1, k_2)  \rightarrow \psi (k_1, k_2 - 4 \pi f)
e^{-i k_1} \nonumber
\\
T_2 &:& \psi(k_1, k_2)  \rightarrow \psi (k_1, k_2) e^{-i k_2}
\nonumber
\\
T_d &:& \psi(k_1, k_2)  \rightarrow \psi (k_1, k_2 - 4 \pi
f)\,\omega e^{-i k_1-i k_2} \nonumber
\\
R_{\pi/3}^{\rm dual} &:& \psi(k_1, k_2)  \rightarrow  \label{eq:triangularktrans} \\
&{}& \hspace{-9mm}
\begin{array}{ll}
 \frac{1}{q} \sum_{m,n=0}^{q-1}
 \psi (k_1 + k_2 +  4 \pi f n,
 \\
 \qquad  -k_1 - 4 \pi m f ) \omega^{m(m-2 n)}
 &,\; q = 2\nu + 1 \\
  \frac{2}{q} \sum_{m,n=0}^{q/2-1}
 \psi (k_1 + k_2 +  4 \pi f n,
 \\
 \qquad -k_1 - 4 \pi m f ) \omega^{m(m - 2n)}
 &,\; q = 4\nu \\
 \frac{2}{q} \sum_{m,n=0}^{q/2-1}
 \psi (k_1 + k_2 +  4 \pi f (n-1/2),
 \\
 \qquad  -k_1 - 4 \pi m f ) \omega^{m(m - 2n + 1)}
 &,\; q = 4\nu + 2 \;,
\end{array}
\nonumber
\end{eqnarray}
where $\nu$ is an integer. Note that we find a different
transformation form for the rotations depending upon the value of
$q$ (mod 4). There is also a compact expression for the rotation
operator valid for general $q$, but this obscures the underlying
symmetries by containing terms which cancel each other in the case
of even $q$.

As in the case of the square lattice in Section~\ref{sec:field},
let us now consider the important vortex fluctuations at momenta
where the spectrum has minima. If $|\Lambda\rangle$ is a state at
a minimum of the spectrum with $T_2 |\Lambda \rangle =
e^{-ik_2^{\ast}}$, then acting successively with $T_1$ on this
state gives us degenerate states $T_1^{\ell} | \Lambda \rangle$
with $T_2$ eigenvalue $e^{-ik_2^{\ast}}\omega^{-2\ell}$. This
procedure actually does not completely span the degenerate
manifold for $q= 4\nu+2$, and we will limit our considerations for
now to the other cases.

\subsection{$q = 2\nu + 1$}

For odd $q$ the eigenvalues $e^{-ik_2^{\ast}}\omega^{-2\ell}$ are
all different, and so the degeneracy of the vortex fields has to
be at least $q$. For the Hofstadter problem on a triangular
lattice we find that there are exactly $q$ minima located at the
wavevectors $(0, 4\pi \ell p/q)$ with $\ell = 0 \dots q-1$.
Analogously to the case of the square lattice in
Section~\ref{sec:field}, we can now proceed and find for the
transformation laws of the vortex fields $\varphi_{\ell}$ at these
minima
\begin{eqnarray}
T_1 &:& \varphi_{\ell} \rightarrow \varphi_{\ell+1} \nonumber \\
T_2 &:& \varphi_{\ell} \rightarrow \varphi_\ell \omega^{-2\ell}  \nonumber \\
T_d &:& \varphi_{\ell} \rightarrow \varphi_{\ell+1} \omega^{-(2\ell+1)} \nonumber \\
R_{\pi/3}^{\rm dual} &:& \varphi_{\ell} \rightarrow
\frac{e^{-i\pi(q-1)/12}}{\sqrt{q}} \sum_{m=0}^{q-1} \varphi_m
\omega^{(\ell -2m)\ell} \;. \label{eq:TTRtriangluarfields}
\end{eqnarray}
It can be verified that these transformation laws obey
Eqs.~(\ref{eq:T1T2Td}) and (\ref{eq:TR3=R3T}).

\subsection{$q = 4\nu$}

If $q$ is a multiple of $4$ we only have $q/2$ different vortex
minima which are located at the wavevectors $(0, 4\pi \ell p/q)$
with $\ell = 0 \dots q/2-1$. The transformations among these
fields are
\begin{eqnarray}
T_1 &:& \varphi_{\ell} \rightarrow \varphi_{\ell+1} \nonumber \\
T_2 &:& \varphi_{\ell} \rightarrow \varphi_\ell \omega^{-2\ell}  \nonumber \\
T_d &:& \varphi_{\ell} \rightarrow \varphi_{\ell+1} \omega^{-(2\ell+1)} \nonumber \\
R_{\pi/3}^{\rm dual} &:& \varphi_{\ell} \rightarrow \frac{e^{-ip
\pi/12}}{\sqrt{q/2}} \sum_{m=0}^{q/2-1} \varphi_m \omega^{(\ell
-2m)\ell} \;. \label{eq:TTRtriangluarfields2}
\end{eqnarray}
These also obey Eqs.~(\ref{eq:T1T2Td}) and (\ref{eq:TR3=R3T}).

%

\section{Vortex band structure and PSG in the ``symmetric gauge''}
\label{sec:symmetric}

\begin{figure}[t]
\ifig[width=0.2\textwidth]{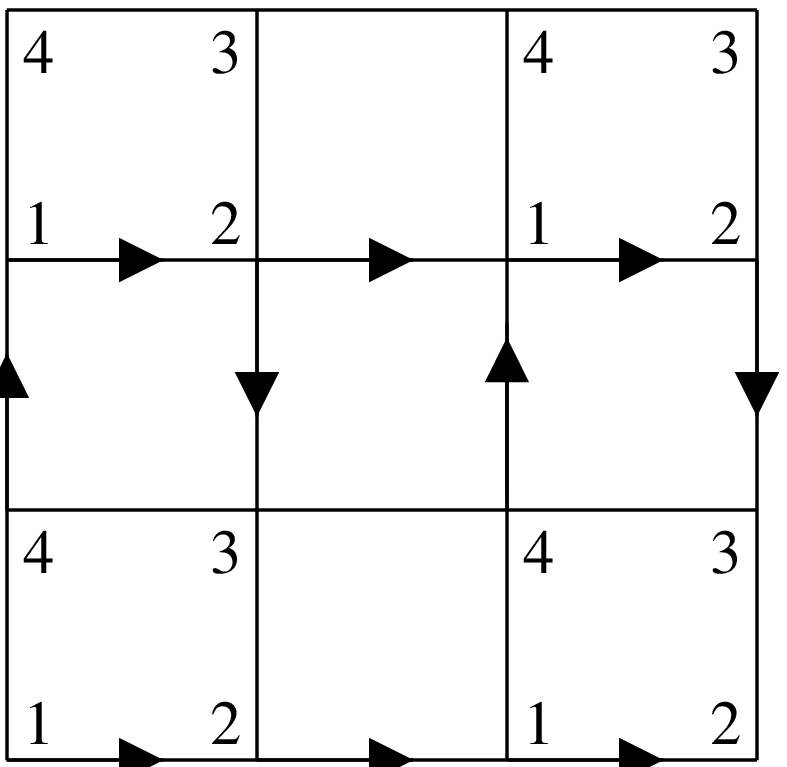}
\caption{Pictorial representation of the ``symmetric gauge'' .
Arrows represent a factor $e^{-i \pi/2}$. Magnetic unit cell
labelling convention is also shown.} \label{symgauge}
\end{figure}

In this appendix we show how to directly obtain a permutative
representation of the PSG for $f=1/4$ by a judicious choice of gauge
for the dual magnetic flux.  We will loosely call this the ``symmetric
gauge''. It is most conveniently expressed pictorially and is shown in
Fig.~\ref{symgauge}, along with the corresponding magnetic unit cell
labelling convention. A unique property of this gauge choice, that
will play a crucial role in the discussion below, is that the {\em
  Bravais net} of the magnetic lattice is invariant under $\pi/2$
rotations.  It is clear that such a gauge choice is not
possible in general, but is at least for filling factor $f=p/q$, satisfying
the condition $q=2,n^2, n=2,3,4,\ldots$ (for the square lattice).  We
do not, however, explore this beyond $q=4$.

Directly diagonalizing the lattice vortex action in Eq.~(\ref{zdh}) we
find 4 low energy vortex modes at wavevectors:
\begin{eqnarray}
\label{eq:3.1}
{\bf k}_0 &=& ( 0, -\pi/2), \nonumber \\
{\bf k}_1 &=& ( \pi, -\pi/2), \nonumber \\
{\bf k}_2 &=& ( \pi, \pi/2), \nonumber \\
{\bf k}_3 &=& ( 0, \pi/2),
\end{eqnarray}
which are defined in the first Brillouin zone of the {\em
magnetic} lattice and we have taken the magnetic lattice constant
to be equal to one. The corresponding normalized eigenvectors are
given by:
\begin{eqnarray}
\label{eq:3.2} v^0 &=& \left(0, 1/2, 1/\sqrt{2}, 1/2 \right),
\nonumber \\
v^1 &=& \left(-i/2, 1/\sqrt{2}, 1/2, 0\right),
\nonumber \\
v^2 &=& \left(1/\sqrt{2}, i/2, 0, 1/2 \right),\nonumber \\
v^3 &=& \left(1/2, 0, 1/2, 1/\sqrt{2} \right),
\end{eqnarray}
As before, concentrating on the long-wavelength vortex dynamics
near the transition, the vortex fields can be written as linear
combinations of the 4 low-energy vortex modes:
\begin{equation}
\label{eq:3.3} \psi_{\alpha} ({\bf R}) = \sum_{\ell=0}^3
v^{\ell}_{\alpha}({\bf R}) \zeta_{\ell},
\end{equation}
where ${\bf R}$ labels magnetic unit cells, $\alpha$ labels sites
within each magnetic unit cell and
\begin{equation}
\label{eq:3.4} v^{\ell}_{\alpha}({\bf R}) = v^{\ell}_{\alpha} e^{i
{\bf k_{\ell}} \cdot {\bf R}}.
\end{equation}
Using Eqs.(\ref{zdh}) and (\ref{eq:3.3}) we can work out the
transformation properties of the low-energy vortex modes
$\varphi_{\ell}$ under the symmetry operations of the microscopic
Hamiltonian (\ref{hubbard}). These transformations turn out to
have the following form:
\begin{eqnarray}
\label{eq:3.5} &&T_x:\, \zeta_0 \rightarrow \zeta_3,\,\,
\zeta_1 \rightarrow \zeta_2,\,\ \zeta_2 \rightarrow -
\zeta_1,\,\,
\zeta_3 \rightarrow \zeta_0, \nonumber \\
&&T_y:\, \zeta_0 \rightarrow \zeta_1,\,\, \zeta_1
\rightarrow \zeta_0,\,\, \zeta_2 \rightarrow i \zeta_3,\,\,
\zeta_3 \rightarrow i \zeta_2,\nonumber \\
&&R_{\pi/2}^{\rm dir}:\, \zeta_0 \rightarrow \zeta_1,\,\,
\zeta_1 \rightarrow i \zeta_2,\,\, \zeta_2 \rightarrow
\zeta_3,\,\,
\zeta_3 \rightarrow \zeta_0,\nonumber \\
&&R_{\pi/2}^{\rm dual}:\, \zeta_0 \rightarrow \zeta_0,\,\,
\zeta_1 \rightarrow - \zeta_3,\,\, \zeta_2 \rightarrow i
\zeta_2,\,\,
\zeta_3 \rightarrow \zeta_1,\nonumber \\
&&I_x^{\rm dir}:\, \zeta_0 \rightarrow i \zeta_1^*, \,\,
\zeta_1 \rightarrow i \zeta_0^*, \,\, \zeta_2 \rightarrow
\zeta_3^*,\,\,
\zeta_3 \rightarrow \zeta_2^*,\nonumber \\
&&I_y^{\rm dir}:\, \zeta_0 \rightarrow \zeta_3^*, \,\,
\zeta_1 \rightarrow \zeta_2^*, \,\, \zeta_2 \rightarrow
\zeta_1^*,\,\,
\zeta_3 \rightarrow \zeta_0^*,\nonumber \\
&&I_x^{\rm dual}:\, \zeta_0 \rightarrow - \zeta_0^*, \,\,
\zeta_1 \rightarrow - \zeta_1^*, \,\, \zeta_2 \rightarrow
\zeta_2^*,\,\,
\zeta_3 \rightarrow \zeta_3^*,\nonumber \\
&&I_y^{\rm dual}:\, \zeta_0 \rightarrow \zeta_0^*, \,\,
\zeta_1 \rightarrow - \zeta_1^*, \,\, \zeta_2 \rightarrow
\zeta_2^*,\,\,
\zeta_3 \rightarrow \zeta_3^*. \nonumber \\
\end{eqnarray}
Here $R_{\pi/2}^{\rm dir}$ denotes a $\pi/2$-rotation about a {\em
  direct} lattice site, $R_{\pi/2}^{\rm dual}$ is a $\pi/2$-rotation
about a {\em dual} lattice site, $I_{x,y}^{\rm dir}$ are reflections
with respect to $x,y$ axes passing through a {\em direct lattice
  site}, while $I_{x,y}^{\rm dual}$ are reflections with respect to
$x,y$ axes passing through a {\em dual lattice site}.  Clearly, the
above rotational and translational transformations are all of the form
of a unitary diagonal matrix multiplying a permutation matrix.  Hence
by composing each with a suitable global (gauge) phase rotation, they
realize a permutative representation of the PSG.

It is straightforward to see that the most general quartic potential
invariant under these transformations has the same form as
Eq.~(\ref{m10}).  Thus one directly obtains in this manner the end
result obtained in the text by first using Landau gauge and then
unitarily transforming to $\zeta_\ell$ variables.

\section{Some ground states and their properties for $q=4$}
\label{sec:some-ground-states}

In this appendix we present a few more details of the structure of the
saddle points of effective potential in Eq.~(\ref{eq:vlambda}) for the
equal amplitude state identified as E in Fig~\ref{figq4a}. Writing
$\zeta_\ell \propto e^{i \vartheta_\ell}$, the relevant term in the
energy density (Euclidean Lagrange density) is (assuming first that
$\lambda < 0$):
\begin{eqnarray}
  \label{eq:3.8} {\cal L}_4 &=&-|\lambda| \left[ \cos
    2(\vartheta_0-\vartheta_1) - \cos 2(\vartheta_1-\vartheta_2)
  \right. \nonumber \\
  &+& \left. \cos 2(\vartheta_2-\vartheta_3) +
    \cos2(\vartheta_3-\vartheta_0) \right].
\end{eqnarray}
In these variables it is not immediately obvious what the ground
state is. We change variables to
\begin{eqnarray}
\label{eq:3.9}
\vartheta_0&=&\tilde \vartheta_0, \nonumber \\
\vartheta_1&=&\tilde \vartheta_1 + \frac{\pi}{8}, \nonumber \\
\vartheta_2&=&\tilde \vartheta_2 - \frac{\pi}{4}, \nonumber \\
\vartheta_3&=&\tilde \vartheta_3 - \frac{\pi}{8}.
\end{eqnarray}
In the new variables the mean-field energy density becomes:
\begin{eqnarray}
\label{eq:3.10} {\cal L}_4 &=&-|\lambda| \left[ \cos 2\left(\tilde \vartheta_0 -
\tilde \vartheta_1 - \frac{\pi}{8}\right) + \cos 2 \left(\tilde
\vartheta_1 - \tilde \vartheta_2  -
\frac{\pi}{8} \right) \right. \nonumber \\
&+& \left.\cos 2\left( \tilde \vartheta_2 - \tilde \vartheta_3 -
\frac{\pi}{8} \right) + \cos 2 \left(\tilde \vartheta_3 - \tilde
\vartheta_0
- \frac{\pi}{8}\right) \right]. \nonumber \\
\end{eqnarray}
It is now clear that the minimum is achieved when
\begin{eqnarray}
\label{eq:3.11}
\tilde \vartheta_0 - \tilde \vartheta_1&=&0,\pi, \nonumber \\
\tilde \vartheta_1 - \tilde \vartheta_2&=&0,\pi, \nonumber \\
\tilde \vartheta_2 - \tilde \vartheta_3&=&0,\pi,
\end{eqnarray}
and
\begin{eqnarray}
\label{eq:3.12} \tilde \vartheta_0 - \tilde
\vartheta_1&=&\frac{\pi}{4},\frac{5 \pi}{4},
\nonumber \\
\tilde \vartheta_1 - \tilde \vartheta_2&=&\frac{\pi}{4}, \frac{5
\pi}{4},
\nonumber \\
\tilde \vartheta_2 - \tilde \vartheta_3&=&\frac{\pi}{4},\frac{5
\pi}{4}.
\end{eqnarray}
Thus the ground state has a 16-fold degeneracy. To see what do
these states look like in real space it is sufficient to
understand how they transform under the symmetry operations of the
microscopic Hamiltonian Eq.~(\ref{hubbard}). Using
Eq.~(\ref{eq:3.5}) one finds that the ground state manifold has
the following properties.
\begin{enumerate}
\item{All 16 states are invariant under $T_x^4$ and $T_y^4$, i.e.
the charge density pattern in the ground state has a period of 4
lattice constants in both $x$- and $y$-directions.}
\item{There
are 2 states that are invariant under rotations by $\pi/2$ about a
given direct lattice site. These states are, in the original variables:
\begin{equation}
\label{eq:3.13} \vartheta_0-\vartheta_1 = - \frac{\pi}{8},\,
\vartheta_1-\vartheta_2 = \frac{3 \pi}{8},\,
\vartheta_2-\vartheta_3 = - \frac{\pi}{8},
\end{equation}
and
\begin{equation}
\label{eq:3.14} \vartheta_0-\vartheta_1 = \frac{7 \pi}{8},\,
\vartheta_1-\vartheta_2 = - \frac{5\pi}{8},\,
\vartheta_2-\vartheta_3 = \frac{7\pi}{8}
\end{equation}
These 2 states are related to each other by $(T_x T_y)^2$
transformation.  More simply, this implies that each state has two
inequivalent centers of rotation within each unit cell.}
\item{The states are also invariant under $I_x$ and $I_y$
    transformations through the centers of rotation.}
\item{All 16 states are related to each other by successive $T_x$ or
    $T_y$ operations.}
\end{enumerate}
It is then clear that the ground state is a charge density wave
(CDW), of the general form shown in Fig.\ref{cdw16}.

Now consider the case $\lambda > 0$. In this case there are 16
degenerate ground states as well. They are given by:
\begin{eqnarray}
\label{eq:3.15} \tilde \vartheta_0 - \tilde
\vartheta_1&=&\frac{\pi}{2},\frac{3 \pi}{2},
\nonumber \\
\tilde \vartheta_1 - \tilde \vartheta_2&=&\frac{\pi}{2}, \frac{3
\pi}{2},
\nonumber \\
\tilde \vartheta_2 - \tilde \vartheta_3&=&\frac{\pi}{2},\frac{3
\pi}{2}.
\end{eqnarray}
and
\begin{eqnarray}
\label{eq:3.16} \tilde \vartheta_0 - \tilde
\vartheta_1&=&-\frac{\pi}{4},\frac{3 \pi}{4},
\nonumber \\
\tilde \vartheta_1 - \tilde \vartheta_2&=&-\frac{\pi}{4}, \frac{3
\pi}{4},
\nonumber \\
\tilde \vartheta_2 - \tilde \vartheta_3&=&-\frac{\pi}{4},\frac{3
\pi}{4}.
\end{eqnarray}

These states have exactly the same symmetry properties as the ones
at $\lambda < 0$ and thus are physically the same states. The two
ground states that are invariant under $R_{\frac{\pi}{2}}$ in this
case are:
\begin{equation}
\label{eq:3.17} \vartheta_0-\vartheta_1 = \frac{3\pi}{8},\,
\vartheta_1-\vartheta_2 = \frac{7\pi}{8},\,
\vartheta_2-\vartheta_3 = \frac{3\pi}{8},
\end{equation}
and
\begin{equation}
\label{eq:3.18} \vartheta_0-\vartheta_1 = - \frac{5\pi}{8},\,
\vartheta_1-\vartheta_2 = - \frac{\pi}{8},\,
\vartheta_2-\vartheta_3 = - \frac{5\pi}{8}.
\end{equation}

\section{Permutative representations of the PSG}
\label{sec:noq3}

As discussed in Sec.~\ref{sec:fract-vort-pict}, the problem of
determining whether a fractionalized boson interpretation of the
$q$-vortex theory at filling $f=p/q$ reduces to a problem in
representation theory.  In particular, we must find
$q$-dimensional matrices $x$, $y$, $r$ representing the generators
$T_x$, $T_y$, $R_{\pi/2}^{\rm dual}$ (for this appendix we neglect
the inversion operations)
\begin{eqnarray}
  \label{eq:permrep}
  x& =& \bar{x} \hat{x}, \nonumber \\
  y& =& \bar{y} \hat{y}, \nonumber\\
  r& =& \bar{r} \hat{r},
\end{eqnarray}
where $\bar{x},\bar{y},\bar{r}$ are diagonal unitary matrices
(i.e. diagonal matrices with all entries of absolute value one),
and $\hat{x},\hat{y},\hat{z}$ are permutation matrices.  We seek
matrices of this form satisfying
\begin{eqnarray}
  \label{eq:alg}
  x y & = & \omega y x , \nonumber\\
  r^{-1} x r & = & y^{-1}, \nonumber\\
  r^{-1} y r & = & x, \nonumber\\
  r^4 & = & 1,
\end{eqnarray}
with $\omega^k=1$ if and only if $k=0 (\mbox{mod q})$.  The form
of Eq.~(\ref{eq:permrep}) requires that the permutation (hat)
parts of the generators satisfy very similar relations to
Eq.~(\ref{eq:alg}), to wit:
\begin{eqnarray}
  \label{eq:algh}
  \hat{x} \hat{y} & = &  \hat{y} \hat{x} , \nonumber\\
  \hat{r}^{-1} \hat{x} \hat{r} & = & \hat{y}^{-1}, \nonumber\\
  \hat{r}^{-1} \hat{y} \hat{r} & = & \hat{x}, \nonumber\\
  \hat{r}^4 & = & 1.
\end{eqnarray}

This leads to a few general observations.  First, recall that any
permutation can be written (decomposed) as a product of disjoint
cycles.  Since $\hat{r}^4=1$, $\hat{r}$ can contain only cycles of
length $1$,$2$, and $4$.  Hence, $\hat{r}^2$ contains only
$1$-cycles and $2$-cycles, and an even number of the latter.
Another interesting identity is
\begin{equation}
  \label{eq:id2}
  r^2 y r^2 = r^{-2} y r^2 = r^{-1} x r = y^{-1}.
\end{equation}
Hence
\begin{eqnarray}
  \label{eq:id3}
&&   r^2 y r^2 y = 1, \nonumber \\
&&   r^2 x r^2 x =1,
\end{eqnarray}
the latter equation being obtained similarly.  Thus $\hat{r}^2
\hat{x}$ can contain only $1$-cycles and $2$-cycles as well.

Finally, to obtain a non-vanishing commutator of $x$ and $y$,
clearly at least one of $\hat{x},\hat{y}$ must not be the identity
permutation.

Unfortunately we have so far been unable to solve this representation
problem generally.  The explicit examples for $q=2,4$ have been
discussed in the text.  In addition, we have been able to prove there
is no such representation for $q=3$.  This proof follows.  First, note
that $\hat{r}$ can contain either one $2$-cycle or be trivial. Suppose
first $\hat{r}=1$.  Then from the third of Eqs.~(\ref{eq:algh}), we
see that $\hat{x}=\hat{y}=\hat{p}$, some permutation which must be
non-trivial according to the above consideration.  But since
$\hat{r}=1$, the second of Eqs.~(\ref{eq:algh}) implies $\hat{p}^2=1$.
Thus, without loss of generality, we can take $\hat{p}=(12)$ (we use
cycle notation, so this is just the permutation of the first two
objects).  Inserting this into $xy=\omega yx$, one finds that, since
the third element is not permuted,
$\bar{x}_{33}\bar{y}_{33}=\omega\bar{x}_{33}\bar{y}_{33}$, which is
impossible. Thus our assumption that $\hat{r}= 1$ must be false.

Hence if such a representation exists, we must take $\hat{r}$ to
be a $2$-cycle, and without loss of generality, we may take
$\hat{r}=(12)$.  Thus $\hat{r}^2=1$, so from Eq.~(\ref{eq:id3}),
$\hat{x}^2=1$ and $\hat{y}^2=1$.  Now at least one of
$\hat{x},\hat{y}$ is non-trivial, so let us assume, again without
loss of generality, that $\hat{x}\neq 1$.  It must be a $2$-cycle
since its square is unity.  There are only two inequivalent
possibilities.  If $\hat{x}=(12)$, then $\hat{x}=\hat{r}$, which
implies $\hat{y}=(12)$ as well.  We have already seen that there
is no solution with $\hat{x}=\hat{y}=(12)$, so this is
inconsistent.  The other inequivalent choice is $\hat{x}=(23)$.
Then we can compute $\hat{y}=\hat{r} \hat{x} \hat{r}^{-1} =(13)$.
But it is easy to see that this violates the first of
Eqs.~(\ref{eq:algh}).

Thus all assumptions for $q=3$ have been proven inconsistent save
one: that no such representation exists.

We have found some explicit examples for $q=8,9$.

For $q=8$, with the notation $\gamma=e^{i \pi/16}$
($\gamma^4=\omega$),
  we find
\begin{eqnarray}
x & = &  \label{eq:xq8}
  \left( \begin{array}[c]{ccccccccc}
0&0&0&0&0&{{\gamma }^9}&0&0 \\
        {{\gamma }^
      {30}}&0&0&0&0&0&0&0 \\
        0&0&0&0&0&0&0&{{\gamma }^{15}} \\
        0&0&{{\gamma }^
      2}&0&0&0&0&0 \\
        0&{{\gamma }^{17}}&0&0&0&0&0&0 \\
        0&0&0&0&1&0&0&0 \\
        0&0&0&{{\gamma }^
      {23}}&0&0&0&0 \\
        0&0&0&0&0&0&1&0
     \end{array} \right),\\
y & = & \left( \begin{array}[c]{ccccccccc} 0&0&0&0&0&0&{{\gamma }^
      {23}}&0 \\
        0&0&0&0&0&0&0&{{\gamma }^
      {17}} \\
        1&0&0&0&0&0&0&0 \\
        0&1&0&0&0&0&0&0 \\
        0&0&{{\gamma }^
      {15}}&0&0&0&0&0 \\
        0&0&0&{{\gamma }^9}&0&0&0&0 \\
        0&0&0&0&{{\gamma }^
      2}&0&0&0 \\
        0&0&0&0&0&{{\gamma }^{30}}&0&0
  \end{array} \right), \\
r & = & \left( \begin{array}[c]{ccccccccc}
0&0&0&0&0&0&1&0 \\
        0&0&0&0&1&0&0&0 \\
        0&0&0&0&0&0&0&1 \\
        0&0&0&0&0&1&0&0 \\

     0&0&1&0&0&0&0&0 \\
        1&0&0&0&0&0&0&0 \\
        0&0&0&1&0&0&0&0 \\
        0&1&0&0&0&0&0&0
  \end{array} \right).
\end{eqnarray}

  For $q=9$, with the notation $\tilde\omega=e^{i \pi/9}$ ($\tilde\omega^2=\omega$),
  we find
\begin{eqnarray}
x & = &  \label{eq:xq9}
  \left( \begin{array}[c]{ccccccccc}
0&0&{{\tilde\omega }^
      5}&0&0&0&0&0&0 \\
        1&0&0&0&0&0&0&0&0 \\
        0&{{\tilde\omega }^
      {16}}&0&0&0&0&0&0&0 \\
        0&0&0&0&0&{{\tilde\omega }^
      9}&0&0&0 \\
        0&0&0&1&0&0&0&0&0 \\
        0&0&0&0&1&0&0&0&0 \\
        0&0&0&0&0&0&0&0&{{\tilde\omega }^
      {13}} \\
        0&0&0&0&0&0&{{\tilde\omega }^2}&0&0 \\
        0&0&0&0&0&0&0&1&0
     \end{array} \right),\\
y & = & \left( \begin{array}[c]{ccccccccc}
       0&0&0&0&0&0&{{\tilde\omega }^
      {13}}&0&0 \\
        0&0&0&0&0&0&0&{{\tilde\omega }^9}&0 \\
        0&0&0&0&0&0&0&0&{{\tilde\omega }^5} \\
        {{\tilde\omega
        }^2}&0&0&0&0&0&0&0&0 \\
        0&1&0&0&0&0&0&0&0 \\
        0&0&1&0&0&0&0&0&0 \\
        0&0&0&1&0&0&0&0&
     0 \\
        0&0&0&0&1&0&0&0&0 \\
        0&0&0&0&0&{{\tilde\omega }^{16}}&0&0&0
  \end{array} \right), \\
r & = & \left( \begin{array}[c]{ccccccccc}
        0&0&0&0&0&0&1&0&0 \\
        0&0&0&1&0&0&0&0&0 \\
        1&0&0&0&0&0&0&0&0 \\
        0&0&0&0&0&
     0&0&1&0 \\
        0&0&0&0&1&0&0&0&0 \\
        0&1&0&0&0&0&0&0&0 \\
        0&0&0&0&0&0&0&0&1 \\
        0&0&0&0&0&
     1&0&0&0 \\
        0&0&1&0&0&0&0&0&0
  \end{array} \right).
\end{eqnarray}

These were obtained as follows.  First, we obtained
$\hat{x},\hat{y},\hat{r}$ from a simple physical construction.  In
particular, we constructed a covering of the sites of the {\sl dual}
square lattice by $q$ sublattices.  For $q=n^2$ ($n=3$ for $q=9$
above), these are chosen by breaking the lattice up into a square grid
of $n\times n$ blocks, labeling the sites in each block identically
from $\ell=1\ldots q$.  Taking the site of the {\sl direct} lattice at
the center of one of these blocks as the origin, one can see that
rotations and translations all act as permutations of the $q$ labels.
These permutations are taken as $\hat{x},\hat{y},\hat{r}$ and by
construction satisfy Eqs.~(\ref{eq:algh}).  For $q=2n^2$ ($n=2$ for $q=8$
above), we break the sites of the dual lattice again up into a square
grid of $n\times n$ blocks, and further break up these blocks into
staggered A and B sublattices.  We label the sites of each block of
the A sublattice identically from $1\ldots n^2$ and the sites of each
block of the B sublattice identically from $n^2+1\ldots 2n^2$.  Now
taking the origin as a point of the direct lattice at the corner of
two A and two B blocks, we once again obtain from physical rotations a
set of permutations on $q$ variables, which solve Eqs.~(\ref{eq:algh}).

We then take $\overline{x},\overline{y},\overline{r}$ as unknowns, and
look for solutions of Eqs.~(\ref{eq:alg}).  These reduce to many
redundant equations for $3q$ unknowns (the phases of each diagonal
entry of $\overline{x},\overline{y},\overline{r}$).  We find for
$q=8,9$ many solutions, of which the above examples are particular
ones.  This procedure can clearly be generalized to arbitrary $n$ as
defined above.  We do not, however, have a proof at present that
solutions for the diagonal unknowns exist in general.  The
overabundance of solutions for $q=8,9$ suggests, however, that they do
exist -- hence the speculation in the text.

\section{Duality for $q$-slave boson gauge theory}
\label{sec:duality-q-slave}

In this appendix, we directly dualize the hamiltonian ${\cal
  H}_{\tilde{q}}={\cal H}_{\tilde{q}}^0+ {\cal H}_{\tilde{q}}^1$ in
Eqs.~(\ref{eq:heff1}),~(\ref{eq:staggpot}).  The first step is to
shift the slave rotor number operators to new variables with an
average zero density:
\begin{equation}
  \label{eq:rotshift}
  \hat{n}_{i\ell} \rightarrow \hat{n}_{i\ell} + p m_{i\ell}.
\end{equation}
Note that, {\sl using $\tilde{q}=q$}, the spatial average of $p
m_{i\ell}$ is $p/q=f$, so that the shifted $\hat{n}_{i\ell}$
fields have zero spatial average.  This shift is allowed and
canonical because $p$ is an integer. After this shift, the
Hamiltonian becomes, up to a constant,
\begin{eqnarray}
  \label{eq:heff2}
&&   {\cal H}_{\tilde{q}}  =  \nonumber \\
  && -t \sum_{i\ell} \cos( \Delta_\alpha
    \hat{\phi}_{i\ell}- {\cal A}_{i\alpha}^\ell + {\cal
      A}_{i\alpha}^{\ell-1} ) \nonumber \\
      && +u
    \left(\hat{n}_{i\ell}-\!\tilde{\mu} \left(m_{i\ell}\!-\!\frac{1}{q} \right) \right)^2\nonumber \\
  & & + \frac{v}{2} \sum_{i\ell} |\vec{\cal E}_i^\ell|^2 -
  K \sum_{a\ell} \textrm{``cos''}(\vec\Delta \times
  \vec{\cal A}^\ell)_a,
\end{eqnarray}
with $\tilde{\mu}= -(\tilde\mu_s-p)/2$.  The Gauss' law constraint
is changed to
\begin{equation}
  \label{eq:gauss2}
  \left[\vec{\Delta} \cdot \vec{\cal E}^\ell\right]_i = \hat{n}_{i,\ell}
  -\hat{n}_{i,\ell+1}+ \epsilon_{i\ell}.
\end{equation}

At this point, we can recover the analysis of
Sec.~\ref{sec:analys-effect-gauge} by assuming the bosons are
gapped, $\hat{n}_{i\ell}\approx 0$, and treating the pure compact
$U(1)^{q-1}$ gauge theory with the static source charges
$\epsilon_{i\ell}$.  Here, we will connect to the dual $q$-vortex
theory directly.  This is accomplished by performing a separate
duality transformation for each of the $q$ fractional boson
fields.  This is conveniently done, as elsewhere in the paper, in
a path integral formulation.  The appropriate world-line form of
the partition function, analogous to Eq.~(\ref{zh}), but with the
gauge fields included is
\begin{eqnarray}
  \label{eq:qbospf}
&& \hspace{-0.4in}  {\cal Z}  =  \sum_{\{ J^\ell , {\cal B}^\ell
\}}
  \int\! d{\cal A}^\ell \, \prod_{i\ell}\delta\left(\Delta \cdot J_i^\ell \right)
  \prod_{a\mu} \delta\left( \sum_\ell {\cal B}^\ell_\mu\right)
  e^{-{\cal S}_q},~
\end{eqnarray}
where
\begin{eqnarray}
  \label{eq:Sq}
  {\cal S}_q & = & \sum_{i\ell}
  \frac{1}{2e^2}\left|J_{i}^\ell-\!\tilde{\mu}\left(m_{i\ell}\!-\!\frac{1}{q}\right)\hat{\tau} \right|^2 \nonumber \\ && -
  i(J_{i}^\ell- J_{i}^{\ell+1} + \epsilon_{i\ell}\hat{\tau})\cdot{\cal A}_{i}^\ell \nonumber \\
  & + & \sum_{a\ell} i \frac{{\cal B}_{a}^\ell}{2\pi}\cdot
  (\Delta\times {\cal A}^\ell)_a + \frac{\tilde{e}^2}{2} |{\cal
    B}_{a}^\ell|^2.
\end{eqnarray}
Here and in the remainder of this section, we will suppress
$3$-vector indices and spatial coordinate labels when possible,
and use the usual vector notations ($\cdot,\times$) for $3$-vector
operations. Physically, $J^\ell=(n_\ell, \vec{J}_\ell)$ is an
integer-valued vector field representing the $3$-current of
fractional boson $\ell$, ${\cal
  A}^\ell = ({\cal A}_0^\ell, \vec{\cal A}^\ell)$ is the $2\pi$-periodic
$3$-vector potential for the $\ell^{th}$ gauge symmetry.  The
${\cal
  B}^\ell$ is a $2\pi\times$ integer-valued field defined on the links
of the dual lattice, which, when summed over, gives a Villain
potential for the $\ell^{th}$ gauge-field kinetic energy.
Finally, $\hat{\tau}=(1,0,0)$ is the unit vector in the time
direction.

To proceed, we solve the divergence constraints on the boson
currents by defining vector potentials $A^\ell_a$ on the dual
lattice,
\begin{equation}
  \label{eq:Aqdef}
  J_i^\ell = \frac{(\Delta\times A^\ell)_i}{2\pi},
\end{equation}
and for convenience we find some fixed non-fluctuating
$2\pi\times$integer fields $\overline{\cal B}^\ell$ on the dual
lattice satisfying
\begin{equation}
  \label{eq:bgdef}
  \frac{(\Delta\times \overline{\cal B}^\ell)_i}{2\pi} =
  \epsilon_{i\ell}\hat{\tau}.
\end{equation}
Since $\sum_\ell \epsilon_{i\ell}=0$, the background fields can
(and should!) be chosen to satisfy $\sum_\ell \overline{B}^\ell =
0$. Inserting these into the partition function, one has
\begin{eqnarray}
  \label{eq:qbospf1}
&& \hspace{-0.4in}  {\cal Z}  =  \sum_{\{ A^\ell , {\cal B}^\ell
\}}
  \int\! d{\cal A}^\ell \,
  \prod_{a\mu} \delta\left( \sum_\ell {\cal B}^\ell_\mu\right)
  e^{-{\cal S}^1_q},
\end{eqnarray}
with
\begin{eqnarray}
  \label{eq:Sq1}
  {\cal S}^1_q & = & \sum_{i\ell}
  \frac{1}{8\pi^2 e^2}\left|(\Delta\times
  A^\ell)_i-2\pi\tilde{\mu}\left(m_{i\ell}\!-\!\frac{1}{q}\right)\hat{\tau} \right|^2
  \nonumber \\ && -
  \frac{i}{2\pi} {\cal A}_i^\ell\cdot [\Delta\times (A^\ell- A^{\ell+1}
  +\overline{\cal B}^\ell-  {\cal B}^\ell)]_i  \nonumber \\
  & + & \sum_{a\ell} \frac{\tilde{e}^2}{2} |{\cal
    B}_{a}^\ell|^2.
\end{eqnarray}
At this point one may integrate out the ${\cal A}^\ell$ fields,
which gives a delta-function constraining
\begin{equation}
  \label{eq:curlcalB}
  \Delta\times{\cal B}^\ell = \Delta\times(A^{\ell}-A^{\ell+1}
  +\overline{B}^\ell).
\end{equation}
The general solution is
\begin{equation}
  \label{eq:calBsol}
  {\cal B}_\mu^\ell = A_\mu^{\ell}-A_\mu^{\ell+1}
  +\overline{B}_\mu^\ell - \Delta_\mu \chi_\ell,
\end{equation}
where $\chi_\ell$ is a $2\pi\times$integer-valued scalar field.
To preserve the constraint $\sum_\ell {\cal B}^\ell=0$, we should
impose $\sum_\ell \chi_\ell=0$ (or at least a constant).  The
partition function becomes
\begin{eqnarray}
  \label{eq:qbospf2}
&& \hspace{-0.4in}  {\cal Z}  =  \sum_{\{ A^\ell , \chi_\ell \}}
  \prod_{a} \delta\left( \sum_\ell \chi_{a\ell}\right)
  e^{-{\cal S}^2_q},
\end{eqnarray}
with
\begin{eqnarray}
  \label{eq:Sq2}
  {\cal S}^2_q & = & \sum_{i\ell}
  \frac{1}{8\pi^2 e^2}\left|(\Delta\times
  A^\ell)_i-2\pi\tilde{\mu}\left(m_{i\ell}\!-\!\frac{1}{q}\right)\hat{\tau} \right|^2
  \nonumber \\ && + \sum_{a\ell}
  \frac{\tilde{e}^2}{2} |A_\mu^\ell- A_\mu^{\ell+1}
  +\overline{\cal B}_\mu^\ell-\Delta_\mu\chi_\ell|^2.
\end{eqnarray}
Next we ``soften'' the $2\pi\times$integer constraints on
$\chi_\ell,A^\ell$, converting the partition sum to an integral,
and adding terms in the action to weight integer values more
heavily (alternatively, this can be done exactly by Poisson
resummation, but we prefer to keep things simple):
\begin{eqnarray}
  \label{eq:qbospf3}
&& \hspace{-0.4in}  {\cal Z}  =  \int\! dA^\ell d\chi_\ell\,
  \prod_{a} \delta\left( \sum_\ell \chi_{a\ell}\right)
  e^{-{\cal S}^3_q},
\end{eqnarray}
with
\begin{eqnarray}
  \label{eq:Sq3}
  {\cal S}^3_q & = & {\cal S}^2_q - \sum_{a\ell} \left[y_{qv} \cos
  A_{a\mu}^\ell +  \lambda \cos \chi_{a\ell}\right].
\end{eqnarray}
At this point it is advantageous to perform a few manipulations in
a row to bring the partition function into a simple form.  Rather
than show the results of the individual steps, we simply give the
procedure, to be executed in order:
\begin{enumerate}
\item Decompose the background fields into longitudinal and
transverse
  parts according to $\overline{\cal B}^\ell_\mu = \Delta_\mu \eta_\ell +
  \breve{\cal B}^\ell_\mu$, where   $\Delta\cdot \breve{\cal B}^\ell=0$,
  and one chooses $\sum_\ell \eta_\ell=0$.
\item Shift $\chi_{a\ell} \rightarrow \chi_{a\ell} -
  \vartheta_{a\ell}+\vartheta_{a,\ell+1}+ \eta_{a\ell}$, and integrate
  the partition function over the $\vartheta_{a\ell}$ so introduced at
  every dual lattice site.
\item Next shift $\vartheta_\ell \rightarrow \vartheta_\ell +
  \sum_{\ell'=\ell}^{q-1} \chi_{\ell'}$.
\item Finally, shift $A^\ell_\mu \rightarrow A^\ell_\mu -
\Delta_\mu
  \vartheta_\ell$.
\end{enumerate}
At this point, all explicit $\chi_\ell$ dependence will be
eliminated from the action, and the integral over $\chi_\ell$ can
be done to give an unimportant constant multiplier.  The rescaled
partition function is then
\begin{eqnarray}
  \label{eq:qbospf4}
&& \hspace{-0.4in}  {\cal Z}  =  \int\! dA^\ell d\vartheta_\ell\,
  e^{-{\cal S}^4_q},
\end{eqnarray}
with
\begin{eqnarray}
  \label{eq:Sq4}
&&   {\cal S}^4_q  =   \sum_{i\ell}
  \frac{1}{8\pi^2 e^2}\left|(\Delta\times
  A^\ell)_i-2\pi\tilde{\mu}\left(m_{i\ell}\!-\!\frac{1}{q}\right)\hat{\tau} \right|^2
  \nonumber \\ && + \sum_{a\ell}
  \frac{\tilde{e}^2}{2} |A_\mu^\ell- A_\mu^{\ell+1} +\breve{\cal
    B}_\mu^\ell|^2  \\
  && - \sum_{a\ell} \left[y_{qv} \cos(\Delta_\mu
  \vartheta_\ell-A_{\mu}^\ell) +   \cos
  (\vartheta_{\ell}-\vartheta_{\ell+1} - \eta_\ell)\right].\nonumber
\end{eqnarray}
We have nearly achieved the desired dual form.  Indeed, due to the
$\tilde{e}^2$ ``Higgs'' term, the only gapless gauge fluctuations
are those with $A^\ell = A$ equal for all $\ell$.  The relative
fluctuations of the different $A^\ell$ fields will be small and
can be neglected. There is in general, however, a mean value
(average over configurations) for each of the $A^\ell$ fields, or
more importantly the fluxes $\Delta\times A^\ell$.  This mean
value is oscillatory and has zero spatial average, since both the
$m_{i\ell}-1/q$ term has zero spatial average as does $\Delta
\times \breve{\cal B}^\ell= 2\pi \epsilon_{i\ell}\hat{\tau}$.
Once can show by a consideration of the vortex band structure
that, {\sl generically}, the uniform vortex modes are essentially
unaffected by this spatially oscillating flux with zero average.
A careful inspection shows, however, that the staggered potential
$\mu_s$ is crucial here.  For the non-generic case $\mu_s=0$, so
that $\tilde\mu=p/2$, one can show that the average flux is
\begin{equation}
  \label{eq:mus0}
  \langle (\Delta \times A^\ell)_i \rangle = 2\pi p \left(m_{i\ell} -
  \frac{1}{q}\right), \qquad \mbox{for $\mu_s=0$}.
\end{equation}
Although this has zero spatial average, this zero value is deceiving,
since the background flux $\Delta\times A^\ell$  differs on
different plaquettes by a multiple of $2\pi$.  Since the vortex
kinetic energy is $2\pi$-periodic, such a ``zero average'' flux
has the effect of a uniform one, and drastically modifies the
vortex spectrum.  In the generic situation with $\mu_s\neq 0$,
however, the differences in fluxes between plaquettes is truly
non-trivial, and it is correct to neglect this oscillating average
flux.

In this case, we may take the continuum limit by setting $A^\ell
\approx A$, $e^{i\vartheta_\ell} \rightarrow
|\overline\varphi|^{-1} \zeta_\ell$, adding potential
terms that keep the magnitude of $\zeta_\ell$
approximately constant, and assuming slow spatial variations of
the $\zeta_\ell$ fields.  One obtains, after
coarse-graining, an effective potential of the same form as
Eq.~(\ref{eq:vpot}) in the main text.


\begin{thebibliography}{99}
\bibitem{arbfoot} In fact, any of these operators can be composed with
  a $U(1)$ gauge transformation without any physical consequences.  If
  we wish the gauge field to transform normally under the space-group
  operations of the PSG, this freedom reduces to the option to compose
  each of the generators with a global $U(1)$ rotation.  This leaves
  Eq.~(\ref{e7}) unchanged, while Eqs.~(\ref{rtr}) need only be satisfied
  up to an arbitrary phase factor.  It is sometimes convenient to use
  this freedom to simplify explicit forms of the representation
  matrices.  Of course, any such form can readily be transformed by
  phase rotations into a solution of Eqs.~(\ref{rtr}).
\end{thebibliography}
\end{document}